\documentclass[11pt,a4paper]{article}
\pdfoutput=1
\usepackage{amsmath,amsfonts,latexsym,amssymb,hhline,stmaryrd,color,verbatim,graphicx,epstopdf,slashed,multirow}
\usepackage{jheppub}
\usepackage{slashed}
\usepackage{lineno}
\modulolinenumbers[5]
\usepackage{color}
\usepackage[utf8]{inputenc}
\usepackage{booktabs}
\usepackage{subfig}
\usepackage{pifont}
\newcommand{\cmark}{\text{\ding{51}}}

\newcommand\sss{\scriptscriptstyle}

\newcommand {\beq} {\begin{equation}}
\newcommand {\eeq} {\end{equation}}
\newcommand {\bea} {\begin{eqnarray}}
\newcommand {\eea} {\end{eqnarray}}

\definecolor{darkred}{rgb}{0.7, 0.0, 0.0}

\def\be{\begin{equation}}
\def\ee{\end{equation}}
\def\bsp#1\esp{\begin{split}#1\end{split}}

\newcommand{\OO}{\ensuremath{\mathcal{O}}}
\newcommand{\pdp}{\ensuremath{\phi^\dagger\phi}}
\renewcommand{\phi}{\ensuremath{\varphi}}

\newcommand{\bpm}{\begin{pmatrix}}      
\newcommand{\epm}{\end{pmatrix}}

\newcommand\mw{m_{\sss W}}
\newcommand\mz{m_{\sss Z}}

\newcommand\mt{m_{\sss t}}
\newcommand\GF{G_{\sss F}}

\newcommand{\Op}[1]{\OO_{\sss #1}}
\newcommand{\Opp}[2]{\OO_{\sss #1}^{\sss #2}}

\newcommand{\red}[1]{ \textcolor{red}{#1} }
 \def\lra#1{\overset{\text{\scriptsize$\leftrightarrow$}}{#1}}
\newcommand{\gth}{g_{\sss th}}
\newcommand{\gwh}{g_{\sss Wh}}
\newcommand{\gzh}{g_{\sss Zh}}
\newcommand{\gztr}{g^{\sss Z}_{t_{\sss R}}}
\newcommand{\gztl}{g^{\sss Z}_{t_{\sss L}}}
\newcommand{\gzbr}{g^{\sss Z}_{b_{\sss R}}}
\newcommand{\gzbl}{g^{\sss Z}_{b_{\sss L}}}
\newcommand{\gbtw}{g_{\sss btW}}
\newcommand{\gta}{g_{\sss t\gamma}}
\newcommand{\gwa}{g_{\sss W\gamma}}
\newcommand{\gwz}{g_{\sss WZ}}
\usepackage{array}
\newcolumntype{P}[1]{>{\centering\arraybackslash}p{#1}}

\preprint{CP3-19-16, MCNET-19-12} 

\title{Top-quark electroweak interactions at high energy}

\author[1,2]{Fabio Maltoni}
\author[1]{\!\!, Luca Mantani}
\author[1]{\!\!, Ken Mimasu}

\emailAdd{fabio.maltoni@uclouvain.be, fabio.maltoni@unibo.it}
\emailAdd{luca.mantani@uclouvain.be}
\emailAdd{ken.mimasu@uclouvain.be}
\affiliation[1]{Centre for Cosmology, Particle Physics and Phenomenology (CP3), Universit\'e catholique de Louvain, B-1348 Louvain-la-Neuve, Belgium}
\affiliation[2]{Dipartimento di Fisica e Astronomia, Universit\`a di Bologna \\and INFN, Sezione di Bologna, 
via Irnerio 46, 40126 Bologna, Italy}

\abstract{ 
Modified interactions in the electroweak sector may lead to scattering amplitudes that grow with energy compared to their Standard Model (SM) counterparts. We present a detailed study of all $2\to2$ scattering amplitudes involving at least one top quark and a pair of EW bosons. We analyse the high energy behaviour of the amplitudes using the Standard Model Effective Field Theory (SMEFT) to parametrise the departures from the SM. We discuss the origin of the energy growth that arise from effective contact interactions by appealing to the Goldstone equivalence theorem and find that the amplitudes obey expected patterns of (non-)interference. The results are connected to unitary-violating behaviour in the framework of anomalous SM interactions. Therein, we identify the appearance of additional growth due to the violation of $SU(2)$ gauge symmetry that leads to substantial differences between the SMEFT and the anomalous couplings approaches. We also discuss the embeddings of the scattering amplitudes into physical collider processes, presenting the parametric SMEFT sensitivity to relevant top quark operators and paying special attention to the extent to which the high energy behaviour of the $2\to2$ amplitude is retained in the actual processes accessible at colliders. The effective $W$ approximation is exploited to gain analytical insight into the embeddings of the $2\to2$ helicity amplitudes. Finally, we provide a compendium of processes detailing numerous directions in which the SMEFT parameter space can be accessed through high energy top quark processes in current and future colliders.
}

\begin{document}
\noindent\today
\maketitle
\section{Introduction \label{sec:intro}}

One of the most fascinating aspects of spontaneously broken, non-Abelian gauge-Yukawa theories, such as the Standard Model (SM), is how the high-energy behaviour of scattering amplitudes seems to magically arise from a set of intricate cancellations between contributions that would otherwise display unacceptable energy growth. Such cancellations, of course, are nothing else than the consequence of the gauge symmetries, which -- while hidden when the low-energy physical degrees of freedom are used as asymptotic states -- constrain the theory.  For example, in the broken phase of the SM, amplitudes involving massive gauge bosons and fermions as external states, might grow with unitarity-violating behaviour if the coupling to the Higgs boson is not included.  The most well known example of such behaviour is the scattering of longitudinally polarised of $W$-bosons, $W_{\sss L}W_{\sss L}\to W_{\sss L}W_{\sss L}$~\cite{LlewellynSmith:1973yud,Lee:1977eg,Lee:1977yc}. Here, the most extreme energy growth occurs in the case where the gauge symmetry is not respected in the self-interaction contributions to the amplitude. If one does not correctly include the quartic gauge interaction as a consequence of the non-Abelian nature of the gauge theory, the amplitude grows with the fourth power of the energy, $E$. When only interactions respecting the gauge symmetry are used, the amplitude continues to display a reduced, but still unacceptable $E^2$ growth. Only with the complete inclusion of the consistent gauge and Higgs interactions does the amplitude cease to grow with energy. This is a generic feature of the Higgs mechanism, which provides a consistent way of generating masses without explicitly violating gauge symmetries. Albeit with a different pattern of cancellations, a similar mechanism is in place for helicity violating amplitudes featuring massive fermions, such as $\bar f_{\sss R} f_{\sss L}\to W_{\sss L}W_{\sss L}$ which, in absence of a Higgs particle exchange, grow linearly with energy $E$ (and proportional to the fermion mass).

More generally, there are many scattering amplitudes which can be found to display neat cancellations as a consequence of gauge symmetry and/or the relationship between the mass of a certain particle and its coupling with the Higgs boson. One can actually exploit the existence of such amplitudes involving massive fermions and gauge/Higgs bosons to investigate how beyond the SM contributions to EW interactions would lead to unitarity violating effects. A particularly interesting class of these effects are proportional to the fermion masses. 
The $2\to 2$ scattering process, $\bar t_{\sss R} t_{\sss L}\to W_{\sss L}W_{\sss L}$, without the Higgs exchange, leads to a violation of unitarity at a scale of the order  $4 \pi v^2/m_t$. This is basis of the famous, yet chimeric, Appelquist-Chanowitz upper bound on the scale of the top-quark mass generation mechanism of about 3 TeV~\cite{Appelquist:1987cf}. A more careful analysis of the Appelquist-Chanowitz argument involves  $\bar t_{\sss R} t_{\sss L}\to n W_{\sss L}$~\cite{Maltoni:2001dc} and gives parametrically lower bounds, converging to $4\pi v$, \emph{i.e.} to the same bound as obtained by $WW \to WW$ scattering for large $n$. However, the Appelquist-Chanowitz bound is parametrically correct for massive Majorana neutrinos, leading to the same unitarity violation as that identified by the Weinberg operator $(\bar L^T\epsilon \phi) C (\phi^T\epsilon L)/\Lambda$~\cite{Maltoni:2000iq}, \emph{i.e.} at energies of order $10^{14}$ GeV, given the experimental lower bounds on the neutrino masses. 
In general, it can be shown that  for any given process unitarity bounds involving fermions scale with inverse powers of the mass and are therefore more useful when the masses involved are large. In this sense, being the most massive particle of the SM, the top quark is the best probe at our disposal for searching for new physics at the TeV scales.  

Top-quark gauge-Yukawa interactions have started to be systematically explored at the LHC. Very recently the observation of $t\bar{t}H$ production at the LHC~\cite{CMS:2018rbc,Sirunyan:2018mvw,Sirunyan:2018shy,Aaboud:2017jvq,Aaboud:2017rss} has confirmed the SM expectation that the top quark couples to the Higgs with a Yukawa coupling of order one. While indirectly expected by the measurements of gluon-fusion Higgs cross sections, the agreement with the SM expectations, has proven the essence of the SM mass generation mechanism and opened the way to a global and  precise determination of EW top-quark couplings. In fact, to date, the EW interactions of the top quark are not very precisely known. The coupling to the $W$-boson is the best measured through its decay into $bW$ while the neutral current interactions are only constrained indirectly or via relatively rare production processes in association with a $Z$-boson or a photon. 

It is therefore mandatory to understand how to best constrain the top-quark EW couplings with the goal to detect deviations which could hint to New Physics at the LHC. A natural strategy is to look for deviations from the precise structure of the SM predictions that may lead to the aforementioned anomalous energy growths in the amplitudes. Such non-unitary behaviour in top quark scattering processes could be observed at high energy collider experiments and would indicate a limited range of validity of the SM description, implying the presence of new physics at higher scales. The Standard Model Effective Field Theory (SMEFT) is therefore an appropriate framework to study these effects. It describes deviations from SM interactions and the associated energy growth of scattering amplitudes with a minimal set of high-scale assumptions.  This picture is rooted in a gauge invariant description of modified interactions through higher dimensional operators that preserve the underlying symmetries of the SM, and offers an additional advantage of being mappable to a large class of theories Beyond the SM (BSM). The higher dimensional operators lead to modified SM vertices as well as the appearance of new Lorentz structures, both of which can introduce energy growth in scattering processes. Many LHC searches for deviations from SM interactions are therefore framed in the context of the SMEFT. However, the additional, well-motivated, structure of the theory due to gauge invariance and its linear realisation, 
means that it does not map to the most general anomalous coupling Lagrangian and may not produce all possible unitarity-violating behaviours. As we will explicitly review, the requirement of gauge invariance implements automatically the possibility that the scale of new physics is well above the EW breaking scale $v$, something that present data seem to indicate.  

In this study, we interest ourselves in the high energy behaviour of a general class of EW scattering amplitudes involving a pair of fermions including at least one top quark and two bosonic EW states, {\it i.e.}, an EW gauge boson or the Higgs boson. We investigate, in depth, the possible sources of unitarity violating behaviour in these amplitudes in the SMEFT framework and contrast this with a general anomalous couplings description. The two complementary approaches allow us to shed light on the various sources of unitarity-violating behaviour, \emph{i.e.}, whether they arise from the spoiling of a cancellation in the SM , are a `pure' higher-dimension effect or whether these two descriptions are simply two facets of the same phenomenon. We then consider the ability of current and future colliders to probe these scattering amplitudes when embedded into physical EW top quark production processes. Two such scattering processes were recently studied in~\cite{Degrande:2018fog}, in the context of single top production in association with either a $Z$ or a Higgs boson. Therein, sources of energy growth were identified in the physical process which could be understood in terms of the $bW\to tZ$ and $bW\to th$ sub-amplitudes. The behaviour for the corresponding operators is not present in other EW top processes such as single-top production or $t\bar{t}Z$ which are not sensitive to the top/EW scattering amplitudes, therefore providing unique constraining potential in the Wilson coefficient space of EW top interactions.

The amplitudes that we consider have also previously been studied in the pure massless limit in the context of deriving unitarity constraints on the SMEFT operators~\cite{Corbett:2014ora,Corbett:2017qgl}. The potential for investigating non-standard top quark interactions through energy growth in these types of scatterings was first discussed in detail in~\cite{Dror:2015nkp}. Therein, the leading high energy behaviour of the fully longitudinal configurations for five of the scatterings that we study in this work are presented. Several of the collider processes that we study are also discussed as candidates for probing these scatterings. The results are then applied to a concrete, phenomenological study of the $t\bar{t}Wj$ process at the LHC, aiming to constrain anomalous top-$Z$ couplings. Our study proceeds in a similar spirit, extending the set of amplitudes and the types of modified interactions considered. 

We consider here the complete set of relevant SMEFT operators dictated by a particular flavour symmetry assumption, computing all helicity configurations and retaining finite EW mass effects. In many cases, the fully longitudinal modes are not the configurations with the largest energy growth and additional unitarity violating behaviours can be uncovered when not neglecting EW scale masses. By also considering the corresponding high-energy behaviour of the SM sub-amplitudes, we can also determine whether a given operator leads to energy growth at linear (interference) or quadratic (square) level in the matrix element (linear growth, of course, implies a quadratic one). This is interesting in view of the recent discussion of helicity selection rules and non-interference in certain $2\to2$ scattering amplitudes involving insertions of dimension-6 operators~\cite{Cheung:2015aba,Azatov:2016sqh}. Our study aims to provide an overview of the potential to probe high energy top-quark scattering processes at colliders. Detailed phenomenological analyses are not performed, with a broader, horizontal approach taken. Rather, we survey a considerable number of scatterings and associated collider processes focusing on quantifying the high energy behaviour, the associated sensitivity to SMEFT operators and discussing several general phenomenological and experimental issues.

The presentation is organised as follows. In Section~\ref{sec:newint}, we review the conventions and flavor assumptions that used in our SMEFT formalism. We then identify the degrees of freedom that we investigate in top-EW scattering processes and summarise existing constraints in the parameter space. Finally, we define the Anomalous Couplings (AC) Lagrangian that we use to parametrise deviations from SM interactions. Section~\ref{sec:highenergytops} presents a detailed discussion of energy growth in top-EW scattering amplitudes scattering amplitudes. We review our calculation methods, discuss the origin of energy growth with regards to interference terms, contact interactions and gauge invariance. We also employ the Equivalent W Approximation (EWA) to investigate the mapping of the high energy behaviour from $2\to2$ scattering amplitudes to a $2\to3$ collider process. The section concludes with a blueprint for the subsequent analysis of the individual scattering amplitudes and the collider processes that may be used to probe their high energy behaviour. Sections~\ref{sec:singletop_scattering} and~\ref{sec:twotop_scattering} present the results of our detailed survey of scattering amplitudes and collider processes. We summarise our findings in Section~\ref{sec:summary} and finally conclude in Section~\ref{sec:conclusions}.

\section{New interactions in the Electroweak symmetry breaking sector \label{sec:newint}}
    \subsection{Standard Model Effective Field Theory\label{subsec:SMEFT}}
We focus primarily on the SMEFT framework to describe BSM interactions between the top quark, the Higgs and the EW gauge bosons, employing the so called Warsaw basis of 
operators~\cite{Buchmuller:1985jz,Grzadkowski:2010es} truncated at dimension six. In order to limit ourselves to the 
interactions of the EW sector and the top quark we impose a flavour symmetry,
 $U(3)_{\ell}\times U(3)_{e}\times U(3)_{d}\times U(2)_{q}\times U(2)_{u}$, on 
 our effective theory such that operators concerning deviations from top/third 
 generation quark interactions can be singled out (see~\cite{AguilarSaavedra:2018nen} for a detailed classification). The labels $\ell,e,d,q,u$  refer to the fermionic representations of the SM: the lepton doublet, right 
 handed lepton, right handed down-type quark, quark doublet and right handed 
 up-type quark, respectively. This symmetry, for example, only allows a SM 
 Yukawa interaction for the top quark, leaving all others massless. The symmetry also 
 implies a flavour-universality for operators involving vector fermion currents, 
 apart from those of the 3rd generation quark doublet and right handed top 
 quark whose corresponding operators can have independent coefficients. One is then left with the SMEFT operators, listed in Table~\ref{tab:operators}, that describe 
 deviations from SM interactions of the EW sector coupled to the top quark.
\begin{table}
{\centering
\renewcommand{\arraystretch}{1.4}
\begin{tabular}{|ll|ll|}
    \hline
     $\Op{W}$&
     $\varepsilon_{\sss IJK}\,W^{\sss I}_{\mu\nu}\,
                             {W^{{\sss J},}}^{\nu\rho}\,
                             {W^{{\sss K},}}^{\mu}_{\rho}$&
     $\Op{t\phi}$&
     $\left(\pdp-\tfrac{v^2}{2}\right)
     \bar{Q}\,t\,\tilde{\phi} + \text{h.c.}$
     \tabularnewline
     $\Op{\phi W}$&
     $\left(\pdp-\tfrac{v^2}{2}\right)W^{\mu\nu}_{\sss I}\,
                                    W_{\mu\nu}^{\sss I}$&
     $\Op{tW}$&
     $i\big(\bar{Q}\sigma^{\mu\nu}\,\tau_{\sss I}\,t\big)\,
     \tilde{\phi}\,W^I_{\mu\nu}
     + \text{h.c.}$
     \tabularnewline
     $\Op{\phi B}$&
     $\left(\pdp-\tfrac{v^2}{2}\right)B^{\mu\nu}\,
                                    B_{\mu\nu}$&
     $\Op{tB}$&
     $i\big(\bar{Q}\sigma^{\mu\nu}\,t\big)
     \,\tilde{\phi}\,B_{\mu\nu}
     + \text{h.c.}$
     \tabularnewline
     \cline{3-4}
     $\Op{\phi WB}$&
     $(\phi^\dagger \tau_{\sss I}\phi)\,B^{\mu\nu}W_{\mu\nu}^{\sss I}\,$&
     $\Op{\phi Q}^{\sss(3)}$&
     $i\big(\phi^\dagger\lra{D}_\mu\,\tau_{\sss I}\phi\big)
     \big(\bar{Q}\,\gamma^\mu\,\tau^{\sss I}Q\big)$
     \tabularnewline

     $\Op{\phi D}$&
     $(\phi^\dagger D^\mu\phi)^\dagger(\phi^\dagger D_\mu\phi)$&
     $\Op{\phi Q}^{\sss(1)}$&
     $i\big(\phi^\dagger\lra{D}_\mu\,\phi\big)
     \big(\bar{Q}\,\gamma^\mu\,Q\big)$
     \tabularnewline
     $\Op{\phi \square}$&
     $(\varphi^\dagger\varphi)\square(\varphi^\dagger\varphi)$ &
     $\Op{\phi t}$&
     $i\big(\phi^\dagger\lra{D}_\mu\,\phi\big)
     \big(\bar{t}\,\gamma^\mu\,t\big)$
      \tabularnewline
      &&     
      $\Op{\phi tb}$&
     $i\big(\tilde{\phi}^\dagger\,{D}_\mu\,\phi\big)
     \big(\bar{t}\,\gamma^\mu\,b\big)
     + \text{h.c.}$
     \tabularnewline
      \hline
      %

     %

 
  \end{tabular}

\caption{\label{tab:operators}
SMEFT operators describing new interactions involving the EW and top quark 
sectors, consistent with a $U(3)^3\times U(2)^2$ flavour symmetry.  
$Q,\,t$ and $b$ denote the third generation components of $q,\,u$ and $d$.
}
}
 \end{table}
 The following conventions are used:
 \begin{align}
   \phi^\dag {\overleftrightarrow D}_\mu \phi&=\phi^\dag D^\mu\phi-(D_\mu\phi)^\dag\phi\\
   \phi^\dag \tau_{\sss K} {\overleftrightarrow D}^\mu \phi&=
   \phi^\dag \tau_{\sss K}D^\mu\phi-(D^\mu\phi)^\dag \tau_{\sss K}\phi \\
   W^{\sss K}_{\mu\nu} &= \partial_\mu W^{\sss K}_\nu 
   - \partial_\nu W^{\sss K}_\mu 
   + g \epsilon_{\sss IJ}{}^{\sss K} \ W^{\sss I}_\mu W^{\sss J}_\nu\\
   B_{\mu\nu} &= \partial_\mu B_\nu - \partial_\nu B_\mu \\
   D_\mu\phi =& \left(\partial_\mu -  i \frac{g}{2} \tau_{\sss K} W_\mu^{\sss K} - i\frac12 g^\prime B_\mu\right)\phi
 \end{align}
 where $\tau_I$ are the Pauli sigma matrices.
$\Op{\phi tb}$ is retained due to its unique right handed charged 
current structure, even though it technically breaks $U(3)_{d}$ down to $U(2)_{d}$. 
This set of operators will be sufficient to describe the high-energy behaviour of 
the $2\to2$ scattering amplitudes under investigation. Some of the operators in question lead to non-canonical kinetic terms for the 
Gauge and Higgs boson after EW symmetry breaking. These effects are absorbed by 
appropriate field redefinitions and propagated into physical quantities employing
the $\mw, \GF, \mz$ input scheme. The Universal FeynRules Output (UFO) and 
{\sc FeynArts} models are based on a common {\sc FeynRules} implementation also 
used in~\cite{Degrande:2018fog}. A public version has recently been made available~\cite{SMEFTatNLO}. 

Our operator set leads to modifications of SM top-quark 
interactions with all EW gauge bosons apart from the photon, whose interactions 
are protected by $U(1)_{\sss EM}$ being a good symmetry of the low energy theory. 
Additionally, weak dipole interactions are also included for the top quark which 
do induce a modified $t\bar{t}\gamma$ vertex. 
On the bosonic side, our operator choice leads to shifts in triple and quartic gauge couplings 
as well as gauge-Higgs interactions along with new Lorentz structures for each 
of these. A general feature of the SMEFT description is that individual operators 
contain more than one interaction vertex. For example, operators with multiple Higgs fields will source several, correlated, interaction vertices with different numbers of Higgs bosons. Operators containing the $SU(2)$ gauge field strength will also lead to vertices with different numbers of gauge fields from its Abelian and non-Abelian components.
Operators that modify 3-point interactions, for example, are always accompanied 
by higher point interactions. This is, of course, a consequence of the hidden, EW gauge symmetry.
The SMEFT therefore crucially differs from  a general anomalous couplings description in predicting correlations between various anomalous couplings fixed by the underlying gauge invariance. In the context 
of our $2\to2$ amplitude study, many operators will affect multiple interaction 
vertices relevant to a particular scattering process, and also contribute 
directly via 4-point contact interactions. These contact terms are often the 
source of maximal energy growth. The de-correlation of the 3-point vertices 
with the associated contact terms is impossible in the dimension-6 truncated 
SMEFT and would necessitate the inclusion of higher dimensional operators. One is therefore able to connect modifications of, e.g., SM vertices with higher-point interactions that would affect new processes and/or induce new energy growth that we can search for at colliders.

The set of operators included in our study is motivated by the possibility of using a class of scattering processes to provide new and phenomenologically interesting sensitivity to the parameter space of SMEFT. To this end, we consider a complete set of these -- given a particular flavor symmetry assumption -- that contribute to the $2\to2$ amplitudes of interest. However, in doing so we also look at some realistic collider processes in which these amplitudes can be embedded. In these cases, it is possible that a further set of operators could contribute to the full process, albeit not via the embedding into the $2\to2$ amplitudes. In particular, operators that modify interactions of the gluon such as the top quark chromomagnetic operator, $\Op{tG}$, the triple gluon operators, $\Op{G}$ or the gluon-Higgs operator, $\Op{\phi G}$,
\begin{align}
    \Op{tG} & = i\big(\bar{Q}\sigma^{\mu\nu}\,T_{\sss A}\,t\big)
     \,\tilde{\phi}\,G^{\sss A}_{\mu\nu}
     + \text{h.c.},\\
     \Op{G} & = g_{\sss S}f_{\sss ABC}\,G^{\sss A}_{\mu\nu}\,
                             {G^{{\sss B},}}^{\nu\rho}\,
                             {G^{{\sss C},}}^{\mu}_{\rho},\\
     \Op{\phi G} & = \left(\pdp-\tfrac{v^2}{2}\right)G^{\mu\nu}_{\sss A}\,
                                             G_{\mu\nu}^{\sss A},
\end{align}
come to mind, along with a number of four-fermion operators involving a mixture of heavy and light quarks. These can affect several processes that are either gluon-initiated, have jets in the final state or involve a pair of top quarks in the final state.

Often, these contributions affect the QCD-induced components of a given process, as opposed to those that actually embed the EW top-quark scattering amplitudes. While quantifying their importance would be important in a global fit exercise, the contributions of these operators to the processes that we consider are not the focus of this sensitivity study, which targets the high energy behaviour EW scattering amplitudes. Moreover, it is likely that other, non-EW processes such as $t\bar{t}$, single-Higgs and multi-jet production would provide better constraints in these directions of parameter space. Finally, many of the aforementioned contributions will favour different regions of phase space of the full processes as those from the EW operators. For example, consider processes that include the emission of a virtual EW gauge boson from a forward jet, such as $tWj$. Not only would an additional jet coming from an insertion of $\Op{G}$ not have the same kinematical features as the VBF-like jet, the high energy region of the embedded $b\,W\to t\,Z$ sub-amplitude also does not probe large momentum flow through the $ggg$ vertex. As such, we expect that due to the comparatively strong constraints from other processes and the non-overlap of interesting phase space, these types of contributions should be subdominant in the collider processes that we study. Ultimately, only dedicated studies will confirm this expectation, which very much depends also on the relative size of the EW and QCD induced components. We leave these for future work. 
\subsubsection{Constraints on dimension-6 operators\label{subsec:constraints}}

In Table~\ref{tab:constraints} we collect some existing constraints on the set of operators of interest to this study. Those not explicitly containing the top quark field can be best constrained by a combination of EW precision measurements, diboson and Higgs production processes. Those featuring a top quark are currently constrained by inclusive and differential top quark processes such as $t\bar{t}$, single top, $W$ helicity fractions and $t\bar{t}+Z/W$. One can observe a significant hierarchy between the constraints on these two classes of operators in the standard Warsaw normalisation. This suggests that the top quark processes studied herein are likely to contribute most meaningfully to the limits on the latter set of operators. 

\setlength{\tabcolsep}{4pt}
\renewcommand{\arraystretch}{1.4}
\begin{table}[h!]
\begin{center} 
{\footnotesize
\begin{tabular}{|c|c|c|c|c|c|}
\hline
\multirow{2}{*}{Operator} & \multicolumn{2}{c|}{Limit on $c_i$ $\big[$TeV$^{-2}$$\big]$} & 
\multirow{2}{*}{Operator} & \multicolumn{2}{c|}{Limit on $c_i$ $\big[$TeV$^{-2}$$\big]$}   
\tabularnewline\cline{2-3}\cline{5-6}
{}                  & Individual    & Marginalised &   
{}                  & Individual    & Marginalised  
\tabularnewline\hline
 $\Op{\phi D}$            & [-0.021,0.0055]~\cite{Ellis:2018gqa}  & [-0.45,0.50]~\cite{Ellis:2018gqa}     &   
 $\Op{t \phi}$            & [-5.3,1.6]~\cite{Hartland:2019bjb}    & [-60,10]~\cite{Hartland:2019bjb}
\tabularnewline\hline
 $\Op{\phi \Box}$         & [-0.78,1.44]~\cite{Ellis:2018gqa}     & [-1.24,16.2]~\cite{Ellis:2018gqa}     &   
 $\Op{tB}$                & [-7.09,4.68]~\cite{Buckley:2015lku}   & $-$
\tabularnewline\hline
 $\Op{\phi B}$            & [-0.0033,0.0031]~\cite{Ellis:2018gqa} & [-0.13,0.21]~\cite{Ellis:2018gqa}     &   
 $\Op{tW}$                & [-0.4,0.2]~\cite{Hartland:2019bjb}   & [-1.8,0.9]~\cite{Hartland:2019bjb}
\tabularnewline\hline
 $\Op{\phi W}$            & [-0.0093,0.011]~\cite{Ellis:2018gqa}  & [-0.50,0.40]~\cite{Ellis:2018gqa}     &   
 $\Op{\phi Q}^{\sss (1)}$ & [-3.10,3.10]~\cite{Buckley:2015lku}   & $-$
\tabularnewline\hline
 $\Op{\phi WB}$           & [-0.0051,0.0020]~\cite{Ellis:2018gqa} & [-0.17,0.33]~\cite{Ellis:2018gqa}     &   
 $\Op{\phi Q}^{\sss (3)}$ & [-0.9,0.6]~\cite{Hartland:2019bjb}   & [-5.5,5.8]~\cite{Hartland:2019bjb}
\tabularnewline\hline
 $\Op{W}$                 & [-0.18,0.18]~\cite{Butter:2016cvz}    & $-$                                   &
 $\Op{\phi t}$            & [-6.4,7.3]~\cite{Hartland:2019bjb}   & [-13,18]~\cite{Hartland:2019bjb}
\tabularnewline\hline
 {}                       & {}                                    & {}                                    &  
 $\Op{\phi tb}$           & [-5.28,5.28]~\cite{Alioli:2017ces}    & [27,8.7]~\cite{Hartland:2019bjb}
\tabularnewline\hline
\end{tabular}
}

\end{center}
\caption{\label{tab:constraints}
Individual and marginalised 95\% confidence intervals on Wilson coefficients 
collected from a selection of global fits to Higgs, top and EW gauge boson data.}
\end{table}
\renewcommand{\arraystretch}{1.}

\subsection{Anomalous couplings\label{subsec:AC}}
Although our main focus is on the SMEFT framework, it is sometimes instructive to also consider a general anomalous couplings (AC) Lagrangian to help understand the effects of the dimension-6 operators. The notion of unitarity cancellations present in the SM arises from this description. Since a given SMEFT operator will generally contribute to many new interaction vertices, including 4-point contact interactions, it is not always obvious to identify the exact sources of energy growth. We therefore employ a general parametrisation of the Lorentz structures present in the SM that are relevant to the scattering processes of interest. 
\begin{align}
    \label{eq:L_AC}
    \nonumber
    \mathcal{L} \supset &-\gth \, \bar{t} \, t \, h + \gwh \, W^\mu W_\mu \, h 
    + \gzh \, Z^\mu  Z_\mu \, h 
    + \gbtw \,( \bar{t} \, \gamma^\mu \, P_L \, b \, W_\mu + \text{h.c})
    \\\nonumber
    &+ \bar{t} \, \gamma^\mu (\gztr \, P_R + \gztl \, P_L) \, t \, Z_\mu 
    + \bar{b} \, \gamma^\mu (\gzbr \, P_R + \gzbl \, P_L) \, t \, Z_\mu 
    - \gta \, \bar{t} \, \gamma^\mu \, t \, A_\mu 
    \\
    &+ \gwa \, (W^\mu \, W^\nu \, \partial_\mu \, A_\nu + \text{perm.})
    + \gwz \, (W^\mu \, W^\nu \, \partial_\mu \, Z_\nu + \text{perm.}),
\end{align}
where the Standard Model values of these couplings are
\begin{equation}
    \label{eq:AC_SM}
    \begin{gathered}
    \gth = \frac{g \, \mt}{2 \, \mw} \, ,\quad
    \gwh = g \, \mw \, , \quad
    \gzh = \frac{g \, \mz}{\cos\theta_W} \, , \\
    \gztr = -\frac{2 \, g \, \sin^2\theta_W}{3 \, \cos\theta_W} \, , \quad
    \gztl = \frac{g}{\cos\theta_W}\left(\frac{1}{2} - \frac{2 \sin^2\theta_W}{3} \right) \, , \\
    \gzbr = \frac{g \, \sin^2\theta_W}{3 \, \cos\theta_W} \, ,\quad
    \gzbl = -\frac{g}{\cos\theta_W}\left(\frac{1}{2} - \frac{\sin^2\theta_W}{3} \right) \, , \\
    \gbtw= \frac{g}{\sqrt{2}} \, ,\quad 
    \gta = \frac{2}{3} \, g \, \sin\theta_W \, , \quad
    \gwa = g \, \sin\theta_W \, , \quad
    \gwz = g \, \cos\theta_W \, .
    \end{gathered}
\end{equation}

\section{High energy top-quark scattering at colliders\label{sec:highenergytops}}
With the general frameworks for modified interactions in the EWSB sector in hand, we proceed to analyse the scattering amplitudes of a generic $2 \to 2$ process $f \, B \to f^\prime \, B^\prime$, where $f,f^\prime = b, t$ and $B,B^\prime= h, W, (Z/\gamma)$ and at least one of the 2 fermions is a top quark. We identify and compute the helicity amplitudes for 10 such scatterings. These can be organised into four categories in terms of the number of top quark external legs and the presence or absence of Higgs bosons, as shown in Table~\ref{tab:amplitude_organisation}.
\begin{table}[h!]
\centering
\begin{tabular}{|r|p{4.5cm}|p{4.5cm}|}
\hline
&Single-top &Two-top ($t\bar{t}$) \tabularnewline\hline
w/o Higgs  &  $b \, W \to t \, (Z/\gamma)$ \hfill(\ref{subsubsec:bwtz_bwta})
& $t \, W \to t \, W$
\hfill (\ref{subsubsec:twtw})\newline 
$t \, (Z/\gamma) \to t \, (Z/\gamma)$ \hfill (\ref{subsubsec:tztz_tzta_tata}) 
\tabularnewline\hline
w/\phantom{o}  Higgs & $b \, W \to t \, h$ 
\hfill (\ref{subsubsec:bwth}) & 
$t \, (Z/\gamma) \to t \, h$ \hfill (\ref{subsubsec:tzth_tath})\newline  
$t \, h \to t \, h$ \hfill (\ref{subsubsec:thth})\tabularnewline\hline
\end{tabular}
\caption{The ten $2\to 2$ scattering amplitudes whose high-energy behaviour we study in this paper, labelled according to the section in which each category is discussed.
\label{tab:amplitude_organisation}}
\end{table}

\begin{figure}[h!]
\centering
\includegraphics[width=0.4\textwidth]{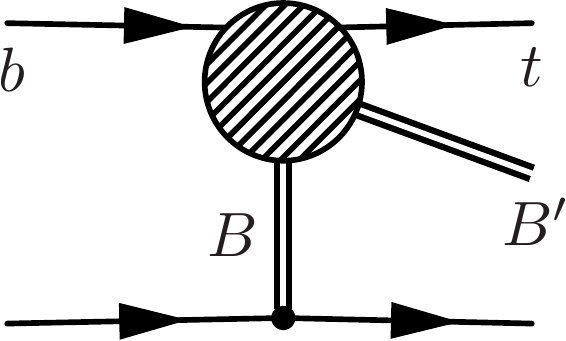}
\caption{\label{fig:topology}
Schematic Feynman diagram for the embedding of an EW top scattering amplitude into a physical, single-top process at a hadron collider. Here $f$ and $f^\prime$ must be a $b$- and $t$-quark respectively, while $B$ and $B^\prime$ can be several combinations of $Z,\,\gamma,\,W$ and $h$.
}
\end{figure}

Figure~\ref{fig:topology} schematically shows how such scattering amplitudes, a single-top one in this case, provide the building blocks for EW collider processes in order to study the high-energy behaviour of the EWSB sector. 
The high-energy limit of the amplitudes in the Mandelstam variables $s~\sim -t\gg v^2$ is taken for each helicity configuration, keeping sub-leading terms proportional to the EW masses, $\mw,\mz$ and $\mt$. We collect our results in Appendix~\ref{app:helamp_tables}, Tables~\ref{tab:bwtz} through~\ref{tab:thth}, summarising the SM prediction and the contribution of each operator from Table~\ref{tab:operators} that displays at least one configuration that grows with energy. As expected from dimension-6 operators, the maximum degree of growth of each amplitude is with $E^2$, while for the SM, they are at most constant with energy.

\subsection{Energy growth and interference \label{subsec:interference}}
At the cross-section level, the SMEFT contributes at leading order in the EFT expansion via the interference of the amplitude with the SM. Therefore, although an EFT amplitude may display energy growth, it is not guaranteed that the corresponding interference term will do the same. Take for example Table~\ref{tab:bwth}, which shows the helicity amplitudes for $b \, W^+\to t \, h$ scattering, previously reported in Ref.~\cite{Degrande:2018fog}. It shows that the current operator, $\Op{\phi Q}^{\sss (3)}$, has an energy-growing contribution to the $(-, 0, -)$ helicity configuration (left-handed $b$ and $t$, longitudinal $W$), proportional to $\sqrt{s(s+t)}$. Given that the SM is constant with energy in this configuration, the interference term in the matrix element squared will also grow with energy. This non-trivial behaviour is particularly interesting as the contribution is linear in the Wilson coefficient and is a leading effect in the EFT expansion. As there are no external transverse gauge bosons, the interference in the high-energy limit is consistent with the helicity selection rules of~\cite{Azatov:2016sqh}. We confirm the behaviour of all amplitudes is consistent with the selection rules insofar as the energy-growing interferences arise in fully longitudinal configurations. In all helicity configurations involving at least one transversely polarised gauge boson the SM amplitude decreases with energy to compensate the growth from the modified interactions such that the interference contribution to the matrix element is constant with energy.

The EFT-squared contribution to the matrix-element is, of course, guaranteed to grow with energy so long as the operator-inserted amplitude does. As constraints on the operators improve, these squared terms are expected to become increasingly small due to the intrinsic $(c_i/\Lambda^2)^2$ suppression. However, in the case of a suppressed interference with the SM at high energies, this hierarchy between the linear and quadratic contributions from the Wilson coefficients is affected beyond the naive power-counting expectation. The relationship between the validity of the EFT description and the relative impact of quadratic dimension-6 contributions is not so clear cut in these cases.

\subsection{Energy growth and contact terms
\label{subsec:contact_terms}}
The maximum energy growth of an amplitude can be guessed by looking at the contact terms generated by higher dimension operators. For a contact term of dimension-$K$, $\mathcal{O}_{\sss K}$, the corresponding coupling must have mass dimension $4-K$. 
\begin{align}
    \mathcal{L}\supset\frac{1}{M^{\sss K-4}}\mathcal{O}_{\sss K},
\end{align}
where $M$ is a generic parameter of mass dimension 1. The scattering amplitude for a $2\to N$ process has mass dimension $2-N$, meaning that the amplitude mediated by an insertion of $\mathcal{O}_{\sss K}$ has a maximum energy scaling that compensates its canonical dimension,
\begin{align}
    \mathcal{M}\propto\frac{1}{M^{K-4}}E^{K-N-2}.
\end{align}
For the specific case of $2\to2$ scattering, the amplitude scales as $E^{K-4}$, since it is dimensionless in 4-dimensional spacetime. However, if massive gauge bosons are involved, this applies only to their transverse polarisations. Since the longitudinal polarisation is related by the Goldstone equivalence theorem~\cite{Cornwall:1974km} to the derivative of a Goldstone boson field, $\frac{\partial_\mu G}{M_{\sss V}}$, with $M_{\sss V}$ the gauge boson mass, it effectively contributes like a dimension 2 degree of freedom. Furthermore, in the case of SMEFT, we know that certain contact interactions are obtained after EW symmetry breaking, via the insertion of the Higgs vacuum-expectation-value (vev), $v$. These consume some of the available mass dimension and therefore reduce the energy growth. A naive formula for the maximum energy dependence of a $2\to N$ scattering amplitude induced by a contact term from a dimension-$K$ operator is then
\begin{equation}
    \label{eq:naive_e_general}
    \mathcal{M} \propto \frac{v^m}{\Lambda^{K-4}} \frac{E^{K-N-m-2+n}}{M_{\sss V} ^n}.
\end{equation}
Here, $n$ is the number of longitudinally polarised massive vector bosons in the external legs, highlighting the fact that longitudinal modes bring additional growth with energy. The parameter $m$ denotes the number of vev-insertions needed to arrive at the contact term of interest. Specifying to $2\to2$ scattering via dimension-6 operators yields
\begin{equation}
    \label{eq:naive_e}
    \mathcal{M} \propto \frac{v^m}{\Lambda^{2}} \frac{E^{2-m+n}}{M_{\sss V} ^n}.
\end{equation}
Sub-leading energy growths can often be understood as helicity-flipped versions of the maximal ones where one `pays' a factor of the final state particle mass to access a different helicity configuration at the cost of a reduced energy growth. 

Maximal energy growth is obtained through the highest point contact interaction generated by a given operator, \emph{i.e.}, by minimising the number of vev-insertions. In the presence of Higgs fields, considering such contact terms in Feynman gauge allows for the estimate of energy growth in scatterings also involving longitudinally polarised gauge bosons. It is instructive to open up the building blocks of our operators involving the Higgs field in Feynman gauge. 
\begin{align}
    \phi = \frac{1}{\sqrt{2}}\begin{pmatrix}
        -iG^+\\
        v+h+i G^{\sss 0}
    \end{pmatrix},\quad
    \tilde{\phi}=i\tau_2\phi^\ast= \frac{1}{\sqrt{2}}\begin{pmatrix}
    v+h-i G^{\sss 0}\\
    iG^-
    \end{pmatrix},\quad
\end{align}
Since we are studying $2\to 2$ scattering, we seek two-point bosonic operators to couple to the fermion currents. The neutral vector current, $\bar{t}\gamma^\mu t$, can couple to
\begin{align}
    i\big(\phi^\dagger\lra{D}_\mu\tau_{\sss 3}\phi\big)&\supset
    (v+h)\lra{\partial}_\mu G^{\sss 0} - 
    2\mz h Z_\mu + 
    iG^-\lra{\partial}_\mu G^+,\\
    i\big(\phi^\dagger\lra{D}_\mu \phi \big)&\supset 
    G^{\sss 0}\lra{\partial}_\mu h + 2\mz h Z_\mu +\big(i G^-\partial_\mu G^+ + 2i\mw G^- W^+_\mu  + \text{ h.c.}\big)
\end{align}
and the neutral scalar current, $\bar{t}_Lt_R$, to
\begin{align}
    (\pdp) \,\tilde{\phi}_0 \supset \frac{v}{2\sqrt{2}}\left(3h^2 + G^{\sss 0}G^{\sss 0} + 2G^+G^- - 2 i G^{\sss 0} h\right).
\end{align}
Here, $\tilde{\phi}_0$, denotes the neutral component of the Higgs doublet.
For the right-handed charged current, $\bar{t}\gamma^\mu b$, we have
\begin{align}
    \nonumber
    i\sqrt{2}(\tilde{\phi}^\dagger\,D_\mu\phi)\supset&
    -(h+i G^{\sss 0})\lra{\partial}_\mu G^+
    -2\mw (h+iG^{\sss 0})\,W_\mu^+
    -2i\mz c^2_{\sss W} G^+Z_\mu\\
    \label{eq:phitDphi_expand}
    &-v(\partial_\mu-ieA_\mu) G^+,
\end{align}
and for the left,
\begin{align}
    \nonumber
    \frac{i}{\sqrt{2}}\big(\phi^\dagger\lra{D}_\mu\tau_{+}\phi\big)\supset &
    (h-i G^{\sss 0})\lra{\partial}_\mu G^+
    +2\mw h\,W^+_\mu
    +2i\mz s_{\sss W}^2 G^+Z_\mu\\
    \label{eq:phiD3phi_expand_charged}
    &+v(\partial_\mu-ieA_\mu) G^+,
\end{align}
where $\tau_{+}=\tau{\sss 1}-i \tau{\sss 2}$.
Finally, the scalar current, $\bar{t}_Rb_L$, couples to the charged component of the doublet,
\begin{align}
    (\pdp) \,\tilde{\phi}_2 \supset -i v h G^+.
\end{align}
Notice the appearance of dimension-5 contact terms involving photons in Equations~\eqref{eq:phiD3phi_expand_charged} and~\eqref{eq:phitDphi_expand}. These appear as part of a manifestly $U(1)_{\sss EM}$ gauge-invariant piece, $v(\partial_\mu-ieA_\mu) G^+\equiv vD_\mu^{\sss EM} G^+$. On their own, these contact interactions indicate the possibility of linear energy growth in scatterings involving a longitudinal, massive gauge boson and a photon. However, one can verify that the putative energy growing component of any such amplitude explicitly breaks $U(1)_{\sss EM}$ (it does not obey the QED Ward identities), and therefore cannot have any physical consequences since QED is a good symmetry of the low every theory. The remaining dimension-5 and -6 contact terms constructed out of the bosonic and fermionic building blocks point to the $E$ and $E^2$ growths induced by a given operator.

Taking another example from Table~\ref{tab:bwth}, we find E$^2$ growth in the $(+, 0, +)$ amplitude from the $\Op{\phi tb}$ operator. In unitary gauge, it contains an effective dimension-5 ($m=1$ in Equation~\eqref{eq:naive_e}) contact term
\begin{align}
    \Op{\phi tb} = i(\tilde{\varphi} D_\mu \varphi)(\bar{t} \gamma^\mu b) +\text{h.c.}\quad \to \quad v\,h \, W^{\sss +} \, \bar{t}_{\sss R} \gamma^\mu b_{\sss R} +\text{h.c.},
\end{align}
whose origin can be traced by Equation~\eqref{eq:phitDphi_expand}.
Although the interaction is of canonical dimension 5, the longitudinal external state ($n=1$) restores an additional power of energy growth and `cancels' the vev insertion. In Feynman gauge, the $E^2$ growth can immediately be seen from an effective dimension-6 contact term with a charged Goldstone boson, also present in Equation~\eqref{eq:phitDphi_expand},
\begin{equation}
    \Op{\phi tb} \quad \to \quad h \, \partial_\mu G^{\sss +} \, \bar{t}_{\sss R} \gamma^\mu b_{\sss R} +\text{h.c.}.
\end{equation}
Flipping the top or $W$ helicities, yields a $\sqrt{-t}m_t$ and $\sqrt{-t}m_W$ dependence in the $(+, 0, -)$ and $(+, +, +)$ configurations respectively.

Since in unitary gauge, the Goldstone modes are not manifest, the origin of energy growth is not always immediately clear. One may, for example, encounter growth in longitudinal configurations from the modifications of interactions of mass dimension 4 or less. These are conventionally interpreted as arising from the spoiling unitarity cancellations in the SM. The anomalous energy growth is then a consequence of modifying the SM relations among interactions fixed by gauge invariance and the Higgs mechanism. This is, for example, one way to understand the linear energy growth coming from the Yukawa operator $\Op{t \phi}$ in Table~\ref{tab:bwth}: by its modification of the relation between the top quark mass and the top-Higgs coupling of the SM. This is the SMEFT analogue of a modified top Yukawa interaction, $\gth$ in the AC framework (Equations~\eqref{eq:L_AC} and~\eqref{eq:AC_SM}). The key difference is the manifestly gauge invariant SMEFT construction that implies additional contact terms. As a consequence, the energy growth due to $\Op{t \phi}$ can again be understood through a dimension-5 contact term arising in Feynman gauge,
\begin{align}
   \Op{t\phi}=\left(\pdp-\tfrac{v^2}{2}\right)
     \bar{Q}\,t\,\tilde{\phi} + \text{h.c.} \quad \to \quad v\,\bar{b}_{\sss L}\,t_{\sss R}\,G^-h +\text{h.c.}.
\end{align}
Due to the 3 Higgs fields in this operator, we see that its contribution to $2\to2$ scatterings is limited to a linear energy growth. Rather, it will grow maximally for 5-point scattering amplitudes and associated processes that embed these~\cite{Henning:2018kys}.

In our study, the majority of operators involving the top quark that we consider lead to maximal growth already at the level of $2\to2$ scattering.  The bosonic operators, $\Op{\phi W}$, $\Op{\phi B}$, $\Op{\phi WB}$, $\Op{\phi W}$, $\Op{\phi D}$ and $\Op{\phi \Box}$ are included because they can contribute to the top quark scatterings that we consider albeit without the possibility for maximal energy growth. They involve at least two Higgs fields and will achieve maximal growth in purely bosonic scattering amplitudes. 

\subsection{Energy growth and gauge invariance\label{subsec:gauge_invariance}}
We have emphasised the fact that the SMEFT respects the spontaneous breaking of the linearly realised electroweak symmetry in the SM. Conversely, ACs do not automatically respect specific relations among couplings nor include additional contact terms that are automatically generated by the dimension-6 operators. As a consequence, we observe that general AC parametrisations can lead to stronger energy growth than predicted by our naive formula. The weak dipole operator, $\Op{tW}$, is a good example of this. In addition to modified dipole-type $t\bar{b}W^-$ and $t\bar{t}Z$ interactions, it generates corresponding contact terms
\begin{equation}
    \label{eq:dipolecontact}
    \Op{tW} =  i\big(\bar{Q}\sigma^{\mu\nu}\,\tau_{\sss I}\,t\big)\,
     \tilde{\phi}\,W^I_{\mu\nu}
     + \text{h.c.}\quad 
     \to \quad gv\,\bar{t}_{\sss L} \sigma^{\mu\nu} t_{\sss R} \, W_\mu^+ W_\nu^-,\, 
     gv\,\bar{b}_{\sss L} \sigma^{\mu\nu} t_{\sss R} \, Z_\mu W_\nu^-,
\end{equation}
from the non-Abelian parts of the gauge field strength.
Such terms directly mediate  $b \, W^+ \to t \, Z$ and $t\,W\to t\,W$ scattering. Using Equation~\ref{eq:naive_e}, with $n=2$ and $m=1$, the maximum energy dependence of the fully longitudinal amplitudes is $E^3$, which appears higher than our expectations from dimension-6 operators. Equivalently, if one omits the contact terms of Eq.~\eqref{eq:dipolecontact}, using the conventional AC description of dipole-type interactions~\cite{AguilarSaavedra:2008zc}, 
\begin{align}
    \mathcal{L}_{\text{dip.}} \supset -\frac{g}{\sqrt{2}}\bar{b} \, \sigma^{\mu\nu}\left( 
    g_{\sss L} P_{\sss L}+g_{\sss R} P_{\sss R}
\right) t \, \partial_\mu \, W_\nu
\end{align}
the energy dependence of, \textit{e.g.}, the $(-1,0,1,0)$ (fully longitudinal) helicity configurations rises from $E \to E^3$. We observe not only a general enhancement of energy growth across the helicity amplitudes, but also that the maximum degree of growth can be higher. However, as shown in Tables~\ref{tab:bwtz} and ~\ref{tab:twtw}, this extreme high-energy behaviour is never produced; it is cancelled in the final result thanks to the $SU(2)$ gauge invariance manifest in the SMEFT description. The field-strength component of this operator sources transversely polarised gauge bosons and the maximally-growing amplitude is therefore in a mixed transverse-longitudinal configuration. It grows with the expected power of energy, $E^2$, and its associated, dimension-6, contact interaction is
\begin{align}
    \label{eq:dipole_contact}
    G\,\bar{f}_{\sss L}\sigma^{\mu\nu}f_{\sss R}\partial_\mu V_
\nu.
\end{align}
We find this maximal behaviour for every scattering that we consider involving at least one massive gauge boson or Higgs.
We are therefore faced with different, incompatible predictions/constraints on the ACs with respect to their SMEFT counterparts. Depending on the process considered, one can obtain different predictions between the anomalous coupling and SMEFT frameworks if the full set of appropriate contact terms is not considered. For instance, considering top decay, the omission of the contact term coming from the weak dipole is not relevant and the two descriptions will give the same result. On the other hand, the study of a process such as $t Z j$ will be influenced by the presence of the contact term, leading to a disagreement in the two approaches. Contact interactions with dynamical Higgs fields are also examples of explicitly higher-dimensional Lorentz structures that arise as a consequence of gauge invariance from operators that modify SM interactions. 

\subsection{Embedding $2\to2$ scatterings in collider processes\label{sec:EWA}}
In this section, we attempt to gain some analytical insight of the mapping between $2\to2$ sub-amplitudes to the full $2 \to 3$ and $2 \to 4$ processes by making use of the Effective $W$ Approximation (EWA)~\cite{Dawson:1984gx,Kunszt:1987tk,Borel:2012by}. 
This exploits the factorisation of certain processes involving virtual, $t$-channel gauge bosons attached to light fermion currents. It approximates them by the convolution of a splitting function for the soft, collinear emission of a real gauge boson from the fermion leg and the on-shell, lower-multiplicity matrix-element. 
Thanks to this method, one can reduce a $2 \to 3$ or a $2 \to 4$ parton-level processes into a lower multiplicity scattering process weighted by splitting functions. This approximation is relevant for several processes that we consider, including  $tXj$, $t\bar{t}Wj$ and $t\bar{t}$ through vector boson fusion (VBF), whose lower multiplicity scatterings correspond to one of the $2\to2$ amplitudes computed in this work. This allows us to investigate how the energy growth of the $2\to2$ helicity amplitudes is propagated to the higher multiplicity final states in specific regions of the phase space. We briefly summarise the EWA result following Ref.~\cite{Borel:2012by} and then move on to its application to high energy top-quark production processes.

We consider a generic process $q X \to q^\prime Y$, where $q$ and $q^\prime$ are massless fermions while $X$ is an unspecified initial state particle and $Y$ can be a generic final state. The $q$-$q^
\prime$ current is connected to X and Y through a $t$-channel, virtual $W$ boson that carries a fraction, $x$, of the longitudinal momentum of the initial fermion. The parametrizations for the particle four-momenta are the following
\begin{equation}
\begin{aligned}
P_q &= (E,\vec{0},E) \, , \\
P_X &= (E_X, \vec{0}, -E) \, ,\\
P_{q\prime} &=  \left(\sqrt{(1-x)^2 E^2 +p_\perp^2},\, \vec{p}_\perp,\, E(1-x)\right) \,.
\end{aligned}
\end{equation}

The $W$ has momentum 
\begin{equation}
\begin{aligned}
K &= P_q - P_{q\prime} = \left(\sqrt{x^2E^2+p_\perp^2 + m^2 - V^2}, - \vec{p}_\perp, xE\right) \, , \\
V^2 &= m^2 -K^2 \approx m^2 +\frac{p_\perp^2}{1-x}\left[1 + O\left(\frac{p_\perp^2}{E^2}\right)\right] \, ,
\end{aligned}
\end{equation}
with $V$ denoting the virtuality of the $W$. The EWA is valid in the regime 

\begin{equation}
    \label{eq:ewa_approx}
E \sim x E \sim (1-x) E, \qquad \frac{m}{E} \ll 1, \qquad \frac{p_\perp}{E} \ll 1 \, ,
\end{equation}
where the initial fermion emits the $W$, experiencing a small angular deflection while losing a large fraction of its longitudinal momentum. The $W$ then participates in the scattering sub-process $W\,X\to Y$ with a characteristic energy of order E.
The general formula for the factorised approximation of the full process is:
\begin{equation}
\frac{d\sigma_{EWA}}{dxdp_\perp}(q X \to q^\prime Y) = \frac{C^2}{2\pi^2}{\sum_{i=+,-,0} f_i \times d\sigma(W_i X \to Y)} \, ,
    \label{eq:EWA}
\end{equation}
with C related to the coupling of the W to the light fermions ($C= g/2 \sqrt{2}$). The equivalent W in the sub-process is treated as an on-shell particle with momentum
\begin{equation}
p_{\sss W} = \left(\sqrt{x^2 E^2 + m^2}, \vec{0}, xE\right) \,,
\end{equation}
and the splitting functions for each $W$ polarization are
\begin{equation}
\label{eq:splitting}
\begin{aligned}
f_+ &= \frac{(1-x)^2}{x}\frac{p_\perp^3}{(m^2(1-x)+p_\perp^2)^2} \, , \\
f_- &= \frac{1}{x}\frac{p_\perp^3}{(m^2(1-x)+p_\perp^2)^2} \, , \\
f_0 &= \frac{(1-x)^2}{x}\frac{2m^2p_\perp}{(m^2(1-x)+p_\perp^2)^2} \, .
\end{aligned}
\end{equation}

We now apply this to a relevant process for this study, namely $t$-channel single top in association with a Higgs (Figure~\ref{fig:topology} with $B,\,B^{\sss\prime} = W,\,h$). This can be used to investigate the high energy behaviour of the $b \, W \, \to \, t \, h$ scattering. This process is particularly simple in that it is only mediated by true `scattering' diagrams that contain the sub-amplitude, since the Higgs can only be radiated from the top and $W$ lines. We therefore do not have to worry about other diagrams that may be present in, \emph{e.g.}, $tZj$ where the Z can be radiated from the light quarks. In this calculation we neglect PDFs and select a representative parton level process, $u \, b \to d \, t \, h$, reducing it to the scattering of a $b$-quark with an equivalent $W$. We are interested in the behaviour of the differential cross section as a function of $\hat{s}$, the invariant mass of the top-Higgs pair, which characterises the energy scale of the $2 \to 2$ sub-process. 

One can express the differential EWA cross section of equation~\eqref{eq:EWA} as
\begin{equation}
d\sigma_{EWA} = \frac{g^2}{16\pi^2}\sum_{i=+,-,0} f_i \frac{1}{4 p_{\sss W} \cdot p_b} (2\pi)^4 |\mathcal{M}^i_{2 \to 2}|^2 \frac{\sqrt{E^2_t - m_t^2} \sin\theta}{(2\pi)^3 2} \frac{1}{(2\pi)^3 2 E_h} d\theta  \, d\phi \, dx \, dp_\perp \, ,
\end{equation}
where the $b$-quark (assumed massless) plays the role of the $X$ initial state and $\theta,\phi$ are the polar and azimuthal angles of the top quark, respectively. The $2 \to 2$ matrix element does not depend on $\phi$, therefore we can trivially integrate it. Changing variables from $x$ to $\hat{s}$ involves the Jacobian
\begin{align}
   J(\hat{s})\equiv \frac{1}{(2E)^2}+\frac{\mw^2}{(\hat{s}-\mw^2)^2},
\end{align}
 and leaves the fully differential cross section  with respect to $\hat{s}$, $p_\perp$ and $\theta$
\begin{equation}
\frac{d\sigma_{EWA}}{d\theta \, dp_\perp \, d\hat{s}} = \frac{g^2}{512\pi^3}\sum_{i=+,-,0} f_i(\hat{s}, p_\perp) \frac{1}{p_{\sss W} \cdot p_b} |\mathcal{M}^i_{2 \to 2}|^2  \frac{\sqrt{E^2_t - m_t^2} \sin\theta}{E_h} J(\hat{s}) \, ,
\end{equation}
where $E_t$ and $E_h$ are the top quark and Higgs boson energies respectively.
%
%

\begin{figure}[t]
\centering
\includegraphics[width=0.32\linewidth]{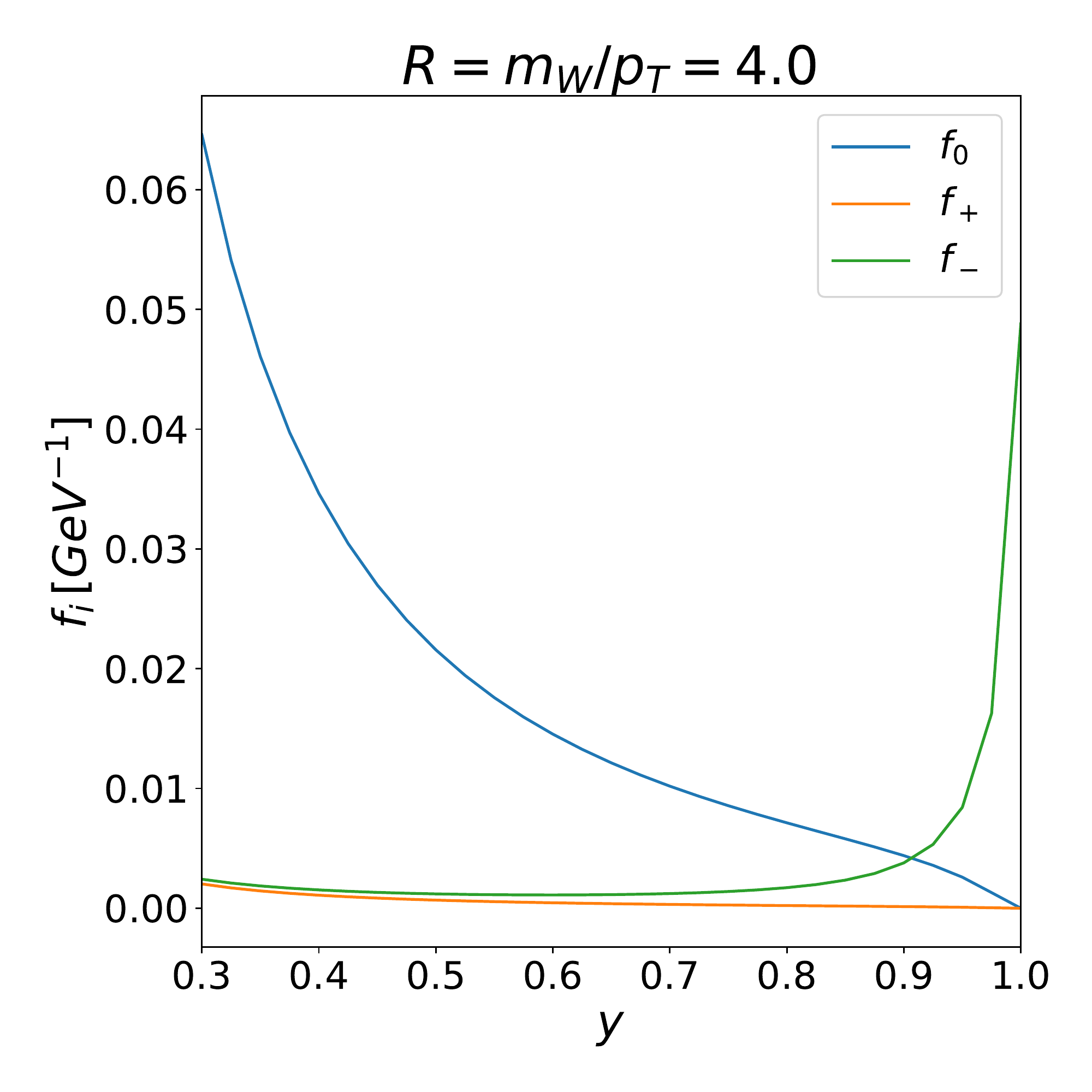}
\includegraphics[width=0.32\linewidth]{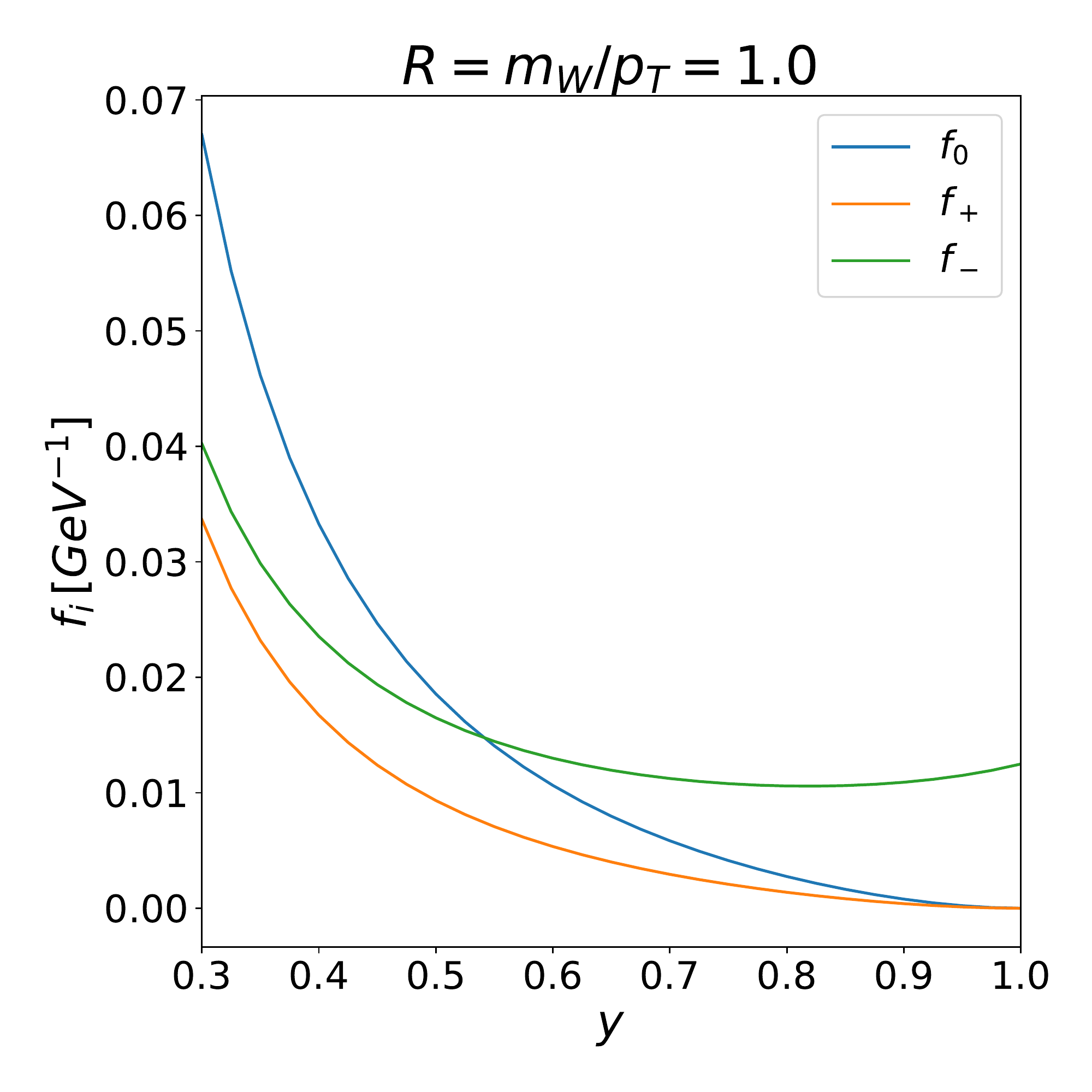}
\includegraphics[width=0.32\linewidth]{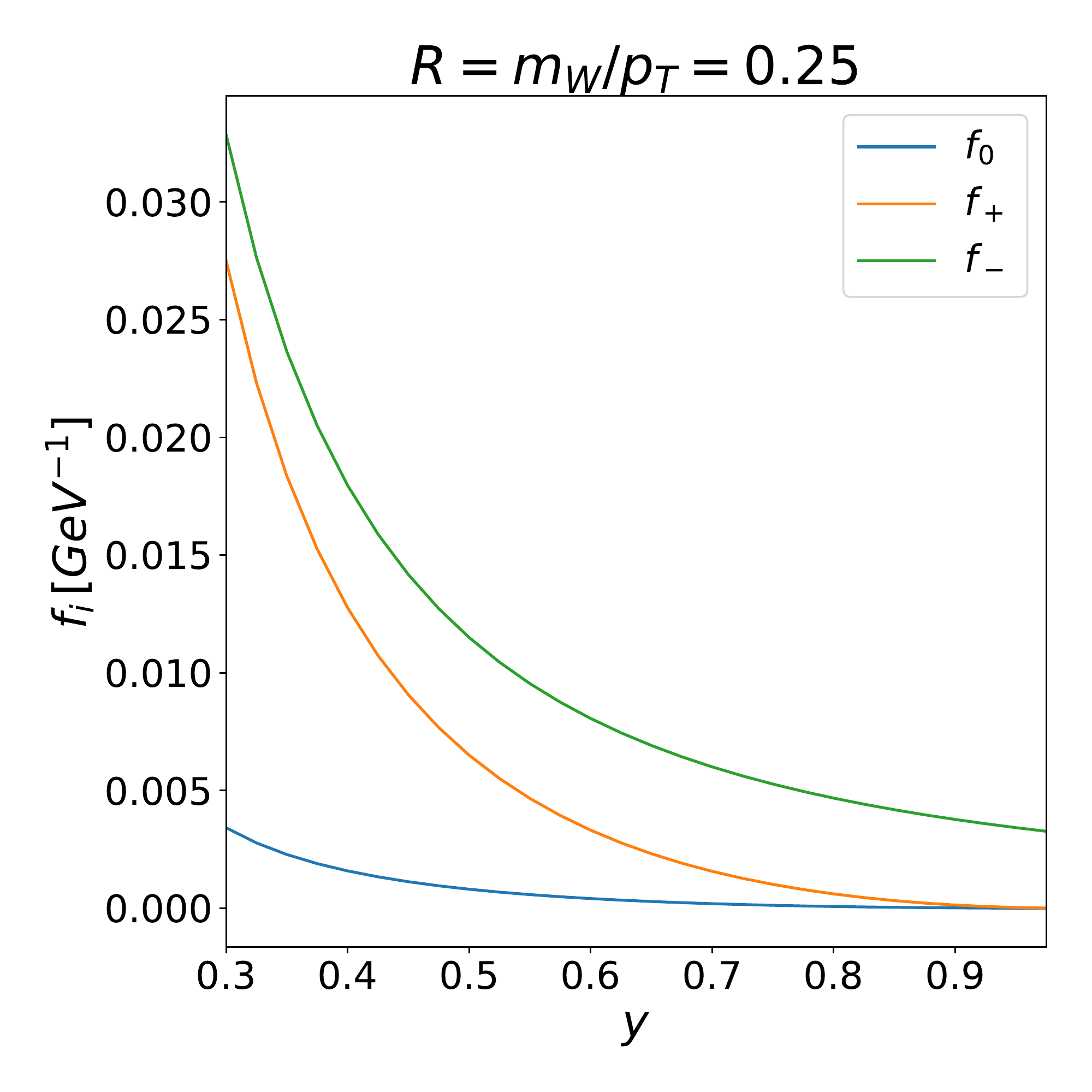}
\caption{Plot of the splitting functions for the equivalent $W$ emission as a function of the hard-scattering energy fraction $y=\sqrt{\hat{s}}/2E$ for three different values of $R=\mw/p_\perp$. \label{fig:splittingFunc}}
\end{figure}

In order to inspect the behaviour of the differential cross section as a function of the hard-scattering energy scale, we consider phase space points that are both in the valid EWA regime and in the high energy regime of the $2\to2$ scattering, $s \sim -t \gg v^2$. We fix the following parameters
\begin{align}
E = 2 \text{ TeV} \, ,\quad \theta = \frac{\pi}{2} \,,
\end{align}
and plot the splitting functions of Equation~\eqref{eq:splitting} for the equivalent $W$ as a function of $y=\sqrt{\hat{s}}/2E$ for three different values of $R=m_W/p_\perp$ in Fig. \ref{fig:splittingFunc}. 
The three polarizations of the $W$ have a very different probability of being emitted as a function of the hard-scattering energy, depending also on the value of $R$. Taking ratios of Eq.~\eqref{eq:splitting} shows that the relative importance of the transverse and longitudinal modes is controlled by $R$, 
\begin{align}
    \frac{f_0}{f_+}=2R^2\,,\quad \frac{f_0}{f_-}=2R^2(1-x)^2.
\end{align}
In particular, the spectrum is dominated at low energies by the longitudinal polarization, when $R>1$. The more $R$ decreases, the less likely the emission of the longitudinal degree of freedom becomes. On the other hand the `$+$' polarization gets more relevant as $R$ decreases, while the `$-$' polarization is always dominant at high energies.
Thus the equivalent, longitudinal $W$ can be accessed through smaller values of $p_\perp$ relative to $\mw$ and intermediate $x$ while the transverse modes prefer larger $p_\perp$. Ref.~\cite{Borel:2012by} identified an additional source of discrepancy between the EWA and full process beyond corrections of order $(\mw/E)^2$ and $(p_\perp/E)^2$, arising in the peculiar case of extremely small $p_\perp$ when $R\sim E/\mw$. One therefore cannot go arbitrarily low in $p_\perp$ in the hopes of accessing a longitudinal, equivalent $W$ in this approximation. The positive helicity and longitudinal polarisations of the $W$ suffer an additional suppression by $(1-x)^2$ in their splitting functions due to helicity conservation in the left-handed current interaction. This leads to a preference for the negative helicity mode at higher values of $x\sim1$.

\begin{figure}[t]
\centering
\includegraphics[width=0.45\linewidth]{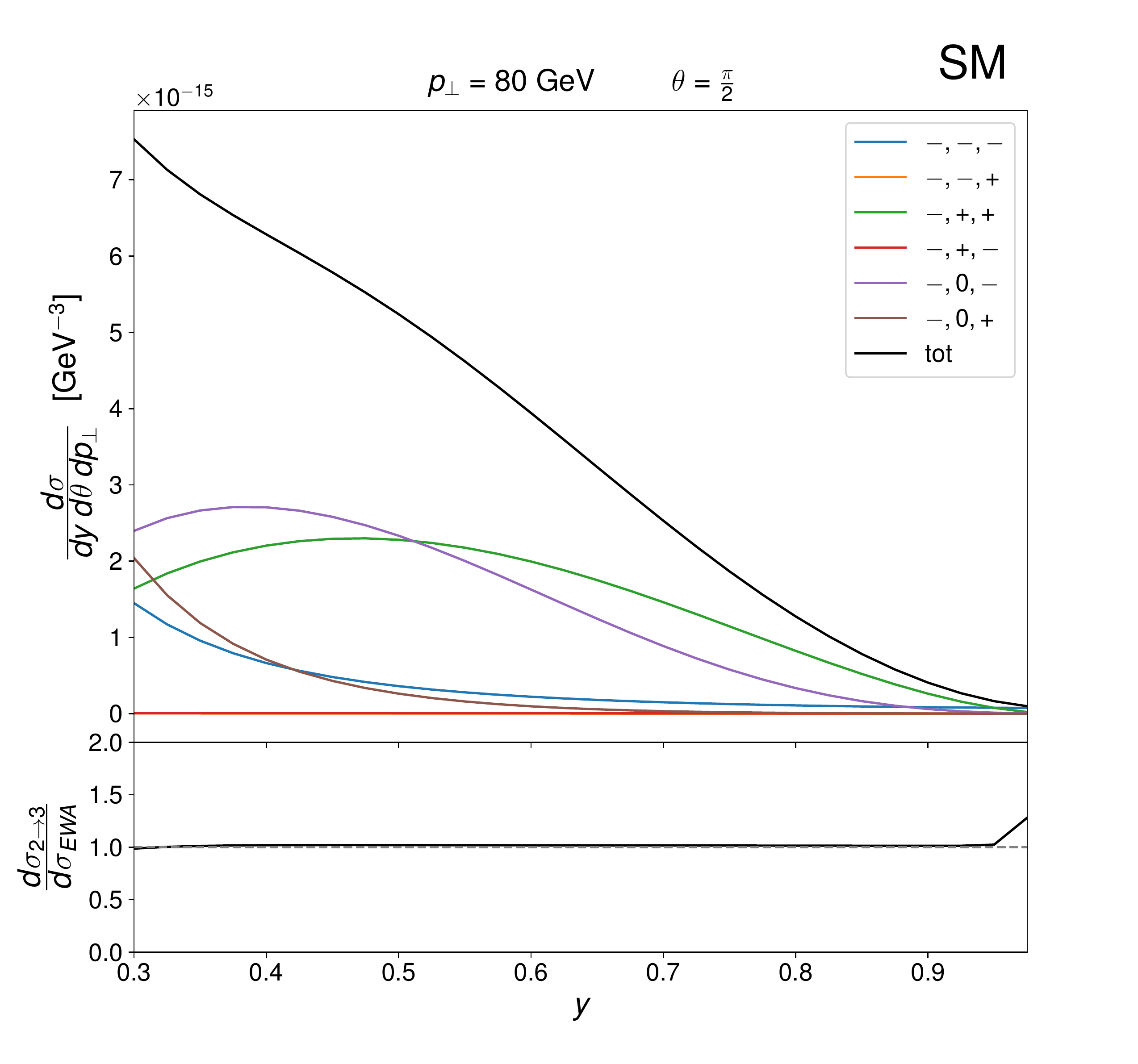}
\caption{The SM contribution in various $(\lambda_{\sss b},\lambda_{\sss W},\lambda_{\sss t})$ helicity configurations to the differential cross section as a function of the hard-scattering energy fraction $y=\sqrt{\hat{s}}/2E$. \label{fig:ewa_SM}}
\end{figure}

\begin{figure}[t]
\centering
\includegraphics[width=0.45\linewidth]{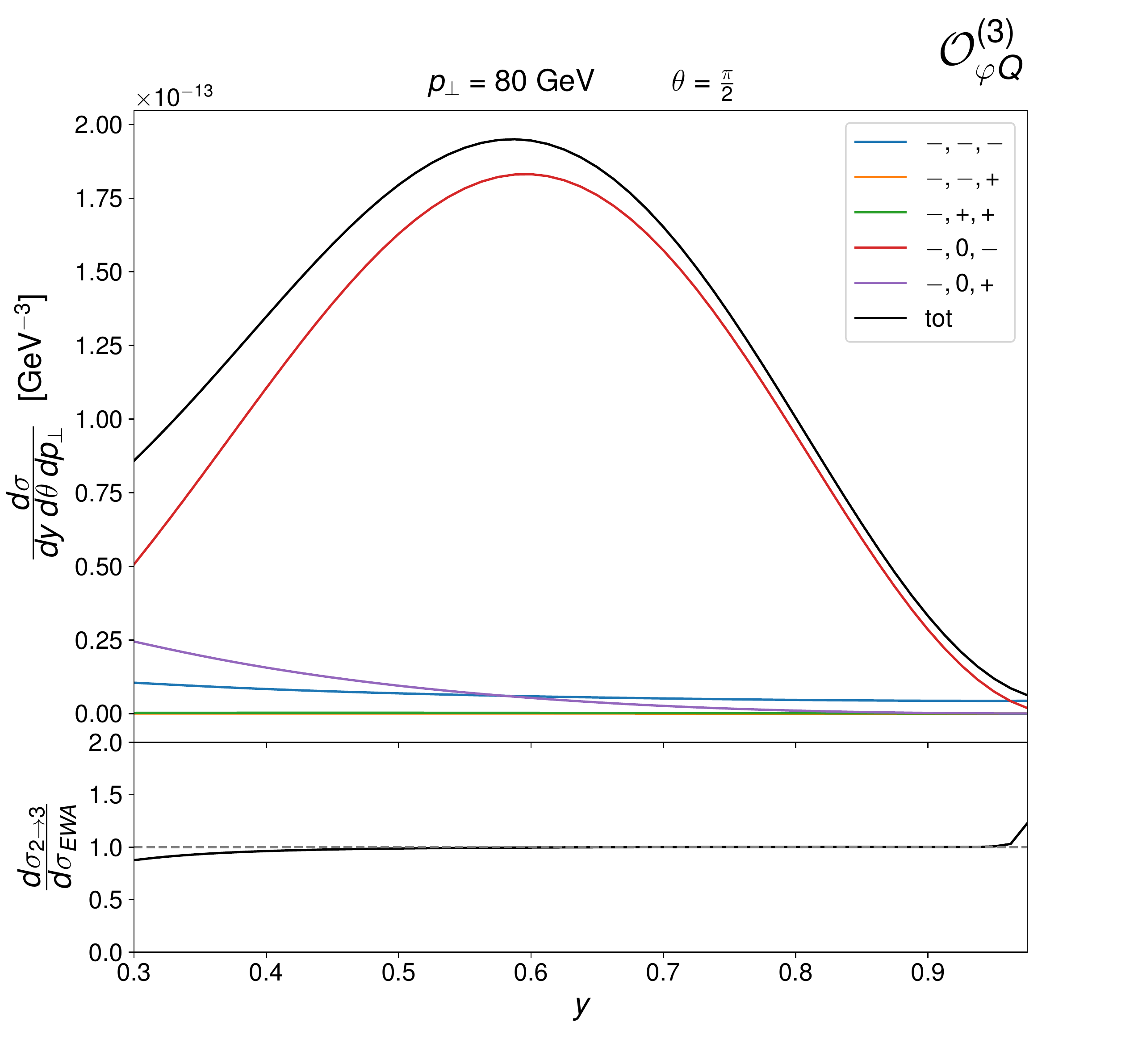}
\includegraphics[width=0.45\linewidth]{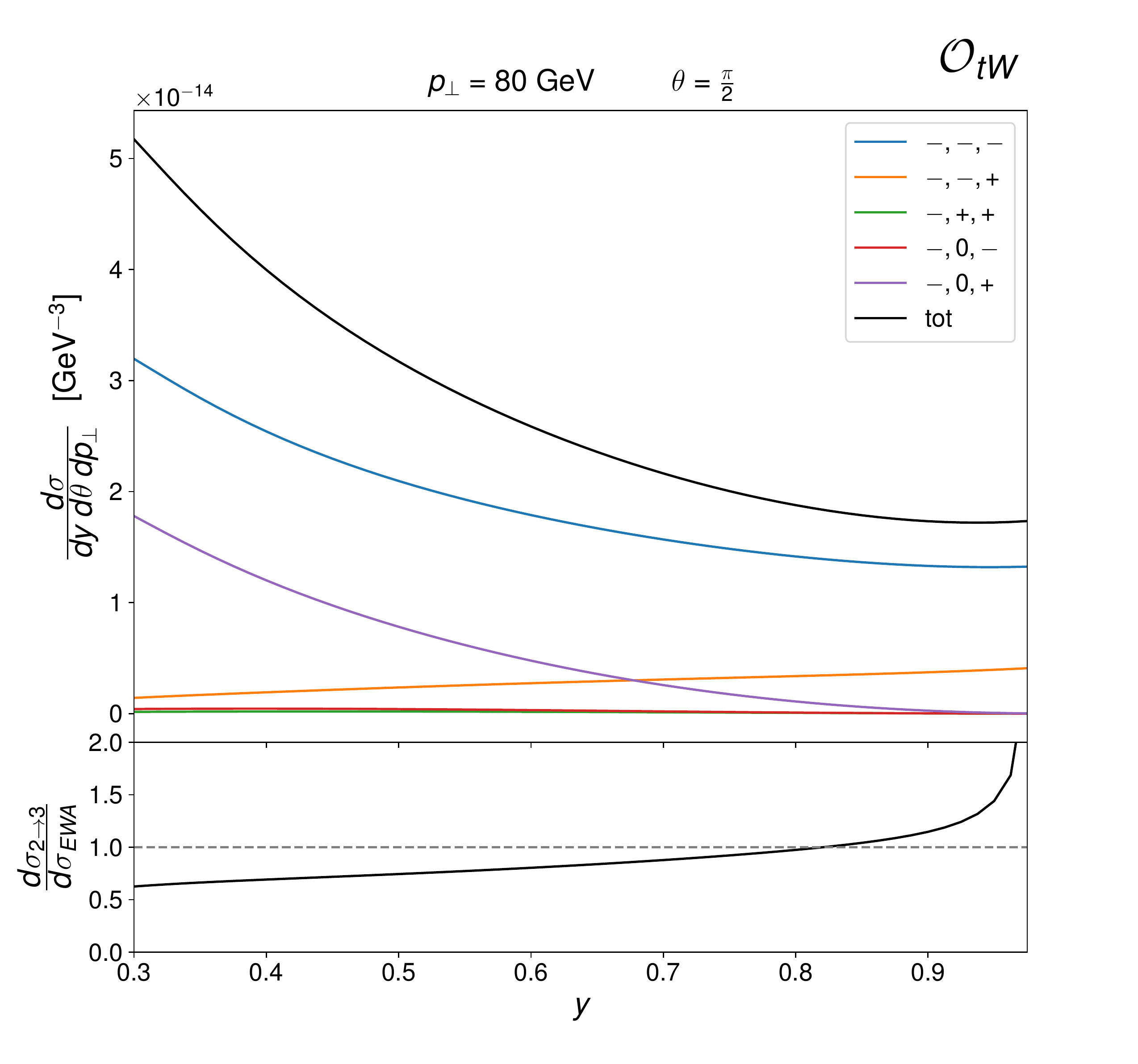} \\
\includegraphics[width=0.45\linewidth]{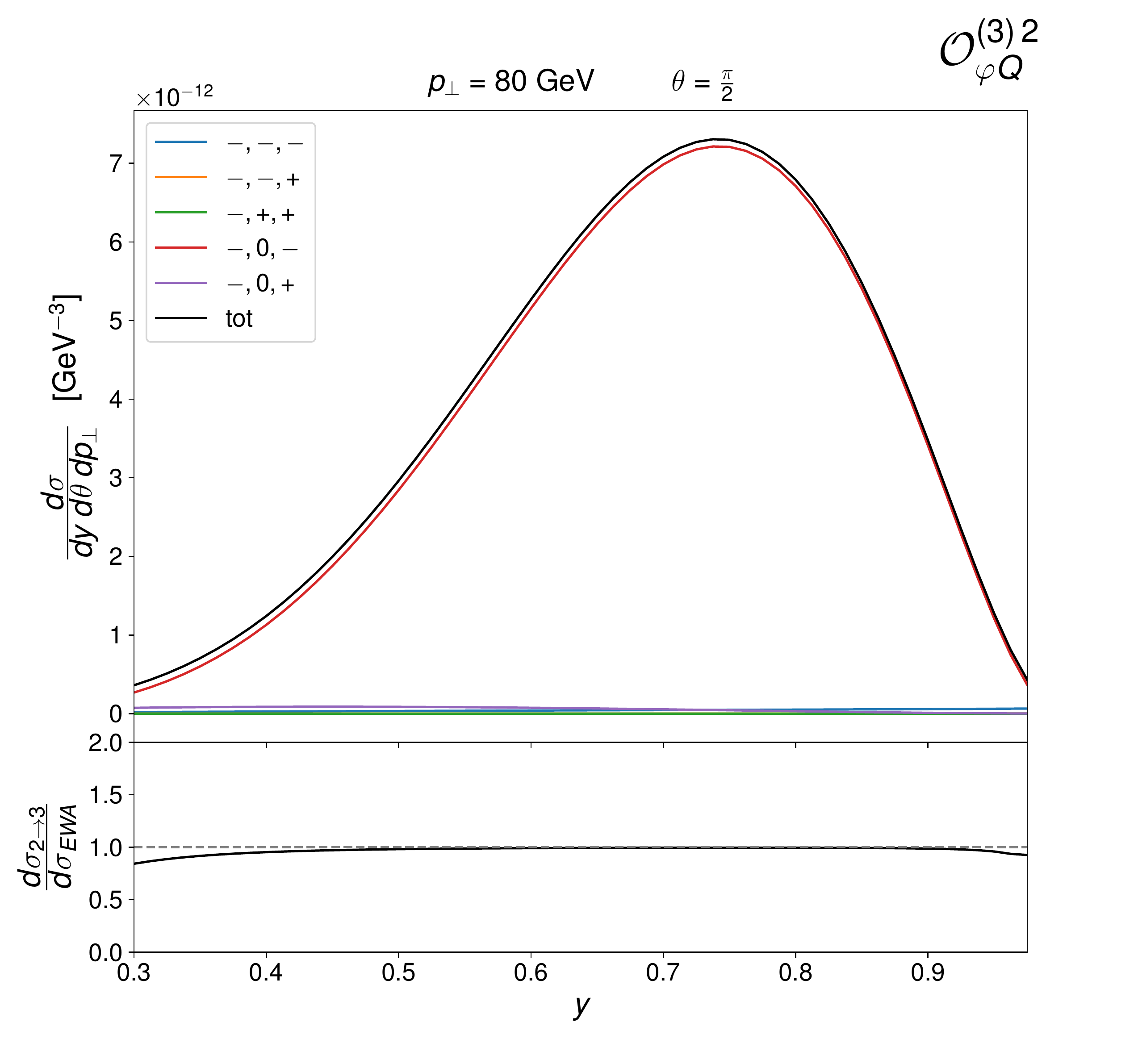}
\includegraphics[width=0.45\linewidth]{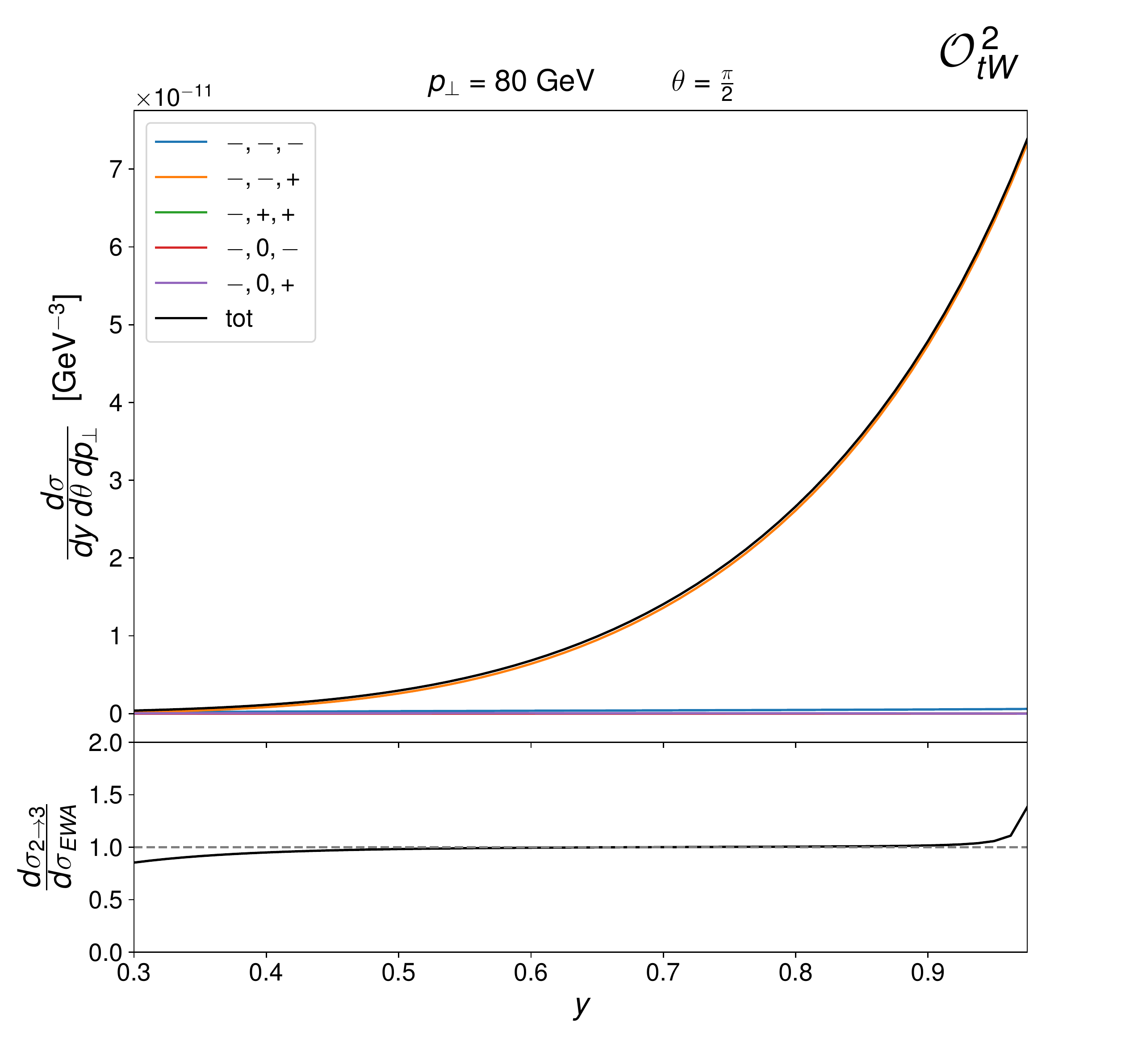}
\caption{Plots of the intereference and square contribution in various $(\lambda_{\sss b},\lambda_{\sss W},\lambda_{\sss t})$ helicity configurations to the differential cross section as a function of the hard-scattering energy fraction $y=\sqrt{\hat{s}}/2E$ for the operators $\Op{\phi Q}^{\sss (3)}$ (on the left) and $\Op{tW}$ (on the right). \label{fig:ewa}}
\end{figure}

The $2 \to 2$ matrix element squared is
\begin{equation}
|\mathcal{M}_{2 \to 2}|^2 = |\mathcal{M}^{SM}_{2 \to 2}+\mathcal{M}^{NP}_{2 \to 2}|^2 = |\mathcal{M}^{SM}_{2 \to 2}|^2+\mathcal{M}^{int}_{2 \to 2}+|\mathcal{M}^{NP}_{2 \to 2}|^2 \, ,
\end{equation}
where $\mathcal{M}^{NP}$ are the matrix elements with dimension six operator insertions, whose high energy behaviour is shown in Appendix~\ref{app:helamp_tables}. Plugging the expressions into the EWA differential cross section formula allows us to assess the energy behaviour of the interference and squared terms in the full process, as well as the relative importance of the different helicity configurations. These are shown in Figures~\ref{fig:ewa_SM} and~\ref{fig:ewa} for the SM and the operators $\Op{tW}$ and $\Op{\phi Q}^{\sss (3)}$, fixing $p_\perp= 80$ GeV, where we have also retained sub-leading energy behaviour in the $2\to2$ helicity amplitudes down to $s^0$. We keep the same values for $E$ and $\theta$ such that the chosen value of $p_\perp$ is close to the $R=1$ profile of the splitting functions shown in the central panel of Figure~\ref{fig:splittingFunc}. The approximation is expected to hold in the intermediate range of $y$, so as to fulfil the conditions of Equation~\eqref{eq:ewa_approx}.

Looking at the $b \, W \to t \, h$ helicity amplitudes in Table~\ref{tab:bwth}, we see that the two dominant helicity configurations for the SM contribution are indeed those which are found to be constant in energy, $(-,0,-)$ and $(-,+,+)$. From the SMEFT, one expects an energy growth for the interference of $\Op{\phi Q}^{\sss (3)}$, coming from the left-handed, longitudinal helicity configuration, $(-,0,-)$, since it coincides with one of the energy-constant SM configurations. However, the weighting of the splitting functions has a noticeable impact depending on which $W$-polarization is involved. The plot shows an initial energy growth for the interference of $\Op{\phi Q}^{\sss (3)}$ which becomes suppressed by $(1-x)^2$ due to the $f_0$ splitting function. In the case of $\Op{tW}$, each energy-growing configuration is compensated by a corresponding inverse power of energy in the SM contribution (see discussion in Section~\ref{subsec:interference}), meaning that no energy-growth is expected in the interference. Indeed, the dominant $(-,-,-)$ helicity configuration, in which the SMEFT amplitude grows linearly with energy, more or less follows the $f_-$ splitting function, implying a lack of energy growth in the $2\to2$ interference term. It decreases with energy in absolute terms but less so than the SM, \emph{i.e} a relative growth compared to the SM is still possible due to the different virtual $W$-polarisation state mediated by this operator. The squared contributions confirm this picture, with the $\Op{\phi Q}^{\sss (3)}$ killed at high $x$ by $f_0$ while the dipole is free to grow given the almost constant behaviour of $f_-$ at high energies. Note that the squared term for $\Op{tW}$ is dominated by a different $W$-polarisation than its interference term.

The subplots of Figs.~\ref{fig:ewa_SM} and~\ref{fig:ewa} display the ratio of the full $2\to3$ differential cross section to the EWA approximation, indicating a very good agreement between the two over the intermediate $x$ range, except for the $\Op{tW}$ interference. The derivation of the EWA cross section is performed considering the full amplitude squared and integrating it over the jet azimuthal angle $\phi$. On the other hand, we are singling out the interference term, which is not positive definite when integrated over the jet angle. This can lead to anomalously small interference terms which are comparable to the EWA error, i.e. the neglected terms in the $p_\perp/E$ and $m/E$ expansion. Therefore the procedure of separating and comparing the individual terms in the Wilson coefficient expansion is \emph{a priori} flawed. We indeed observe a cancellation over the $\phi$ angle for $\Op{tW}$ while we do not observe it for $\Op{\phi Q}^{\sss (3)}$. Nevertheless, the validity of EWA is guaranteed if one considers the full amplitude square in both cases.

This study allows us to draw non-trivial conclusions on which operators we expect to be more relevant at high energy. Despite the common assumption that the longitudinal modes dominate at high energy we notice that, although present, they are suppressed by the splitting functions as the hard scattering energy increases. As a result, in the high-energy regime, the transverse polarizations provide the dominant component of the scattering amplitude. Operators affecting those modes, \emph{e.g.} the dipoles, are likely to give the most sizable effects.

\subsection{Blueprint for the analysis\label{sec:blueprint}}
For each category of scattering amplitudes, we begin by analysing in detail the results of the $2\to2$ helicity amplitude computation.  We then turn to addressing how the energy growing behaviour can be probed
by physical processes at collider experiments. We consider whether looking at high-energy EW top processes can improve our understanding 
of the interactions of the EWSB sector. In other words, from an SMEFT perspective, 
we explore what new information/sensitivity on the Wilson coefficients can be obtained from collider 
measurements of such processes. In particular we investigate 
the degree to which the behaviour of the $2\to2$ amplitude is preserved when 
going to $2\to3$ or $2\to4$ processes that could occur at colliders. 

Notice that Table~\ref{tab:constraints} displays a clear 
hierarchy between the current sensitivity to the operators with and without 
a top quark field. Although this statement is specific to our choice of basis and 
normalisation, one can intuitively understand that the operators not 
explicitly including a top quark field will generally affect other 
interactions that could be better constrained by EW precision, diboson and Higgs 
measurements. We therefore focus on the sensitivity of these 
processes to the remaining set of operators:
\begin{align}
\Op{t\phi}, \Op{tB}, \Op{tW}, \Op{\phi Q}^{\sss(1)}, \Op{\phi Q}^{\sss(3)}, \Op{\phi t} \text{ and } \Op{\phi tb}.
\end{align}
We will limit the discussion of the high energy behaviour of the dipole operators given that we know they are expected to give maximal energy growth almost everywhere due to~\eqref{eq:dipole_contact}. One general point worth making is that we often find the dipole operators to show some energy growth compared to the SM in their interference contributions. This confirms the behaviour uncovered in our analysis of the EWA in the previous section and highlights the complex interplay between the various helicity configurations involved in the high energy regime of a given process.

We consider the 
High-Luminosity LHC phase as well as at a future hadron collider at 27
TeV centre of mass energy and a high energy $e^+e^-$ collider operating at 
380 GeV, 1.5 TeV and 3 TeV. All cross section computations are performed with {\sc MadGraph5\_aMC@NLO}~\cite{Alwall:2011uj,Alwall:2014hca} using the SMEFT {\sc UFO} model based on the aforementioned {\sc FeynRules} implementation. The interference 
and squared contributions of each operator, $\mathcal{O}_i$, are computed relative 
to the pure EW SM contribution, and represented by the ratios $r_i$ and 
$r_{i,i}$ respectively.  The 
relative impact is shown for each operator with the Wilson coefficient 
set to 1 TeV$^{-1}$. 
In order to determine the high 
energy behaviour of each process, we compute both the inclusive production cross 
sections and the cross sections in a restricted, high energy region of phase 
space determined by a process dependent cut on a particular kinematical variable. The corresponding relative impacts are denoted by $r^\text{tot}$ and $r^\text{HE}$, respectively. 
A substantial growth in the relative impact of the operator from inclusive to high-energy phase space can indicate the 
presence of unitarity violating behaviour due to the effective operator insertions.

In the case of the interference contribution, the assessment is more subtle due 
to the non-positive-definite nature of this term in the matrix element. The sign 
of the matrix element can change over the phase space and result in cancellations 
upon integration. This can lead to `anomalously' small contributions 
compared to other, similar operators and obfuscate the high energy behaviour 
present in the EFT contribution to the amplitude. We will identify these occurrences 
in the following survey of processes. Whenever a charged particle is present in the final
state we always include contributions for both particle and antiparticle in the numerical
results. In the case where there is a corresponding QCD-induced 
contribution, which does not probe the EW $2\to2$ scatterings, this rate is also 
quoted. We also consider the relative impact of the operator on the EW processes 
when saturating the individual limits quoted in Table~\ref{tab:constraints} to 
give an idea of whether each process may provide additional sensitivity to a 
particular operator given existing constraints.

In some cases, we consider processes with a light jet in the final state. These are defined {\it excluding} $b$-jets. The main reason behind this is, on one side, that processes with final state $b$'s can be considered experimentally distinct from processes involving gluons or light jets (even though we are technically employing the 5-flavour scheme for our numerical simulations) and on the other, that $b$-quark final states can also introduce dependence on single-top $2\to2$ subprocesses, {\it i.e.}, $b W\to t X$, not present in all other channels. These types of processes are therefore also phenomenologically distinct from the perspective of constraining the high energy behaviour of EW top quark scattering.  Notwithstanding the omission of outgoing $b$-quarks, such final states can often be obtained as a real QCD radiation from an underlying Born-level process. These are not infrared finite on their own given that they form a subset of the NLO QCD corrections to said process.  Often, these  can admit contributions from non-EW SMEFT operators that modify QCD interactions, as discussed in Section~\ref{subsec:SMEFT}. An example of this is $tWj$, which has an order $\alpha_{\sss EW}^3$ contribution related to $b\,W\to t\,Z$ scattering as well as an order $\alpha_{\sss S}^2\alpha_{\sss EW}$ component coming from real QCD corrections to $tW$ production. We can nevertheless study these EW components independently, bearing in mind that they will always have an additional QCD-induced background. When computing predictions for these processes, a suitable jet $p_T$ cut must be applied to avoid the singular phase space region. These cuts are chosen in a process-dependent way, by studying the stability of the ratio between the EW- and QCD-induced versions of the process, choosing a cut for which this ratio has roughly plateaued and the absolute size of the QCD component of the cross section ($tWj$ at $\mathcal{O}(\alpha_{\sss S}^2\alpha_{\sss EW}$)) is lower than that of the underlying Born process ($tW$).

\section{Single-top scattering\label{sec:singletop_scattering}}
This set of scatterings is characterised by the presence of $b$, $W$ and $t$ external legs accompanied by a fourth, electrically neutral state.  They enter in single-top collider processes, summarised in Table~\ref{tab:amp_proc_table_singletop}, which in the 5-flavor scheme are initiated by $b$-quarks in the proton. 
\begin{table}[h!]
\centering
\begin{tabular}{|P{2cm}|P{1cm}|P{1cm}|P{1cm}|P{1cm}|P{1cm}|P{1cm}|P{1cm}|}
     \hline
                              & $tWj$  & $tZj$ & $t\gamma j$ & $tWZ$ & $tW\gamma$ & $thj$ & $thW$
     \tabularnewline\hline             
                                       
     $b\,W\to t\,Z$           & \cmark & \cmark &             & \cmark &          &        &        
     \tabularnewline\hline                                                      
                                                                                  
     $b\,W\to t\,\gamma$      & \cmark &        & \cmark      &        &  \cmark  &        &         
     \tabularnewline\hline                                                      
                                                                                 
     $b\,W\to t\,h$           &        &        &             &        &           & \cmark & \cmark
     \tabularnewline\hline
\end{tabular}
\caption{ The set of single-top $2\to2$ scattering amplitudes considered in this work mapped to the collider processes in which they are embedded.
\label{tab:amp_proc_table_singletop}}
\end{table}
Theres is almost complete separation between the different subamplitudes and the processes that can probe them, with only $b\,W\to t\, Z$ and $b\,W\to t\, \gamma$ contributing simultaneously to $tWj$.

Since the amplitudes always involve a $W$-boson, in the SM the $b$-quark external leg is always left-handed. Helicity configurations with a right-handed $b$ are suppressed by the $b$-quark mass which we set to zero in accordance with the 5-flavor scheme. The right-handed charged current operator, $\Op{\phi tb}$, is the only operator that mediates right-handed $b$ external legs considered in our study. Since the corresponding SM amplitude is absent, this operator has no interference contribution, entering only quadratically in the Wilson coefficient. 

Table~\ref{tab:helamp_summary_singletop} summarises the maximum energy growths induced by our set of SMEFT operators, taken from the full helicity configuration results of Tables~\ref{tab:bwtz}--\ref{tab:bwth}. The red entries denote an energy-growing interference term with the SM. This behaviour is only present for the left handed SU(2) triplet current operator, $\Opp{\phi Q}{(3)}$. Notice that the amplitude involving a photon is only affected by operators that induce new Lorentz structures with respect to the SM. QED current interactions are not affected in the SMEFT as opposed to the $Z$ or $W$ boson couplings, which are shifted by the current operators ($\Opp{\phi Q}{(1)}$, $\Opp{\phi Q}{(1)}$, $\Op{\phi t}$ and $\Op{\phi tb}$).  The photon only receives modified interactions from a linear combination of the weak dipole operators ($\Op{tB}$ and $\Op{tW}$), higher derivative triple-gauge couplings ($\Op{W}$), and dimension-5 gauge-Higgs interactions after EW symmetry breaking ($\Op{\phi WB}$, $\Op{\phi WB}$ and $\Op{\phi WB}$). 
\begin{table}[h!]
{\footnotesize
\setlength{\tabcolsep}{4pt}
\renewcommand{\arraystretch}{1.2}
\begin{center}
  \begin{tabular}{|c|c|c|c|c|c|c|c|c|c|c|c|c|c|}
 \hline
                          & $\Op{\phi D}$ & $\Op{\phi \Box}$ &  $\Op{\phi B}$ & $\Op{\phi W}$  & $\Op{\phi WB}$  & $\Op{W}$ & $\Op{t \phi}$ & $\Op{tB}$ & $\Op{tW}$ & $\Op{\phi Q}^{\sss (1)}$ & $\Op{\phi Q}^{\sss (3)}$ &  $\Op{\phi t}$ & $\Op{\phi tb}$ 
 \tabularnewline\hline
 
 $b\,W\to t\,Z$           & $E$           & $-$              & $-$            & $-$            & $E$             & $E^2$    & $-$           & $E^2$     & $E^2$     & $E$                      & \red{$E^2$}              & $E$            & $E^2$           
 \tabularnewline\hline

 $b\,W\to t\,\gamma$      & $-$           & $-$              & $-$            & $-$            & $E$             & $E^2$    & $-$           & $E^2$     & $E^2$     & $-$                      & $-$                      & $-$            & $-$           
 \tabularnewline\hline
 
 $b\,W\to t\,h$           & $-$           & $-$              & $-$            & $E$            & $-$             & $-$      & $E$           & $-$       & $E^2$     & $-$                      & \red{$E^2$}              & $-$            & $E^2$            
 \tabularnewline\hline
 
  \end{tabular}
\end{center}
\renewcommand{\arraystretch}{1.}
}

\caption{\label{tab:helamp_summary_singletop}
Maximal energy growths induced by each operator on the set of single top scattering amplitudes considered. `$-$' denotes either no contribution or no energy growth and the red entries denote the fact that the interference between the SMEFT and the SM amplitudes also grows with energy.
}
\end{table}
\subsection{Without the Higgs\label{subsec:onetop_nohiggs}}
\subsubsection{$b\,W\to t\,Z$ \& $b\,W\to t\,\gamma$ scattering\label{subsubsec:bwtz_bwta}}
	\begin{figure}[h!]
		\centering
		\subfloat[]{
        \includegraphics[width=0.25\linewidth]{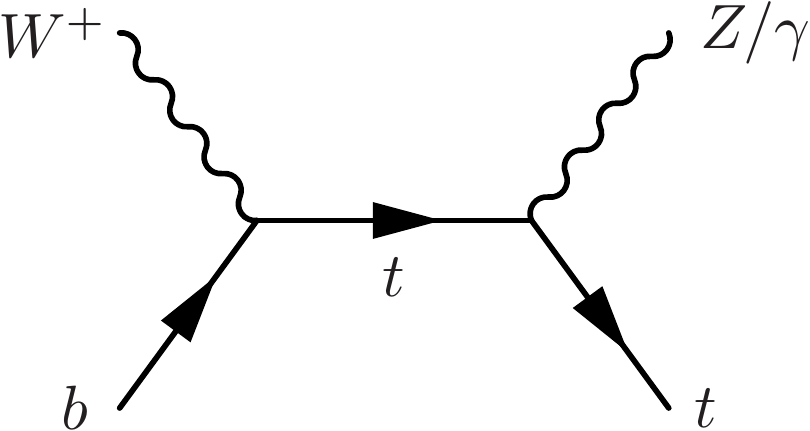}
        }  \subfloat[]{
        \includegraphics[width=0.25\linewidth]{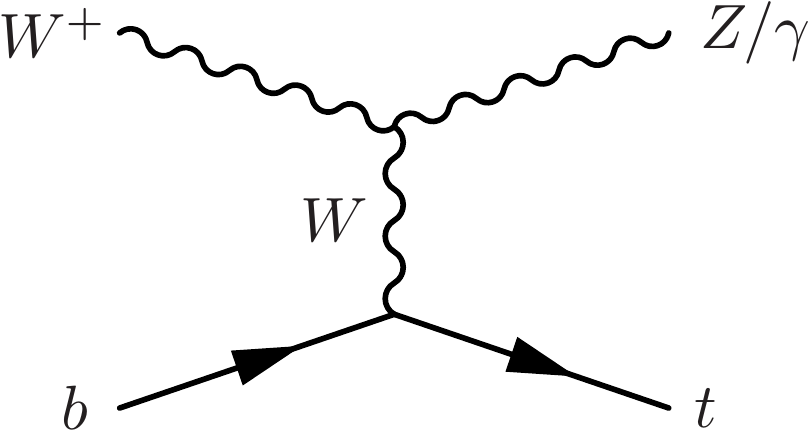}
        }
        \subfloat[]{
        \includegraphics[width=0.25\linewidth]{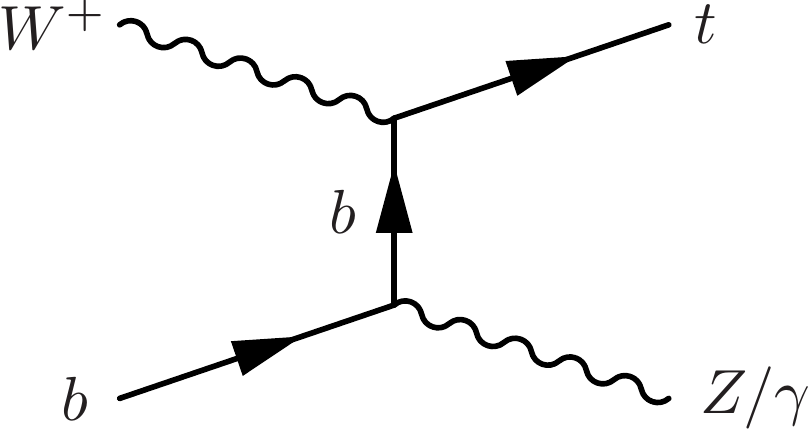}
        }
		\caption{SM diagrams for the $b W \to t Z$ subprocess. \label{fig:bwtz_sm}}
	\end{figure}
\noindent  This process probes the coupling of the top and bottom quarks with the $Z$ and with the $W$, as can be seen in Fig. \ref{fig:bwtz_sm}, depicting the contributing unitary gauge Feynman diagrams. The $btW$ vertex is present in every diagram, meaning that a modification to the SM coupling cannot lead to non-unitary behaviour, only scaling the amplitude by an overall factor. From the AC perspective, one can modify the $WWZ$ and the left- and right-handed $ttZ$ and $bbZ$ couplings. In the SM these interactions are related by Gauge invariance and the Higgs mechanism. The left handed $b$ and $t$ couplings are connected since they form an $SU(2)$ doublet. Modifying these 2 couplings independently constitutes a source of EW symmetry breaking beyond the SM and, as a consequence, leads to unitarity violation. The contribution to the left-handed, fully longitudinal helicity configuration, $(-,0,-,0)$, grows with $E^2$ and is proportional to 
    \begin{equation}
        \label{eq:bwtz_max}
        \sqrt{s(s+t)}\,(\gzbl - \gztl + \gwz) \,.
    \end{equation}
The simultaneous appearance of $\gzbl,\,\gztl$ and $\gwz$ highlights the fact that both the gauge-fermion and gauge self-interactions arise from the same non-Abelian gauge theory. Independent modifications of the gauge coupling strength in each sector violate the gauge symmetry and lead to non-unitary behaviour. There are sub-leading growths when going from the fully longitudinal to the mixed transverse-longitudinal configurations, $(-,-,-,0)$ or $(-,0,-,-)$, where either the $W$ or $Z$ boson polarisation is changed ($0\to-$),
\begin{equation}
        \label{eq:bwtz_max_flip}
        \sqrt{-t}\,(\gzbl - \gztl + \gwz) \,.
\end{equation}
They are subject to the same cancellation condition as the previous amplitude but appear at the cost of a factor $\sim v/\sqrt{s}$. 

A further linear growth is present in the helicity flipping $(-,0,+,0)$ configurations proportional to
\begin{equation}
        \label{eq:bwtz_subl}
    \sqrt{-t}\,(2m_W^2(\gzbl - \gztr + \gwz) - \gwz m_Z^2) \,,
\end{equation}
that displays a quite non-trivial cancellation in the SM. It involves Gauge charges of two different fermion representations, gauge boson self-interactions and the gauge boson masses. The fact that the fermions involved are those that participate in the Yukawa interaction with the Higgs field suggests that this cancellation is a consequence of the Higgs mechanism, even though the particle does not explicitly participate in the scattering.

In the SMEFT predictions, one can verify from Table~\ref{tab:bwtz} that when both weak bosons are transversely polarised, the only source of energy growth comes from operators that introduce new Lorentz structures. Following the discussion of Section~\ref{subsec:contact_terms}, the other sources of maximal growth arising from the right- and left-handed current operators can be mapped to the dimension-6 contact terms
\begin{alignat}{2}
    \label{eq:contact_ophitb}
\Op{\phi tb} &= i(\tilde{\varphi} D_\mu \varphi)(\bar{t} \gamma^\mu b) +\text{h.c.}\quad 
&\to\quad &
G^{\sss 0} \, \partial_\mu G^{\sss +} \, \bar{t}_{\sss R} \gamma^\mu b_{\sss R} 
+\text{h.c.},\\
    \label{eq:contact_o3phiQ}
\Opp{\phi Q}{(3)} &= i\big(\phi^\dagger\lra{D}_\mu\,\tau_{\sss I}\phi\big)
\big(\bar{Q}\,\gamma^\mu\,\tau^{\sss I}Q\big) \quad
&\to\quad &
G^{\sss 0}  \, \lra{\partial}_\mu G^{\sss +} \, \bar{t}_{\sss L} \gamma^\mu b_{\sss L}
+\text{h.c.}.
\end{alignat}
In unitary gauge the effects of these operators map to the AC analysis above. Although both $\Opp{\phi Q}{(1)}$ and $\Opp{\phi Q}{(3)}$ both modify $\gzbl$  and $\gztl$, only the latter contributes to Eq.~\eqref{eq:bwtz_max}. The former is an SU(2) singlet current which does not `split' the ($t$,$b$) doublet. All three neutral current operators contribute to the subleading term of Eq.~\eqref{eq:bwtz_subl}, along with $\Op{\phi D}$, that modifies the $Z$ boson mass eigenstate, shifting the last term in the expression.

Table~\ref{tab:bwta} shows the corresponding helicity amplitudes for the $b\,W \to t\,\gamma$ process. Apart from the expected maximal growth from $\Op{tB}$ and $\Op{tW}$, we also observe subleading, linear growths from the two bosonic operators $\Op{\phi WB}$ and $\Op{W}$ in configurations with both longitudinal and linear $W$ polarisations.
\subsubsection{$tWj$\label{subsubsec:tWj}}
\begin{figure}[h!]
  \centering 
  \includegraphics[width=0.35\linewidth]{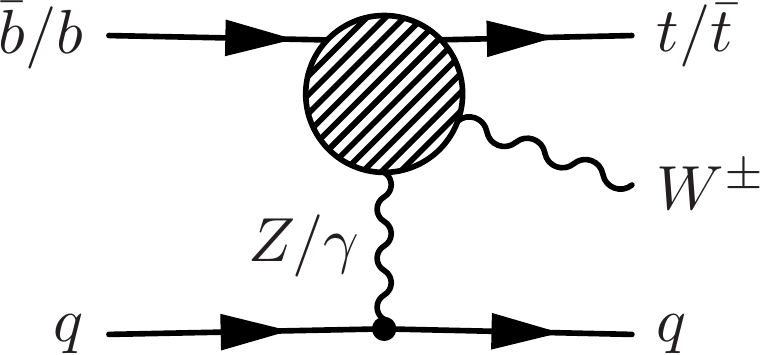}
  \hspace{1cm}
  \includegraphics[width=0.35\linewidth]{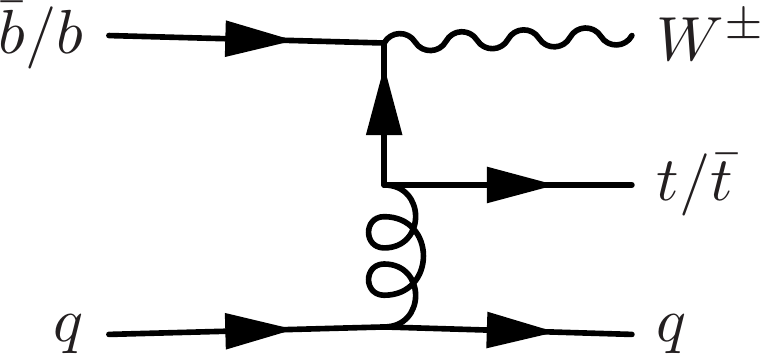}
  \caption{ 
    \emph{Left}: schematic Feynman diagram for the EW-induced $tWj$ process and its embeddings of the $b\,W\to t\,Z$ and $b\,W\to t\,Z$ subamplitudes. \emph{Right}: sample Feynman diagram for the QCD-induced component of $tWj$, arising as real-QCD radiation from the underlying $tW$ production process.
  \label{fig:diag_tWj}}
\end{figure} 

Depicted by the Feynman diagrams in Figure~\ref{fig:diag_tWj}, $tWj$ can be considered a `hybrid' between the $t$-channel and $tW$ single-top production mechanisms. On one hand, the purely EW induced process is similar to the $tZj$ process, simply interchanging the virtual and final state gauge bosons and has a cross section around 1 pb. On the other, it has a significant QCD-induced component arising due to real radiation from $tW$ production, which we compute to be of order 20 pb for a jet $p_T$ cut of 40 GeV. As discussed in Section~\ref{subsec:SMEFT}, this component can be affected by modified QCD interactions, albeit mainly in different regions of phase space from the high energy limit of the EW sub-amplitude. Furthermore, and perhaps crucially, the $tWj$ final state only differs after top decay from $t\bar{t}$ by a single $b$-tag. A $t\bar{t}$ event with one $b$-quark mis-tagged as a light jet would have an identical final state to $tWj$, suggesting that significant kinematical discrimination techniques would have to be employed to isolate the EW component of this process. Concerning EW sub-amplitudes, $tWj$ contains both the $bW\to tZ$ and $bW\to t\gamma$ scatterings.

\begin{figure}[h!]
  \centering \includegraphics[width=\linewidth]{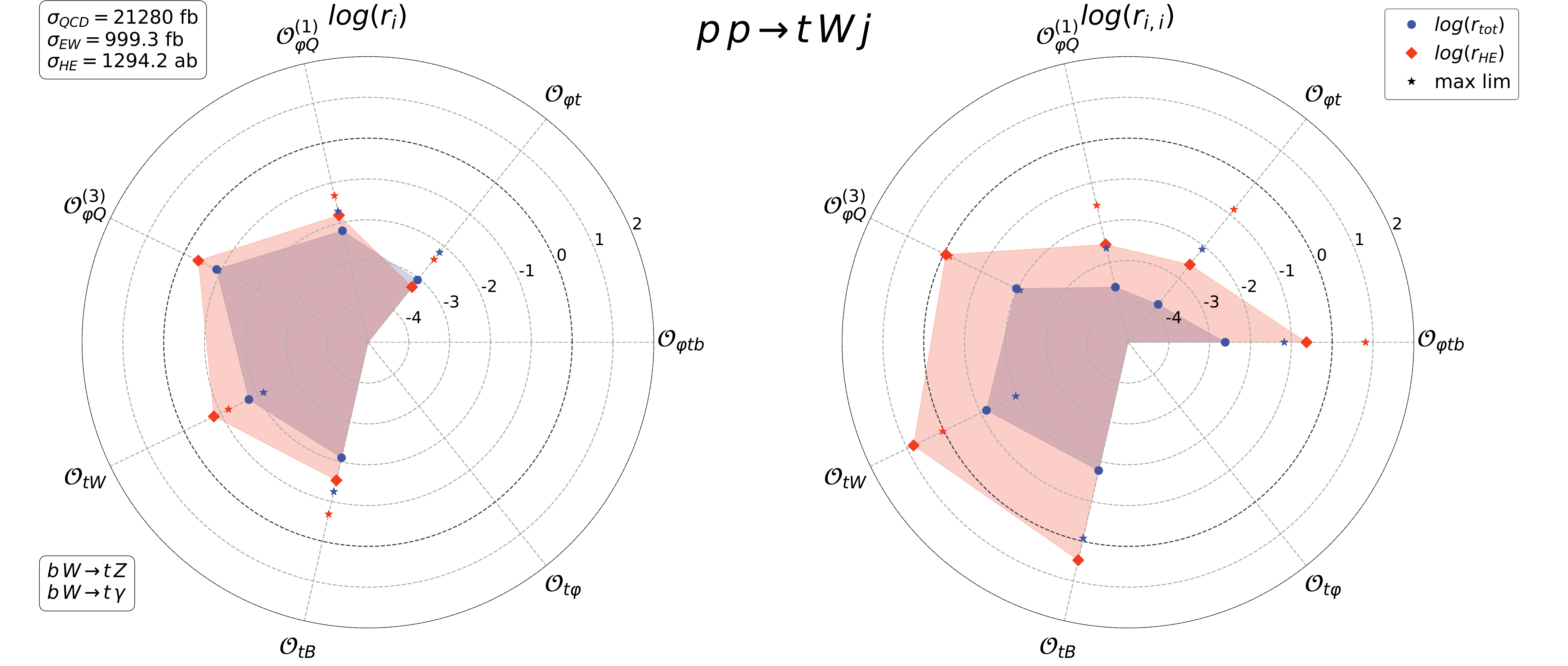}
  \caption{ 
    Radar plot for the $p \, p \to t \, W \, j$ process at the 13 TeV LHC.
  The left and right figures show the impact of each operator, for a Wilson coefficient of 1 TeV$^{-2}$, at linear (absolute value of SM--EFT interference) and quadratic (EFT amplitude squared) level in each Wilson coefficient, relative to the SM EW contribution, $\sigma_{\sss EW}$. The inclusive QCD-induced cross section, $\sigma_{\sss QCD}$, is also shown. The relative impacts at inclusive level and in a high energy region of phase space are depicted by the blue and red dots, respectively (see main text). The stars denote the corresponding prediction when saturating the individual limits on the coefficients summarised in Table~\ref{tab:constraints}.
  \label{fig:radar_twj_LHC13}}
\end{figure} 

We present the results of our sensitivity analysis in a compact format by making use of radar plots of the kind
shown in Fig~\ref{fig:radar_twj_LHC13}. In the top left corner the inclusive
EW and QCD cross sections for the process are displayed. The bottom left corner recalls the list of the 
$2 \to 2$ sub-amplitudes probed. We plot the ratios $r_i$ and $r_{i,i}$ on a logarithmic scale
for each of the selected operators, highlighting the circle corresponding to a $100\%$ relative
contribution to the SM. Since the $r_i$ are non-positive-definite, we display their absolute value. The blue dots show the value of the ratios at the inclusive level, $r^\text{tot}$, while the
red dots in the high energy region of phase space, $r^\text{HE}$. For the $t \, W\, j$ final state for example, 
we impose a $p_T$ requirement of 500 GeV on both the top quark and the associated boson to access the high energy region. The stars
provide information on the absolute size of the impact when the current limits on the Wilson
coefficients (Table~\ref{tab:constraints}) are saturated. The shading is added simply as a visual aid.

By looking at the radar plot for $t W  j$ one can 
see that not only is the process more sensitive to certain operators than
others, but also that the high energy cut enhances some more than others. A common feature to all processes, clearly noticeable in Fig~\ref{fig:radar_twj_LHC13},
is the fact that the high energy cut enhances the square contribution more  than the
interference, as naively expected by the energy behaviour of the $2 \to 2$ sub-amplitudes. The constraining power for this process to the right handed neutral current operator, $\Op{\phi t}$ seems feeble,
while the dipoles have a more significant impact both at the interference and square level. For the interference, the high energy phase space cut uncovers some growth in the relative impact of the operators, with the exception of $\Op{\phi t}$, that seems to go down with energy. This is indicative of a cancellation in phase space over this contribution from this operator, given that the $\Opp{\phi Q}{(1)}$, which has the same expected high energy behaviour in Table~\ref{tab:bwtz}, does show a mild growth. This is further supported by looking at the squared contribution of both operators, which appear to behave in a similar way. The triplet current operator, $\Opp{\phi Q}{(3)}$, shows a similar interfering growth to $\Opp{\phi Q}{(1)}$, even though it has a higher expected degree of energy growth. The higher growth with respect to the other two current operators is displayed by its squared contribution. Saturating the limits on the operators yields effects up to $O(0.1-1)$ on the inclusive rate for the dipoles and $\Opp{\phi Q}{(3)}$ that could be significantly enhanced, should the high energy region of this process be observed.

\subsubsection{$tZj$ \& $t\gamma j$\label{subsubsec:tZj_taj}}
\begin{figure}[h!]
  \centering 
  \includegraphics[width=0.35\linewidth]{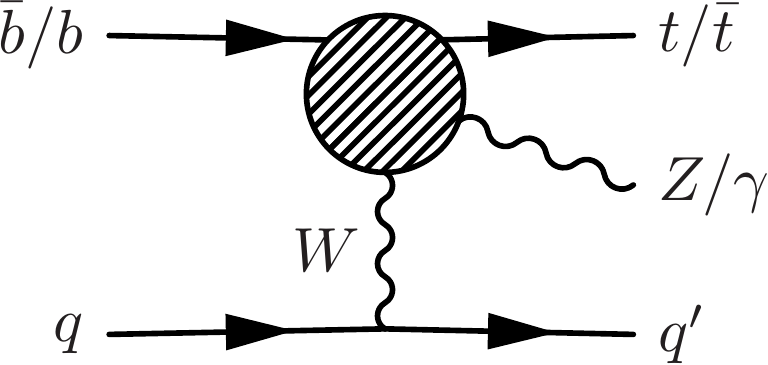}
  \hspace{1cm}
  \includegraphics[width=0.35\linewidth]{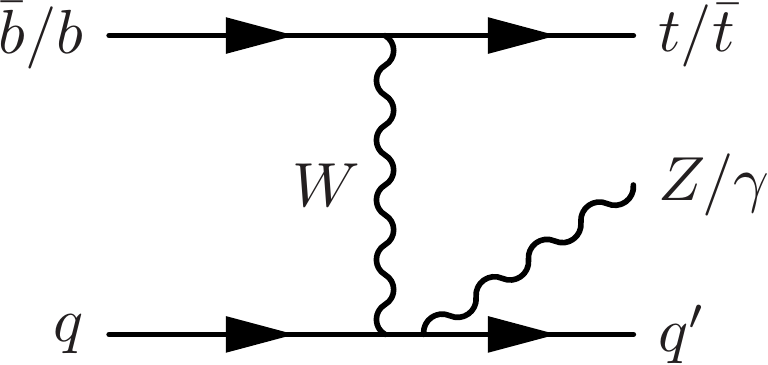}
  \caption{ 
    \emph{Left}: schematic Feynman diagram for the EW-induced $tZj$/$t\gamma j$ processes and the embeddings of the $b\,W\to t\,Z$ and $b\,W\to t\,Z$ subamplitudes. \emph{Right}: sample Feynman diagram for the topologies which do not embed the subamplitudes of interest.
  \label{fig:diag_tZj_taj}}
\end{figure} 
These processes correspond to the additional emission of a $Z$ boson or photon from the $t$-channel single-top process (see Figure~\ref{fig:diag_tZj_taj}). The $tZj$ process was recently observed for the first time at the LHC~\cite{Aaboud:2017ylb,Sirunyan:2017nbr,Sirunyan:2018zgs} while evidence for $t\gamma j$~\cite{Sirunyan:2018bsr} was also observed last year. While they both admit the radiation of the gauge boson from any of the four fermion legs, only a subset of these diagrams identify with the embedding of the associated $b\,W\to t\,Z$  or $b\,W\to t\,\gamma$ sub-amplitudes (left vs. right diagrams  in Figure~\ref{fig:diag_tZj_taj}). In the SM, $t Z j$ has a cross section of around 600 fb while $t\gamma j$ is about a factor of 2--3 larger. The high energy behaviour of $tZj$ was studied with respect to the $2\to 2$ scatterings in a preliminary way in Ref.~\cite{Degrande:2018fog} and was found to show promising sensitivity to top quark interactions in the SMEFT. This process is additionally affected by four-quark operators but not the chromomagnetic or triple gluon ones (at leading order). Furthermore, it was argued in Ref.~\cite{Degrande:2018fog} that single-top production is most likely to best constrain these Wilson coefficients.

\begin{figure}[h!]
  \centering \includegraphics[width=\linewidth]{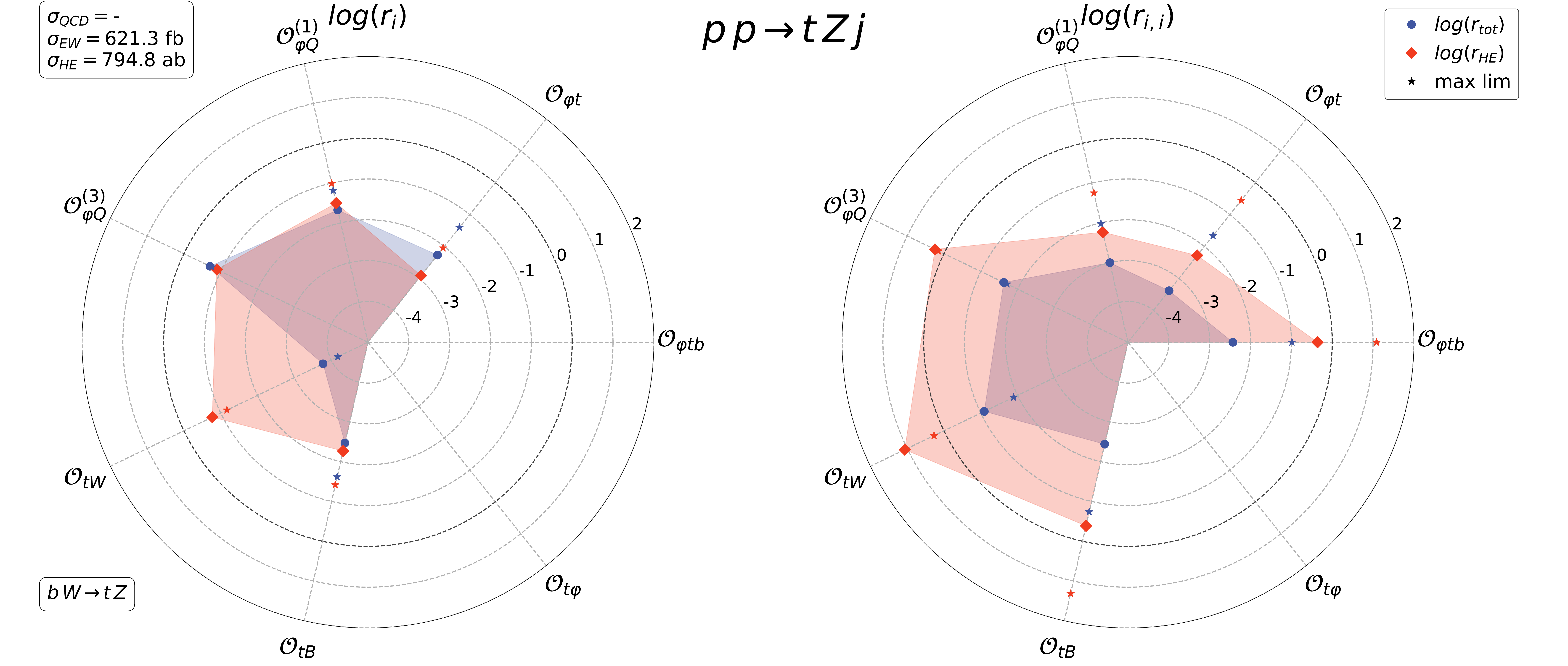}
  \centering \includegraphics[width=\linewidth]{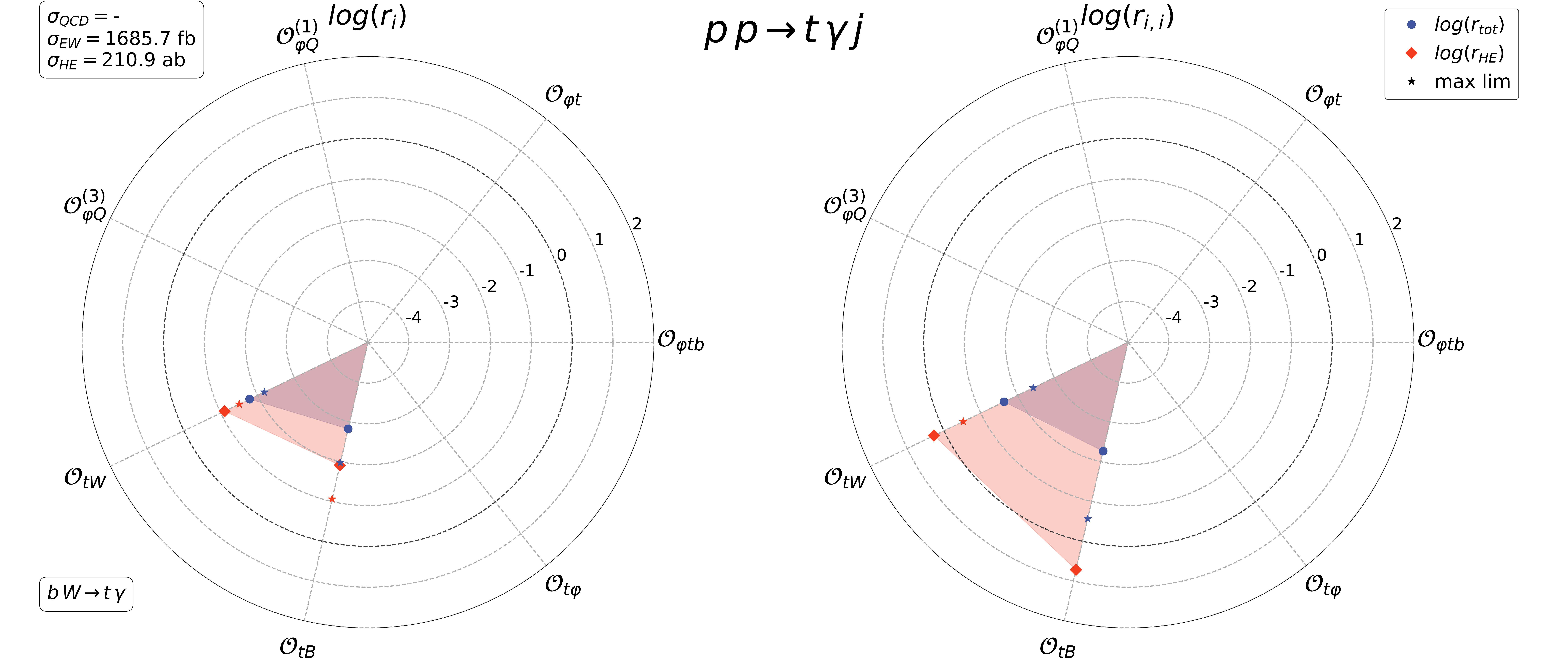}
  \caption{ 
    Radar plots for the $p \, p \to t \, Z \, j$ and $p \, p \to t \, \gamma \, j$ processes at the 13 TeV LHC, see
  Figure~\ref{fig:radar_twj_LHC13} and Section~\ref{subsubsec:tWj} of the main text for a detailed description.
  \label{fig:radar_tzj_taj_LHC13}}
\end{figure} 

$tZj$ is particularly interesting as its corresponding $b\,W\to t\,Z$ subprocess is sensitive to 6 of the 7 operators that we study in detail here, with the exception of the Yukawa operator. The upper plot of Figure~\ref{fig:radar_tzj_taj_LHC13} reports our results for the 13 TeV LHC in the operator space. It shows that energy-growing interference contribution expected for $\Opp{\phi Q}{(3)}$, discussed in Section~\ref{subsubsec:bwtz_bwta}, is not observed at inclusive level. Subsequent investigation revealed that the cross section for the production of purely longitudinal $Z$ bosons does display the expected energy growing interference. This suggests that the full process is dominated by the transverse modes, even after the high-energy cut. The only significant high energy behaviour to be noted at interference level is the behaviour of $\Op{tW}$, that displays a huge energy enhancement suggestive of a cancellation at inclusive level. This is not present in the $t\gamma j$ process, which is only sensitive to the two dipole operators, that display the expected energy growing interference. The neutral current operators show a milder energy growth, as expected, with the right handed $\Op{\phi t}$ even displaying a decrease with energy.  

The squared contributions from the SMEFT operators, however, exhibit the strong growth with energy in line with the expectations from the $2 \to 2$ computations. The operators that mediate a $G^+\partial G^0\,\bar{t}b$ contact interaction, $\Opp{\phi Q}{(3)}$ and $\Op{\phi tb}$, grow significantly, as do the dipoles, while the two other contact operators show slightly less pronounced but nonetheless significant growth. The relatively large cross section of both processes will most likely permit differential measurements to be performed that could provide access to the high energy behaviour of the EW sub-amplitudes. 
\subsubsection{$tZW$ \& $t\gamma W$\label{subsubsec:tWZ_tWa}}
\begin{figure}[h!]
  \centering 
  \includegraphics[width=0.35\linewidth]{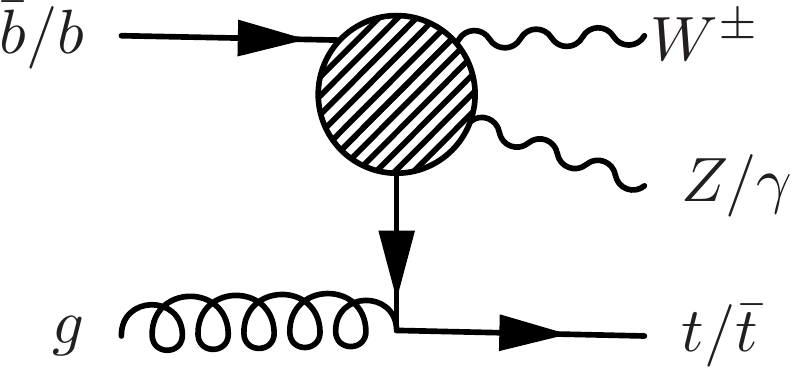}
    \hspace{1cm}
  \includegraphics[width=0.35\linewidth]{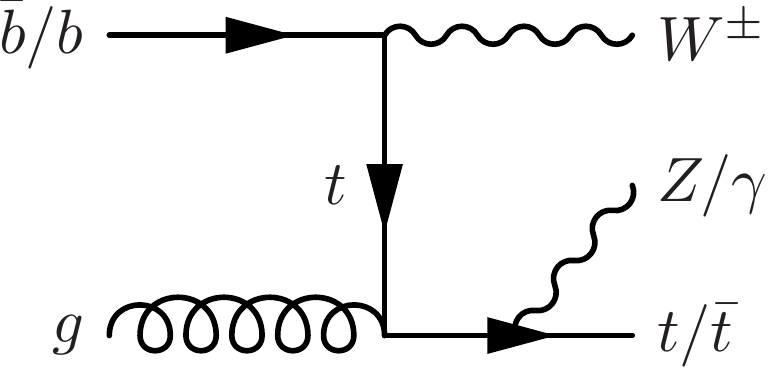}
  \caption{ 
\emph{Left}: schematic Feynman diagram for the EW-induced $tZW$/$t\gamma W$ processes and the embeddings of the $b\,W\to t\,Z$ and $b\,W\to t\,Z$ subamplitudes. \emph{Right}: sample Feynman diagram for the topologies which do not embed the subamplitudes of interest.  
\label{fig:diag_tZW_taW}}
\end{figure} 

In analogy with $tZj/t\gamma j$ to $t$-channel single-top, $tZW$ and $t\gamma W$ arise from the additional emission of a $Z$ boson or photon from the $tW$ single-top process (see Figure~\ref{fig:diag_tZW_taW}). They are sensitive to the same sub-amplitudes as their $t$-channel counterparts. Again, there exist some topologies, in which the neutral boson is radiated from the top, that do not embed the EW-top quark scatterings. The relative smallness of $tW$ compared to $t$-channel single-top production is propagated to the relative rates of $t(Z/\gamma)j$ and $t(Z/\gamma)W$. One important difference between the two sets of processes, however, is the fact that the bosons participating in the $2\to 2$ scattering are both in the final state in this case. This means that their polarisation can in principle be measured. The contribution of each polarisation to the full $2\to3$ rate is separate and does not depend on its interaction with the light quark current, in contrast to $tZj$ and $tHj$. A significant experimental caveat, however, is the similarity of the $tXW$ final state to that of $t\bar{t}X$. After decaying the top quarks, the two only differ by a single $b$-quark. This is conceptually identical to the overlap between the $tW$ and $t\bar{t}$ final states. It may be experimentally challenging to distinguish the two processes, the latter of which has much larger cross section owing to the QCD-induced component. However, the fact that evidence for the $tW$ single-top channel was observed already at 7 TeV gives us hope that $tXW$ is not out of reach at the LHC.

\begin{figure}[h!]
  \centering \includegraphics[width=\linewidth]{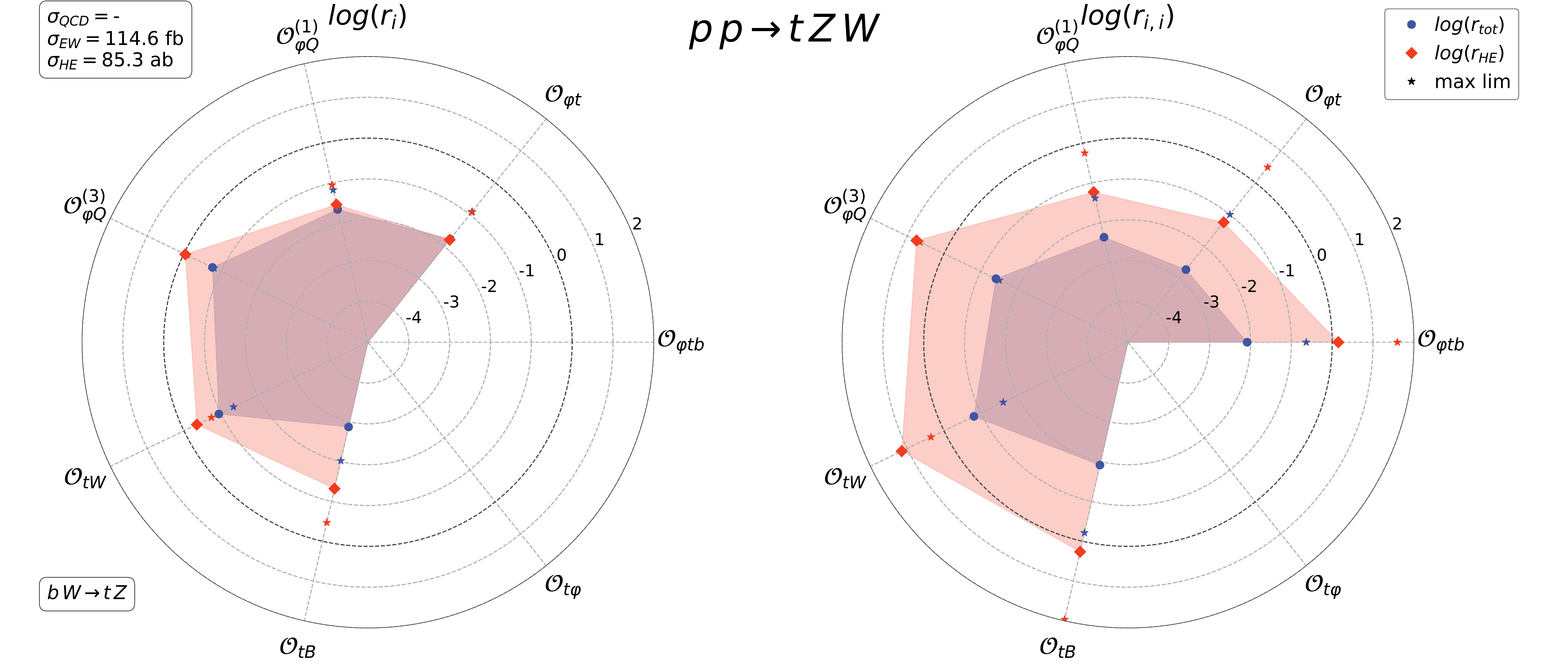}\\[1.5ex] \includegraphics[width=\linewidth]{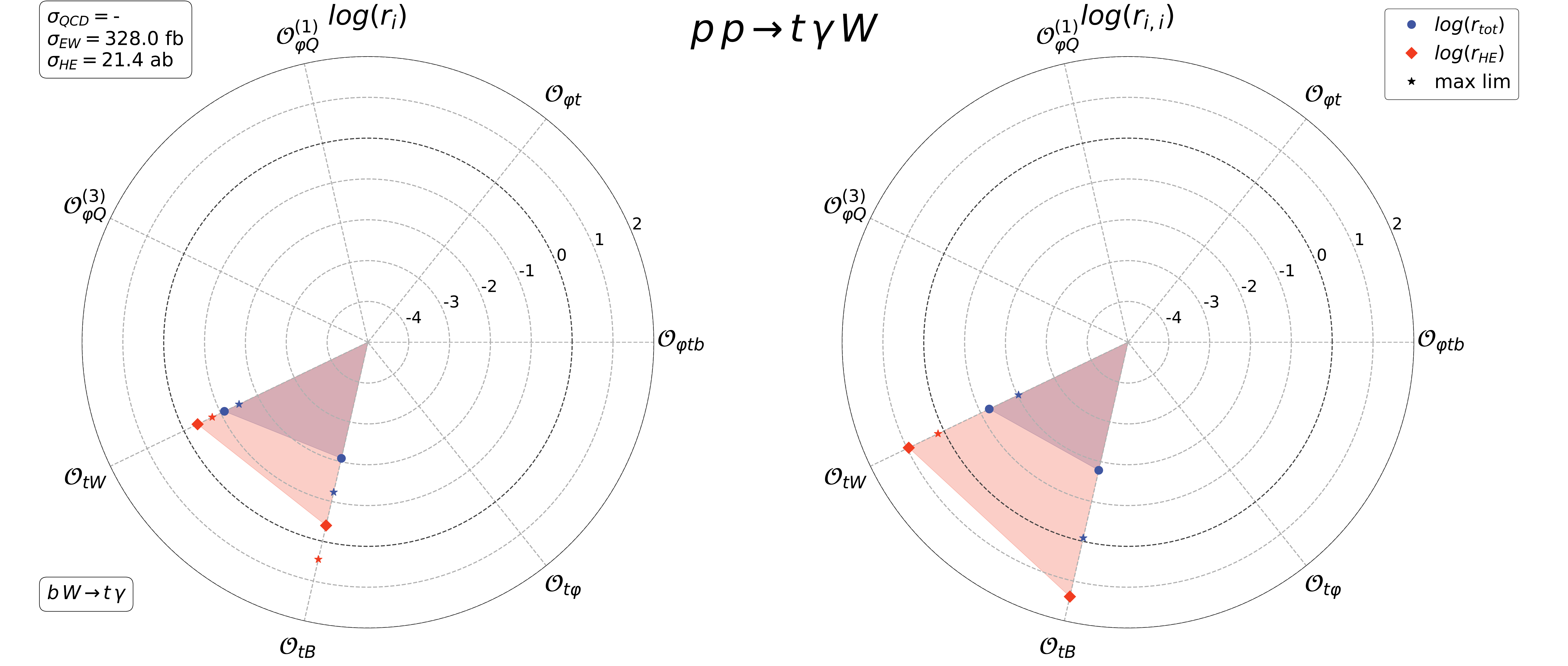}
  \caption{Radar plot for the process $p \, p \to t \, Z \, W$ and $p \, p \to t \, \gamma\, W $ at the 13 TeV LHC, see
  Figure~\ref{fig:radar_tzj_taj_LHC13} and Section~\ref{subsubsec:tWj} of the main text for a detailed description.
 \label{fig:radar_tzw_taw_LHC13}}
\end{figure}

This study has revealed $tZW$ production to be a particularly interesting candidate for probing the high energy behaviour of $b\,W\to t\,Z$ scattering. Figure~\ref{fig:radar_tzw_taw_LHC13},
demonstrates a clear improvement of inclusive sensitivity and energy growth compared to $tZj$ (Figure~\ref{fig:radar_tzj_taj_LHC13}), across the board. The process exhibits the expected energy growth at interference level for $\Opp{\phi Q}{(3)}$, without any evidence of a cancellation. The cancellation of the $\Op{tW}$ interference term in the inclusive cross section is also not present, leading to an improved overall sensitivity. The energy growth observed for this term is not naively expected from the $2\to 2$ analysis. There does, however, appear to be a cancellation in the inclusive $\Op{tB}$ interference contribution.  The energy growth of the squared terms is an enhanced version of the $tZj$ process. From a theoretical perspective, we find this process to be extremely promising. A similar improvement in the relative impact of the dipole operators is also found for $t\gamma W$ compared to $t\gamma j$. Furthermore, the only non-EW operator that can contribute to it is $\Op{tG}$ and, in particular, it is not directly affected by four-quark operators. On a practical level, the corresponding cross section is about five times smaller than $tZj$, meaning that it may take more time to observe this process and the possibility of differential measurements is yet unclear. Overall, we conclude that it would be very valuable in the context of globally constraining the SMEFT in the top sector to take on the challenge of measuring this process and distinguishing it from $t\bar{t}Z$. 
\subsection{With the Higgs\label{subsec:onetop_yeshiggs}}
\subsubsection{$b\,W\to t\,h$ scattering\label{subsubsec:bwth}}
	\begin{figure}[h!]
		\centering
        \subfloat[]{
        \includegraphics[width=0.25\linewidth]{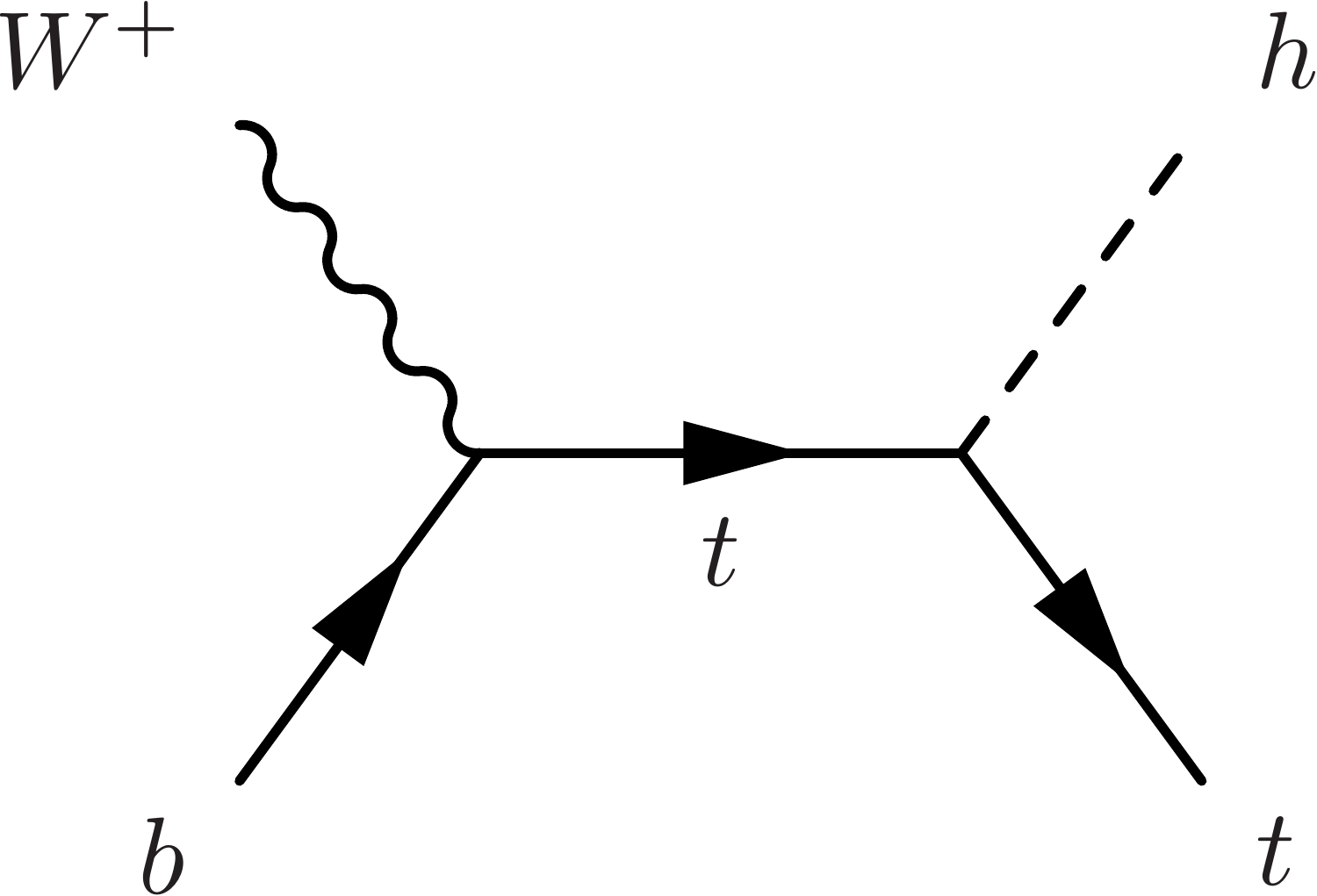}
        } 
        \subfloat[]{
        \includegraphics[width=0.25\linewidth]{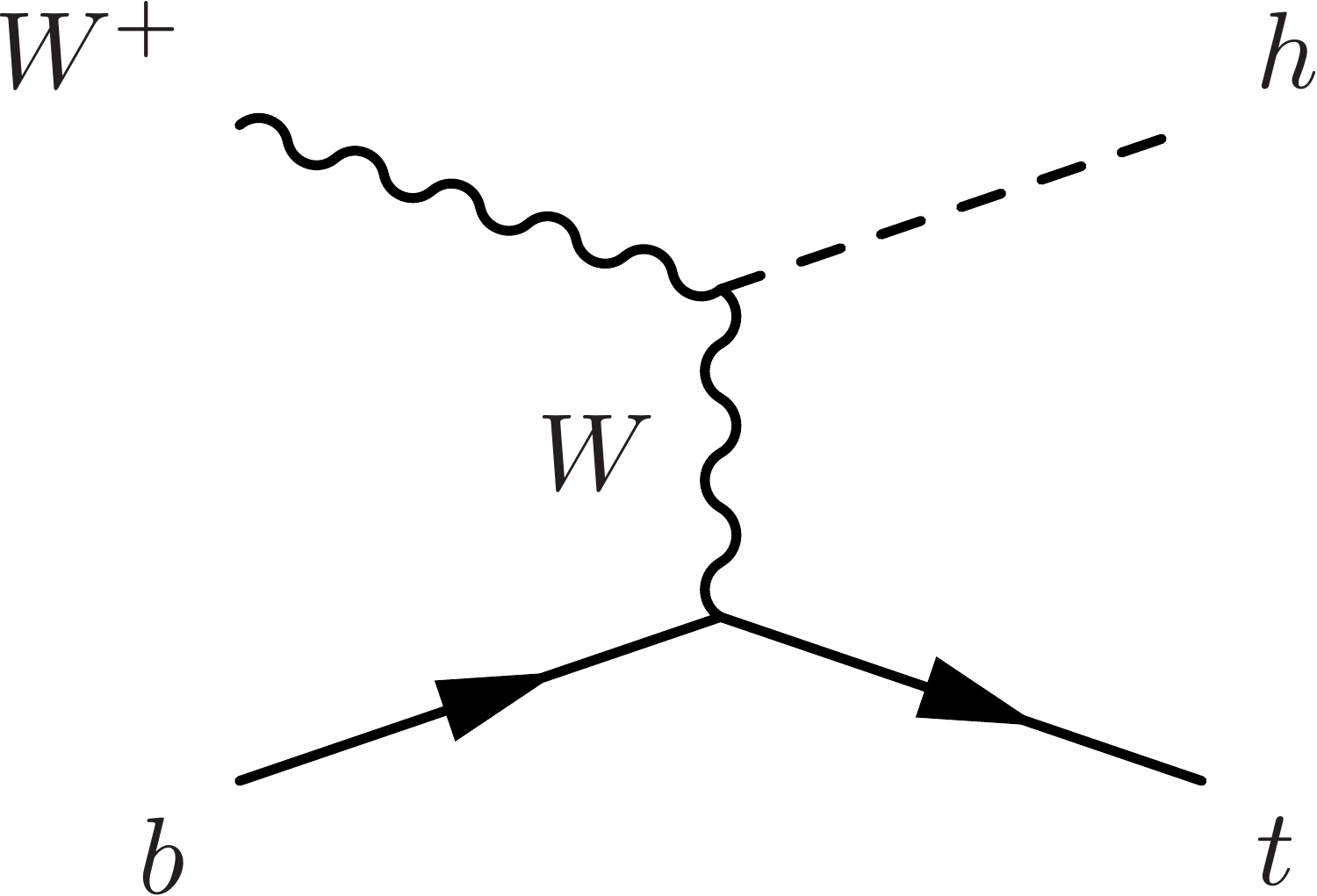}
        }
        \subfloat[]{
        \includegraphics[width=0.25\linewidth]{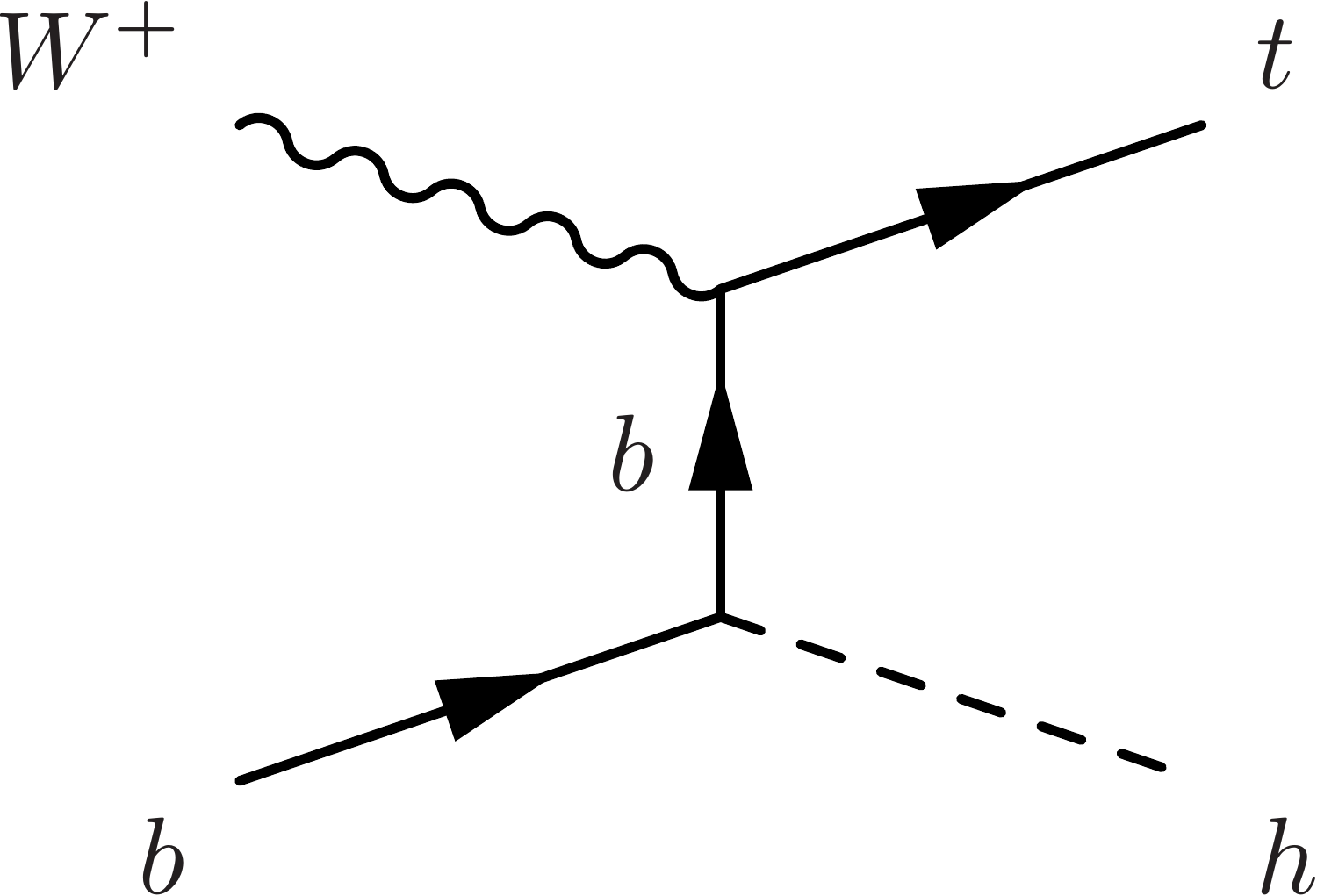}
        }
 
		\caption{SM diagrams for the $b W \to t h$ subprocess. \label{bwth_sm}}
	\end{figure}
\noindent This process probes the coupling of the top quark with the Higgs and with the $W$, as can be seen in Fig. \ref{bwth_sm}, depicting the contributing unitary gauge Feynman diagrams. Contrary to $b\,W\to t\,Z$, the third diagram can be neglected in principle, since the $b$-quark is nearly massless and therefore its coupling to the Higgs negligible. As with  $b\,W\to t\,Z$, the $Wtb$ vertex is present in every diagram, and leads to an overall rescaling of the rate. 

The helicity amplitude table for this process is given in Table~\ref{tab:bwth}. From the SMEFT perspective the table can be interpreted in terms of contact terms and helicity flips, as previously explained. The leading energy growth proportional to $\sqrt{s(s+t)}$ comes from the $\Opp{\phi Q}{(3)}$ contribution to the longitudinal, left handed configuration, $(-1,0,-1)$. An analogous term exists for the right-handed charged current operator, $\Op{\phi tb}$. These are attributed to the dimension-6 contact terms, analogous to Equations~\eqref{eq:contact_o3phiQ} and~\eqref{eq:contact_ophitb},
 in which the Higgs boson takes the place of the neutral Goldstone boson,
\begin{alignat}{2}
    \label{eq:contact_ophitb_2}
\Op{\phi tb} &= i(\tilde{\varphi} D_\mu \varphi)(\bar{t} \gamma^\mu b) +\text{h.c.}\quad 
&\to\quad &
h \, \partial_\mu G^{\sss +} \, \bar{t}_{\sss R} \gamma^\mu b_{\sss R} 
+\text{h.c.},\\
    \label{eq:contact_o3phiQ_2}
\Opp{\phi Q}{(3)} &= i\big(\phi^\dagger\lra{D}_\mu\,\tau_{\sss I}\phi\big)
\big(\bar{Q}\,\gamma^\mu\,\tau^{\sss I}Q\big) \quad
&\to\quad &
h \, \lra{\partial}_\mu G^{\sss +} \, \bar{t}_{\sss L} \gamma^\mu b_{\sss L}
+\text{h.c.}.
\end{alignat}
There are no AC analogues to these interactions, being a pure contact terms induced by the gauge-invariant, dimension-6 operators. 
We note the continued appearance of maximal growth due to the dipole operator, $\Op{tW}$, as well as a linear growth from $\Op{\phi W}$. The former induces a contact term analogous to Equation~\eqref{eq:dipole_contact}, with a Higgs boson in place of the neutral Goldstone boson. The latter arises due to the higher-derivative, dimension-5 $WWh$ interaction that it induces.

The $tth$ coupling and $WWh$ in the SM are related such that there is a cancellation of energy growth~\cite{Maltoni:2001hu,Farina:2012xp}. Specifically, there is a contribution to the helicity configuration $(-1,0,1)$, 
	\begin{equation}
        \label{eq:bwth_subl}
		\sqrt{-t}\,(2\mw^2 y_t - \gwh m_t),
	\end{equation}
which cancels when assigning the SM values to the couplings. There is no higher degree energy growth due to ACs for this process. Mapping the SMEFT to these ACs, we see that the Yukawa operator $\Op{t \phi}=(\phi^\dagger\phi-\frac{v^2}{2})\bar{Q}t\tilde{\phi}$, which modifies the $tth$ coupling, indeed gives a contribution proportional to $\sqrt{-t}$ in the same helicity configuration. This operator modifies both the $tth$ vertex and the mass of the top, but does so with different contributions due to combinatorics
. If the alterations were equal, one would have no energy growing behaviour. Our definition of the operator modifies $y_t$, but not $m_W, m_t$ or $\gwh$. The exact anomalous coupling that is modified is basis dependent, \textit{i.e.}, trading away the Yukawa operator with a Higgs doublet field redefinition would induce new operators that modify $m_W$ and $\gwh$ instead. Alternatively, redefining away the term proportional to $-v^2/2$ leads to a simultaneous modification of $y_t$ and $m_t$. All cases lead to the same physical effect. As mentioned in Section~\ref{subsec:contact_terms}, the energy growth can also be understood in terms of a $btG^+h$ contact term in Feynman gauge. 
\subsubsection{$thj$\label{subsubsec:thj}}
\begin{figure}[h!]
  \centering 
  \includegraphics[width=0.35\linewidth]{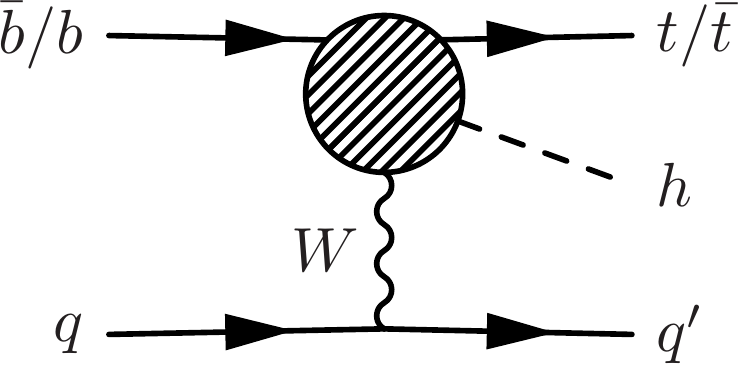}
  \caption{ 
Schematic Feynman diagram for the EW-induced $thj$ process embedding the $b\,W\to t\,h$ subamplitude.  
\label{fig:diag_thj}}
\end{figure} 
Similarly to $tZj$ and $t\gamma j$, $thj$ arises from the emission of a Higgs boson from the $t$-channel single-top process (see Figure~\ref{fig:diag_thj}).
It has the advantage of the Higgs boson only being radiated from the top quark leg, meaning that it is an especially clean probe of the $b\,W\to t\, h$ sub-amplitude, that is embedded in every diagram. This process was also studied in Ref.~\cite{Degrande:2018fog}, in which it was found to obey the expected high energy behaviour of the $2\to2$ subprocess and promised a good sensitivity to the relevant SMEFT operators. It is often cited as a probe of the sign of the top quark Yukawa coupling due to the unitarity non-cancellations that a `flipped' scenario would incur, as seen in Equation~\eqref{eq:bwth_subl}. However, such a modification of the coupling constitutes an $O(1)$ effect that does not fall in the remit of SMEFT, which assumes parametrically small deviations from SM interactions. Concerning non-EW operators, the discussion is identical to $tZj$ (Section~\ref{subsubsec:tZj_taj}).

\begin{figure}[h!]
  \centering  \includegraphics[width=\linewidth]{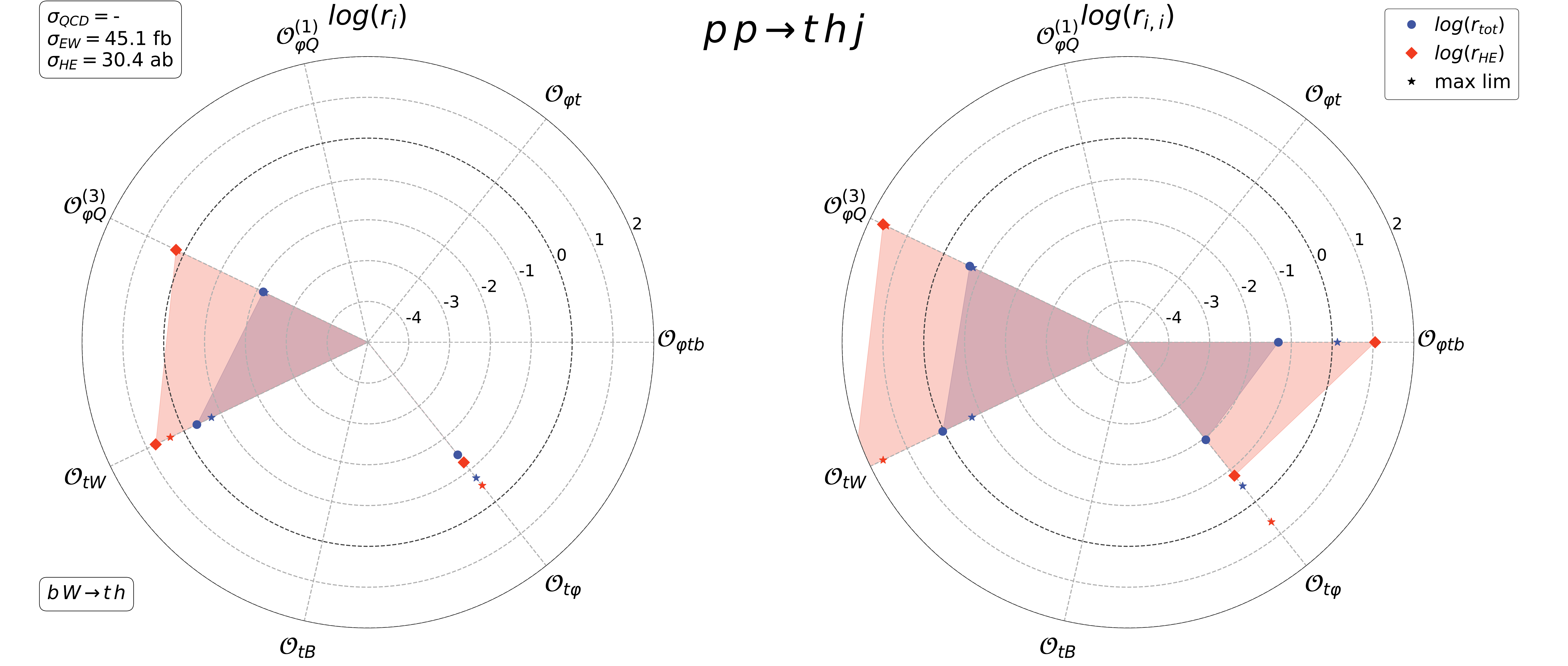}
  \caption{Radar plot for the process $p \, p \to t \, h \, j$ at the 13 TeV LHC, see Figure~\ref{fig:radar_tzj_taj_LHC13} and Section~\ref{subsubsec:tWj} of the main text for a detailed description.
 \label{fig:radar_thj_LHC13}}
\end{figure}
As seen in Figure~\ref{fig:radar_thj_LHC13}, contrary to the case of $tZj$, $thj$ displays an apparent growth at interference level for $\Opp{\phi Q}{(3)}$. This is, in fact, enhanced by a cancellation at inclusive level which is consistent with the fact that the absolute size of the inclusive interference piece (blue dot, lower left plot) is smaller than the squared one (blue dot, lower right plot). Although this is a normalisation dependent statement for a given process, comparison to other processes such as $tZj$ or $tZW$ (Figures~\ref{fig:radar_tzj_taj_LHC13} and \ref{fig:radar_tzw_taw_LHC13}) show that this relative suppression is not universal and therefore indicates the presence of a cancellation in $thj$. Analogous energy growth is found for $\Op{\phi tb}$, as expected from the presence of the same contact interaction as $\Opp{\phi Q}{(3)}$, which can only be seen in the quadratic term due to lack of interference in the 5-flavor scheme. We also observe a very strong dependence on $\Op{tW}$, whose impact grows with energy already at interference level (see discussion in Section~\ref{sec:EWA}). Saturating the limits on this coefficient still produce $O(1)$ effects on the inclusive rate, meaning that the measurement of this process at the LHC would considerably improve the current constraints on this operator. Finally, sensitivity to $\Op{t \phi}$ and $\Op{\phi tb}$ are found with $O(1)$ effects when saturating existing limits. Considerable energy growth is found from the unitarity violating behaviour highlighted in Section~\ref{subsubsec:bwth}. 

As discussed in  Ref.~\cite{Degrande:2018fog}, the relatively small cross-section coupled with the challenging final state means that it is unlikely that this process will be accessed at differential level. Existing searches for this mode suffer from a significant overlap with the much larger $t\bar{t}h$ process, meaning that current sensitivity is at the level of 25 times the SM prediction while the totally opposite sign Yukawa scenario has been excluded~\cite{Sirunyan:2018lzm}. The inclusive measurement of this channel and the differential measurement of the $tZj$ counterpart are therefore argued to be complementary ways to access the operators that we consider.
\subsubsection{$thW$\label{subsubsec:thW}}
\begin{figure}[h!]
  \centering 
  \includegraphics[width=0.35\linewidth]{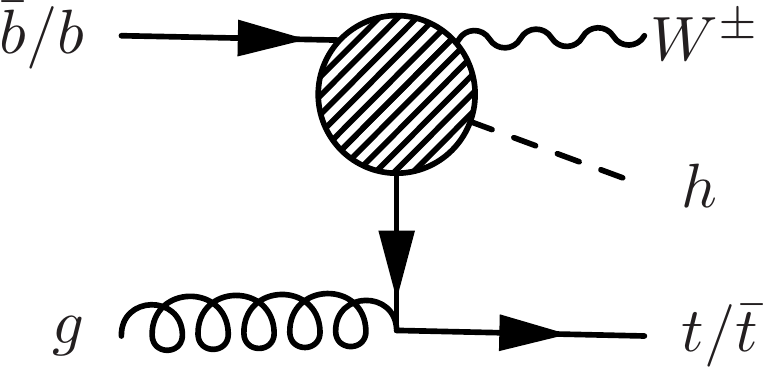}
  \hspace{1cm}
  \includegraphics[width=0.35\linewidth]{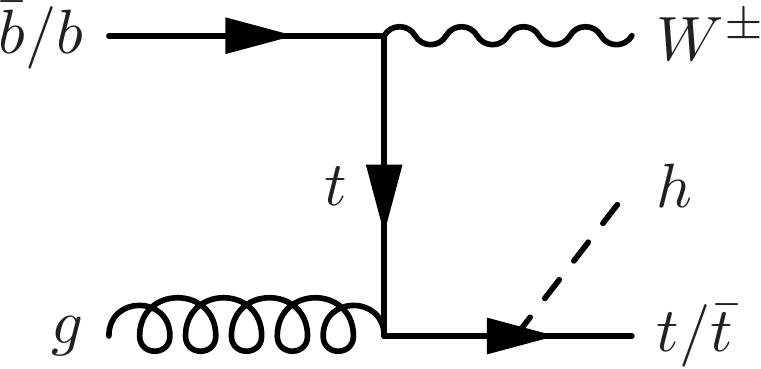}
  \caption{ 
\emph{Left}: schematic Feynman diagram for the $thW$ process embedding the $b\,W\to t\,h$ subamplitude. \emph{Right}: Feynman diagram for the contribution to $thW$ that does not embed the subamplitude of interest. 
\label{fig:diag_thW}}
\end{figure} 
Like $tZW$ and $t\gamma W$, $thW$ is based on $tW$ single top production with the additional emission of a Higgs. Unlike $thj$, however, it admits a topology that does not embed $b\,W\to t\,h$, shown on the right of Figure~\ref{fig:diag_thW}. Like its $Z/\gamma$ analogues, this process resembles the $t\bar{t}h$ final state up to a single $b$-jet, meaning that it is challenging to distinguish the two experimentally. Again, the non-EW operators affect it in a similar way to $tZW$, with the addition of a possible $\Op{\phi G}$ insertion allowing the Higgs to be emitted from the initial state gluon. 

\begin{figure}[h!]
  \centering  \includegraphics[width=\linewidth]{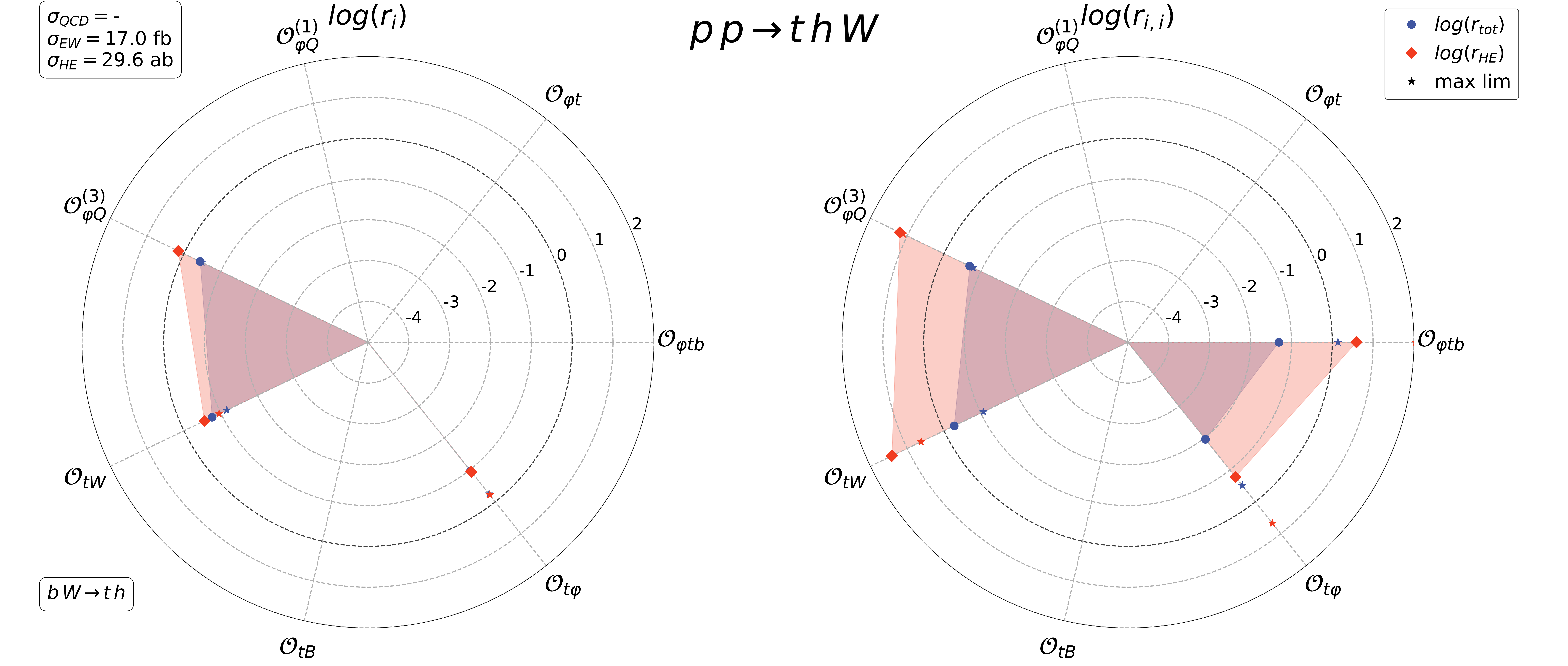}
  \caption{Radar plot for the process $p \, p \to t \, h \, j$ at the 13 TeV LHC, see Figure~\ref{fig:radar_twj_LHC13} and Section~\ref{subsubsec:tWj} of the main text for a detailed description.
 \label{fig:radar_thw_LHC13}}
\end{figure}
Interestingly, from a sensitivity perspective, $thW$ does not appear superior to $thj$ in the way that $tZW$ is to $tZj$. Figure~\ref{fig:radar_thw_LHC13} displays a slightly milder sensitivity to the set of operators entering in $b\,W\to t\,h$, albeit without the cancellation in the $\Opp{\phi Q}{(3)}$ interference term for the inclusive rate. The evidence from this process and $thj$ shows that the expected interfering growth in the subamplitude appears to be translated to the full $2\to3$ kinematics. There is, instead, a slight cancellation in the $\Op{tW}$ interference term. The global energy growing behaviour is analogous to $thj$, as expected from the fact that they contain the same subamplitude. Given that the total cross section is even smaller, we conclude that this process is not likely to be isolated at the LHC. It could however, be searched for in combination with $thj$ since they both have very similar sensitivity profiles. Indeed, the latest LHC search for single-top in association with a Higgs boson~\cite{Sirunyan:2018lzm} considers both of these channels as part of their signal region.

\section{$t\bar{t}$ scattering\label{sec:twotop_scattering}}
We now move to the scatterings involving two top quarks. The remaining external legs must consequently form an electrically neutral system comprised of a pair of neutral EW bosons or $W^+W^-$. The corresponding collider processes, have a top anti-top pair in the final state produced in association with one or more bosonic states and possibly forward jets. This means that $\Op{tG}$, $\Op{G}$\footnote{Except for $t\bar{t}W$, whose QCD-induced contribution does not have a gluon-initiated component, meaning $\Op{G}$ cannot contribute. } and four-fermion operators can always contribute to these processes in addition to those embedded in the $2\to2$ amplitudes. However, as discussed in Section~\ref{subsec:SMEFT}, these will always be better constrained by simpler processes such as $t\bar{t}$ and multijet. When there is a Higgs boson in the final state, $\Op{\phi G}$ insertions can also contribute to this process, allowing it to be emitted from a gluon line. Only the Vector Boson Fusion (VBF) process does not involve external bosonic states. 
\subsection{Without the Higgs\label{subsec:twotop_nohiggs}}
Table~\ref{tab:amp_proc_table_twotop_nohiggs} summarises the set of two-top amplitudes not involving the Higgs alongside the collider processes that embed them. As in the case of single-top scattering, there is always at least one process that uniquely probes a given scattering amplitude. $t\bar{t}Z(j)$, $t\bar{t}\gamma(j)$ are sensitive to more than one amplitude, due to the overlap between virtual $Z/\gamma$ exchange. At hadron colliders, VBF embeds all of the above amplitudes while at lepton colliders, $t\,W\to t\,W$  can be separately probed from the rest since it changes the final state accompanying the top pair to a neutrino pair.
\begin{table}[h!]
\centering
\begin{tabular}{|p{2cm}|P{1.1cm}|P{1.1cm}|P{1.1cm}|P{1.1cm}|P{1.1cm}|P{1.1cm}|P{1.1cm}|P{1.1cm}|}
     \hline
     
                              &$t\bar{t}W(j)$ & $t\bar{t}WW$ & $t\bar{t}Z(j)$ & $t\bar{t}\gamma (j)$ & $t\bar{t}\gamma\gamma$ & $t\bar{t}\gamma Z$ & $t\bar{t}ZZ$ & $VBF$ 
     \tabularnewline\hline             
                                       
     $t\,W\to t\,W$           & \cmark        & \cmark       &                &                      &                        &                    &              & \cmark
     \tabularnewline\hline

     $t\,Z\to t\,Z$           &               &              & \cmark         &                      &                        &                    & \cmark       & \cmark 
     \tabularnewline\hline

     $t\,Z\to t\,\gamma$      &               &              & \cmark         & \cmark               &                        & \cmark             &              & \cmark
     \tabularnewline\hline
 
     $t\,\gamma\to t\,\gamma$ &               &              &                & \cmark               & \cmark                 &                    &              & \cmark
     \tabularnewline\hline
\end{tabular}
 
\caption{The set of two-top $2\to2$ scattering amplitudes without Higgs bosons considered in this work mapped to the collider processes in which they are embedded.
\label{tab:amp_proc_table_twotop_nohiggs}}
\end{table}

A summary of the maximal energy growths obtained in our helicity amplitude computations, taken from Tables~\ref{tab:twtw}--\ref{tab:tata}, is shown in Table~\ref{tab:helamp_summary_twotop_nohiggs}. A clear favourite emerges in $t\,W\to t\,W$ scattering, which displays maximal and interfering energy growth for all current operators. It has equal or better energy growth for all other operators apart from $\Op{\phi B}$. In contrast, the other two amplitudes show at most linear growth in all cases barring the dipole operators, which have a tendency to grow maximally everywhere. 
\begin{table}[h!]
{\footnotesize
\setlength{\tabcolsep}{4pt}
\renewcommand{\arraystretch}{1.2}
\begin{center}
  \begin{tabular}{|c|c|c|c|c|c|c|c|c|c|c|c|c|c|}
 \hline
                          & $\Op{\phi D}$ & $\Op{\phi \Box}$ &  $\Op{\phi B}$ & $\Op{\phi W}$  & $\Op{\phi WB}$  & $\Op{W}$ & $\Op{t \phi}$ & $\Op{tB}$ & $\Op{tW}$ & $\Op{\phi Q}^{\sss (1)}$ & $\Op{\phi Q}^{\sss (3)}$ &  $\Op{\phi t}$ 
 \tabularnewline\hline

 $t\,W\to t\,W$           & $E$           & $E$              & $-$            & $E$            & $E$             & $E^2$    & $E$           & $E$       & $E^2$     & \red{$E^2$}              & \red{$E^2$}              & \red{$E^2$}               
 \tabularnewline\hline
 
 $t\,Z\to t\,Z$           & $E$           & $E$              & $E$            & $E$            & $E$             & $-$      & $E$           & $E^2$     & $E^2$     & $E$                      & $E$                      & $E$                       
 \tabularnewline\hline
 
 $t\,Z\to t\,\gamma$      & $-$           & $-$              & $E$            & $E$            & $E$             & $-$      & $-$           & $E^2$     & $E^2$     & $-$                      & $-$                      & $-$                       
 \tabularnewline\hline
 
 $t\,\gamma\to t\,\gamma$ & $-$           & $-$              & $E$            & $E$            & $E$             & $-$      & $-$           & $E$       & $E$       & $-$                      & $-$                      & $-$                       
 \tabularnewline\hline
 
%
%
 
  \end{tabular}
\end{center}
\renewcommand{\arraystretch}{1.}
}

\caption{\label{tab:helamp_summary_twotop_nohiggs}
Same as Table~\ref{tab:helamp_summary_singletop} for two top scattering amplitudes without the Higgs. See Tables~\ref{tab:twtw}--\ref{tab:tata} for the full helicity amplitude results.
}
\end{table}
\subsubsection{$t\,W\to t\,W$ scattering\label{subsubsec:twtw}}
\begin{figure}[h!]
        \centering
        \subfloat[]{
        \includegraphics[width=0.25\linewidth]{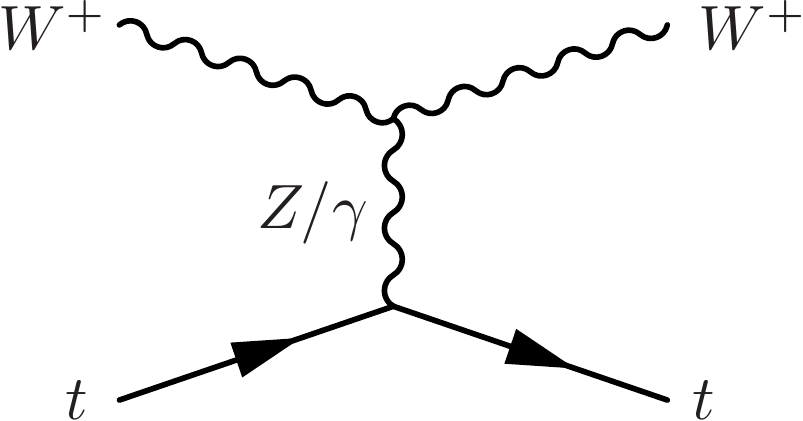}
        }  \subfloat[]{
        \includegraphics[width=0.25\linewidth]{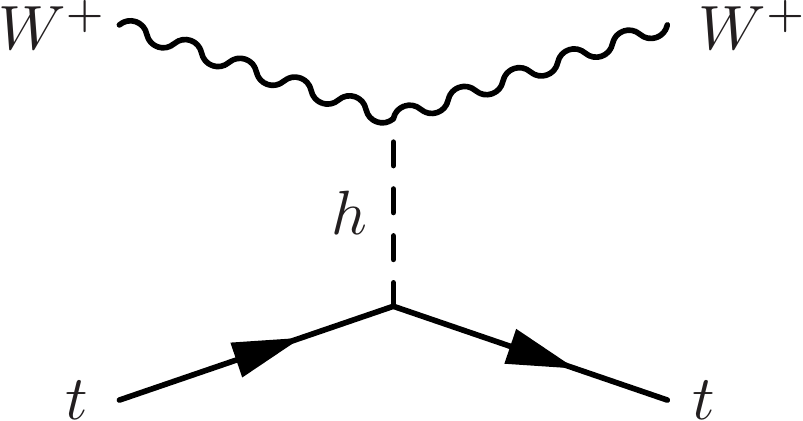}
        }
        \subfloat[]{
        \includegraphics[width=0.25\linewidth]{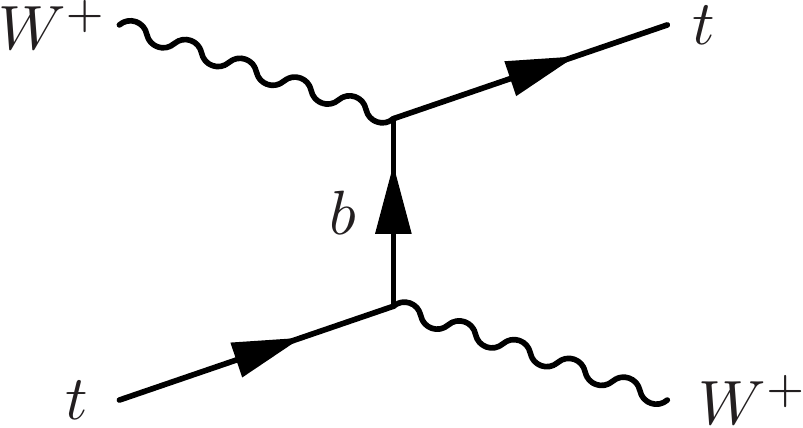}
        }
        \caption{SM diagrams for the $t W \to t W$ subprocess. \label{twtw_sm}}
    \end{figure} 
This scattering has a very rich coupling structure, involving the top-$Z$, top-$\gamma$, triple-gauge, top-Higgs,  $W$-Higgs and  $Wtb$ couplings (see Fig.~\ref{twtw_sm}). The AC calculation reveals several unitarity cancellations with $E^2$ growth, analogous to $b\,W\to t\,Z$ scattering. These are found in the fully longitudinal left- or right-handed helicity configurations 
    \begin{align}
        \label{eq:twtw_max_left}
        (-,0,-,0)\propto &\sqrt{s(s+t)}\,(\gbtw^2 - \gta \, \gwa - \gwz \, \gztl) \,,\\
        (+,0,+,0)\propto &\label{eq:twtw_max_right}
        \sqrt{s(s+t)}\,(\gta \, \gwa + \gwz \, \gztr) \,.
    \end{align}
These clearly probe the gauge structure of the theory and are sensitive to additional sources of spontaneous symmetry breaking. They are accompanied by corresponding linear ($m_{\sss V} \sqrt{-t}$) growths when one gauge boson helicity is `flipped', \emph{e.g.}, $(0\to-)$ for the left-handed configuration and $(0\to+)$ for the right-handed. Among the scatterings that we investigate, this amplitude turns out to be the only one that is explicitly sensitive to modifications of the $Wtb$ vertex. 

The remaining linear growths arise in the chirality-flipping, longitudinal channel $(\pm,0,\mp,0)$ and are proportional to
\begin{align}
    \sqrt{-t}(\gth \gwh-(2 \gta \gwa+\gwz (\gztl+\gztr)) \mt).
\end{align}
One can see that it probes the details of Higgs mechanism, as it contains both the kinematical top mass and its coupling to the Higgs.

Mapping to the SMEFT, Table~\ref{tab:twtw} is understood in analogy to the previous processes, in terms of contact terms with Goldstones or dimension-5 operators that introduce a new Lorentz structure. For example, the two maximal growth channels mentioned above are induced by the left or right-handed operators, $\Opp{\phi Q}{(3)}$ and $\Opp{\phi Q}{(1)}$ or $\Op{\phi t}$ that all source $G^\pm\partial_\mu G_\mp\,\bar{t}\gamma^\mu t$ contact interactions. 
\subsubsection{$t\bar{t}W(j)$\label{subsubsec:ttw_ttwj}}
\begin{figure}[h!]
  \centering 
  $\vcenter{\hbox{\includegraphics[width=0.35\linewidth]{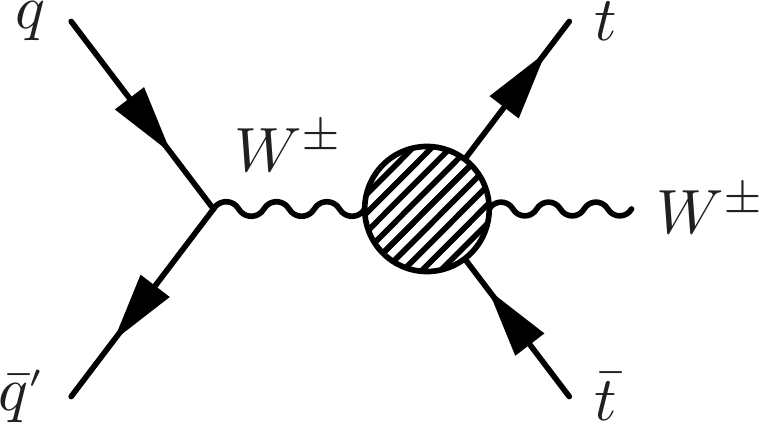}}}$
  \hspace{1cm}
  $\vcenter{\hbox{\includegraphics[width=0.28\linewidth]{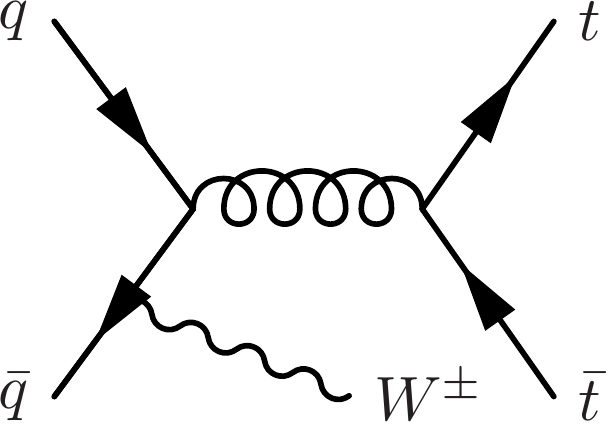}}}$
  \caption{ 
    \emph{Left}: schematic Feynman diagram for the EW-induced $t\bar{t}W$ process and its embedding of the $t\,W\to t\,W$ subamplitude. \emph{Right}: sample Feynman diagram for the QCD-induced $t\bar{t}W$, which does not probe modified top-EW interactions.
  \label{fig:diag_ttW}}
\end{figure} 
As seen in Figure~\ref{fig:diag_ttW}, $t\bar{t}W$ embeds the $t\,W\to t\,W$ sub-amplitude. In the SM, a $W$-boson is produced in the s-channel and subsequently emits a Higgs or $Z$ boson that splits into the $t\bar{t}$ final state. The process is unique among the class of $t\bar{t}X$ processes in that the corresponding QCD induced process cannot probe modified EW interactions in the top sector since the $W$ boson is uniquely emitted from the initial state quark leg. The presence of a highly off-shell intermediate state implies a $\sim1/s$ factor in the amplitudes that may negate potential energy growth from the subamplitude.
\begin{figure}[h!]
  \centering \includegraphics[height=0.2\paperheight]{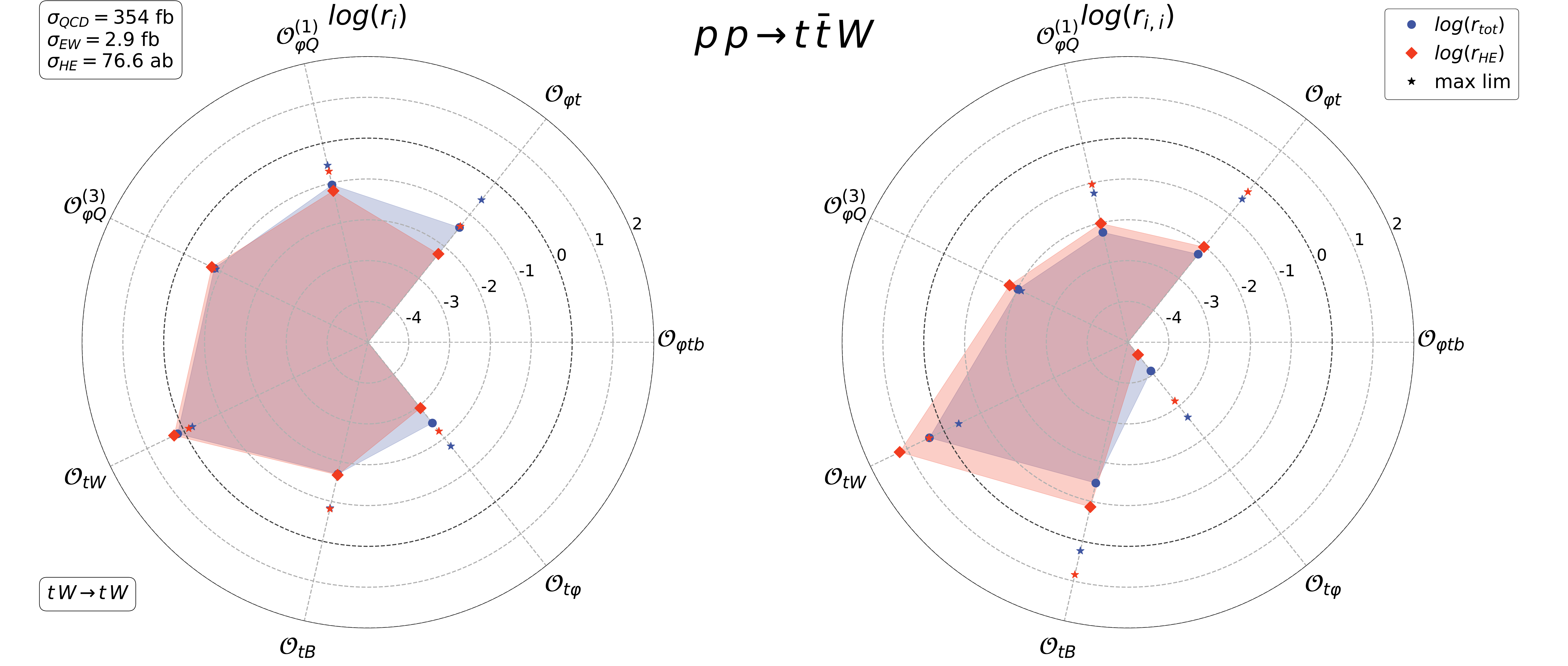}
  \hspace{1cm} \includegraphics[height=0.2\paperheight]{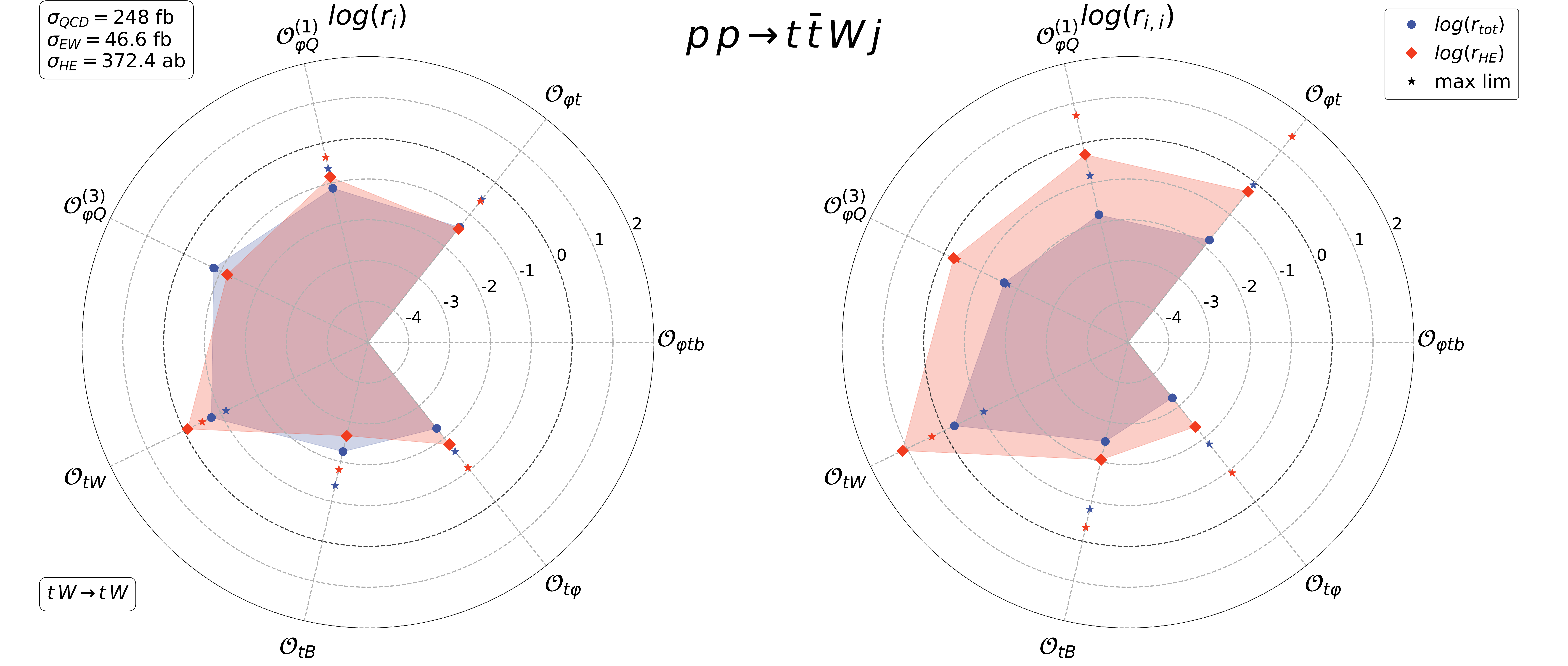}
  \caption{ 
   Radar plot for the $p \, p \to t \, \bar{t} \, W$ (\emph{upper}) and  $p \, p \to t \, \bar{t} \, Wj$ (\emph{lower}) processes at the 13 TeV LHC, 
 see
  Figure~\ref{fig:radar_twj_LHC13} and Section~\ref{subsubsec:tWj} of the main text for a detailed description.
  \label{fig:radar_ttw_LHC13}}
\end{figure}

The upper panel of Figure~\ref{fig:radar_ttw_LHC13}, shows that all operators contribute at or below the 10\% level for $c_i=1$ apart from $\Op{tW}$ that can produce $O(1)$ effects. We define our high energy region by imposing a $p_T$ cut of 500 GeV on both the top and the anti-top. No energy growth is observed at interference level and the effects at squared level for the current operators is very mild, while some limited growth is present in the case of the weak dipole operators. Combined with the fact that the QCD induced contribution is $\mathcal{O}(100)$ times bigger than the EW one (354 vs. 2.9 fb), this process is not likely to be among the most interesting candidates for probing high energy EW scatterings of top quarks. Only $\Op{tW}$ leads to the required enhancement of the EW process such that it becomes significant with respect to the QCD one. 

\begin{figure}[h!]
  \centering 
$\vcenter{\hbox{\includegraphics[width=0.28\linewidth]{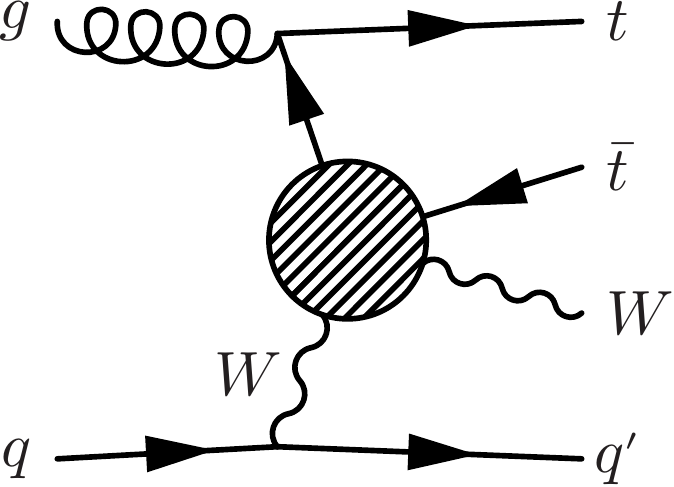}}}$
  \hspace{1cm} $\vcenter{\hbox{\includegraphics[width=0.28\linewidth]{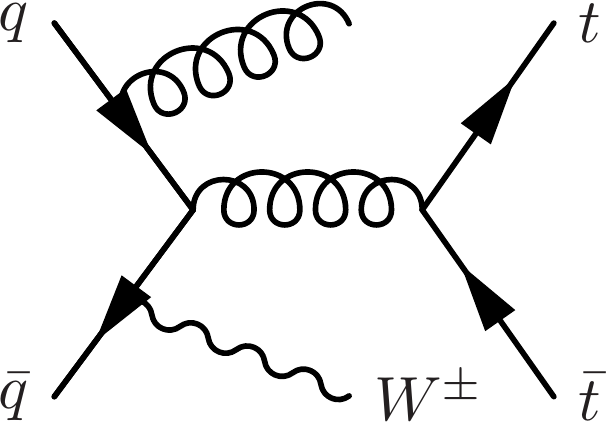}}}$
  \caption{ 
    \emph{Left}: schematic Feynman diagram for the EW-induced $t\bar{t}Wj$ process and its embedding of the $t\,W\to t\,W$ subamplitude. \emph{Right}: sample Feynman diagram for the QCD-induced $t\bar{t}Wj$, which does not probe modified top-EW interactions.
  \label{fig:diag_ttWj}}
 \end{figure}
  
The lower panel of Figure~\ref{fig:radar_ttw_LHC13} demonstrates that the $ttWj$ process is a more favourable environment for accessing $t\,W\to t\,W$ scattering. This process was studied in Ref.~\cite{Dror:2015nkp} as a probe of $t\,W$ scattering where some promising sensitivity was demonstrated. As shown in Figure~\ref{fig:diag_ttWj}, requiring an additional jet in the final state completely changes the topology of the EW process, effectively turning it on its side. The kinematic consequence of this modification is that the previously off-shell, $s$-channel $W$-boson is now in a $t$-channel topology. The rate of the process is enhanced by a factor 15 and it is able to access the high energy region unsuppressed. We define this region by imposing a 500 GeV $p_T$ cut on the $W$-boson as well as requiring that only one of either the top or anti-top had a $p_T$ greater than 500 GeV. This seemingly contrived cut is a result of the fact that either top or the anti-top can `participate' in the hard $2 \to 2$ sub-amplitude in this topology. The dominant, QCD-induced counterpart consists of the real radiation from the original QCD $t\bar{t}W$ (Figure~\ref{fig:diag_ttWj}). Interestingly, the relative size of the EW and QCD pieces changes drastically when going from $t\bar{t}W$ (1:120) to $t\bar{t}Wj$ (1:5). 

While the inclusive relative impacts are similar to $t\bar{t}W$ ($O(0.1-1)$), the presence of the energy growth from $t\,W\to t\,W$ is clearly visible, particularly at quadratic order in the Wilson coefficients. Comparing to Table~\ref{tab:helamp_summary_twotop_nohiggs}, not all of the interfering growth is manifested by our high energy cuts. Only $\Opp{\phi Q}{(1)}$ displays signs of the expected growth at this order. The fact that  $\Op{t\phi}$ and $\Op{tW}$ also show similar growths even though they are not expected from the $2\to2$ analysis suggests that the sensitivity to unitarity violating behaviour at interference-level is limited for this process. The squared contribution, however, shows growth for all operators, matching the expected dependence of Table~\ref{tab:helamp_summary_twotop_nohiggs}. The interesting sensitivity profile of this process combined with the improved ratio of QCD to EW contributions makes this process an interesting candidate for measuring top-EW scatterings at the LHC.
\subsubsection{$t\bar{t}W^+W^-$}
\begin{figure}[h!]
  \centering 
\includegraphics[width=0.28\linewidth]{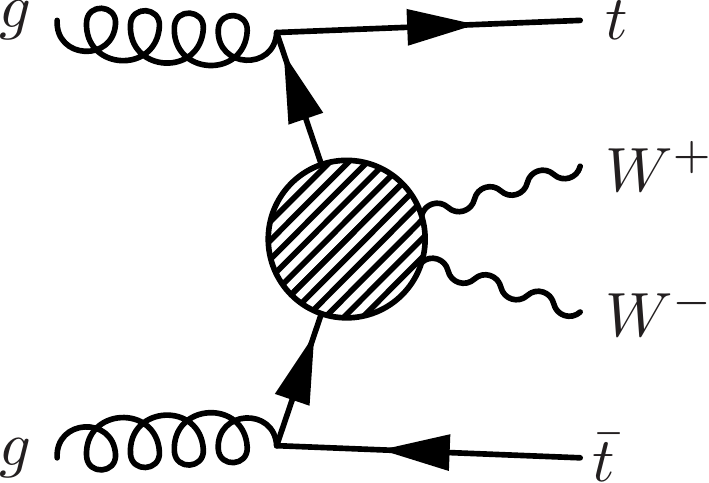}
 \caption{ 
Schematic Feynman diagram for the EW-induced $t\bar{t}W^+W^-$ process and its embedding of the $t\,W\to t\,W$ subamplitude. 
  \label{fig:diag_ttWW}}
 \end{figure}
The alternative to requiring an extra jet is to directly produce the second gauge boson participating in the scattering. This activates the $gg$ channel which was not previously accessed by the other two processes in this section. Figure~\ref{fig:diag_ttWW} depicts this contribution. There is also no irreducible `background' from processes with higher QCD and lower EW coupling orders. However, the high kinematic threshold for the process results in much smaller cross sections. For this reason we also compute the predictions at a 27 TeV $pp$ collider for this and other related $t\bar{t}XY$ processes. The possibility to study top quark dipole interactions through these processes has been studied in Ref~\cite{Etesami:2017ufk}.  The $WW$-invariant mass should serve as a proxy for the partonic $s$ of the subamplitude. We define the high energy region by imposing a 500 GeV $p_T$ cut on both $W$'s in order to access the large $s,-t$ limit of $t\, W\to t\,W$ scattering.

\begin{figure}[h!]
  \centering \includegraphics[height=0.2\paperheight]{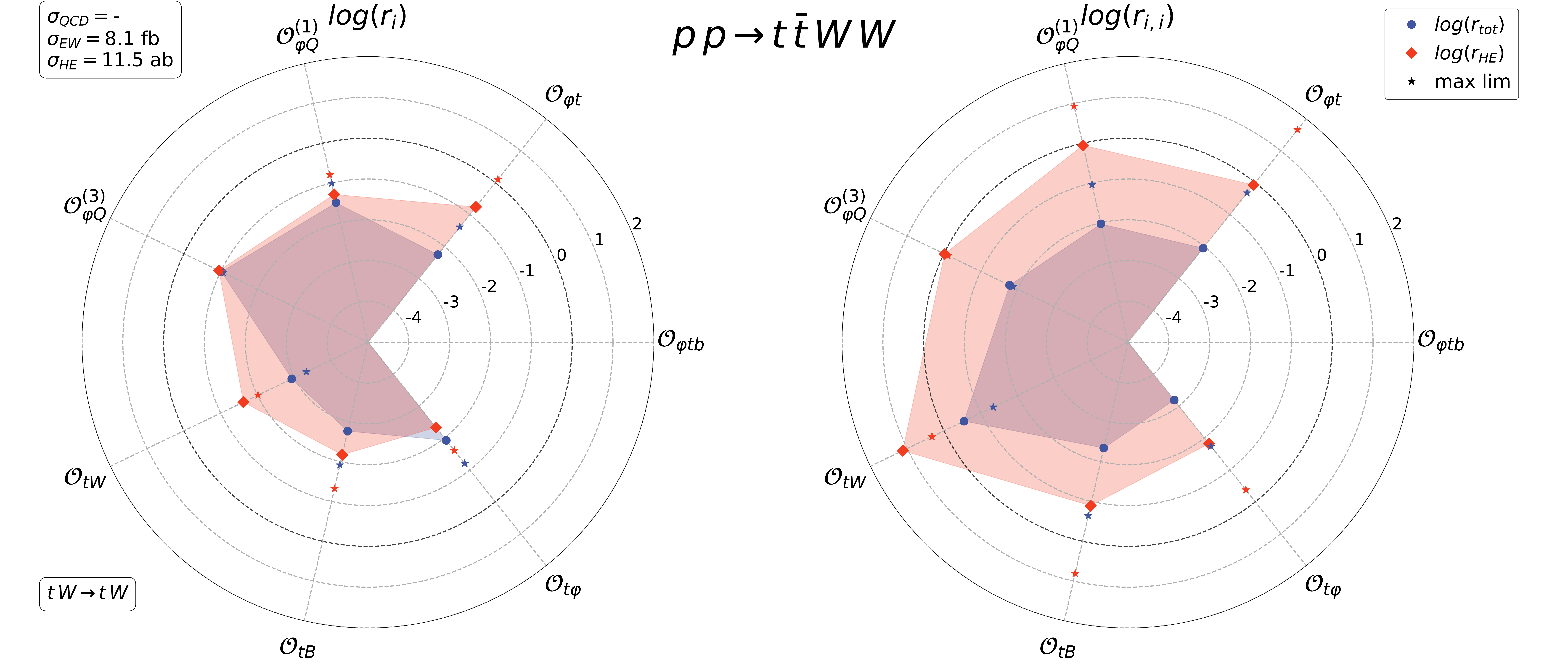}
  \caption{ 
   Radar plot for the $p \, p \to t \, \bar{t} \, W^+ W^-$ process at the 13 TeV LHC, 
 see
  Figure~\ref{fig:radar_twj_LHC13} and Section~\ref{subsubsec:tWj} of the main text for a detailed description.
  \label{fig:radar_ttww_LHC13}}
\end{figure}

Figure~\ref{fig:radar_ttww_LHC13} confirms the small, 8.1 $fb$ rate at the 13 TeV LHC. This grows to 53 $fb$ at a 27 TeV collider. The $gg$-initiated component of the process accounts for about half of the total rate at 13 TeV and two-thirds at 27 TeV. The relative sensitivity at inclusive level is quite mild, contributing 1-10\% effects in the interference terms with evidence of phase space cancellations in $\Op{tW}$  and $\Op{\phi t}$. Saturating the limits, one can expect total contributions at most of order one. Energy growth in this channel is, however, pronounced. The squared contributions grow even more than $t\bar{t}Wj$. We observe very similar sensitivity results for 27 TeV. One can expect this channel to be an interesting probe of strong EW-scattering at future colliders, particularly if the high energy region is accessed by differential measurements. 
\subsubsection{$t\,Z\to t\,Z$, $t\,Z\to t\,\gamma$ \& $t\,\gamma\to t\,\gamma$ scattering\label{subsubsec:tztz_tzta_tata}}

    \begin{figure}[h!]
        \centering
        \subfloat[]{
        \includegraphics[width=0.25\linewidth]{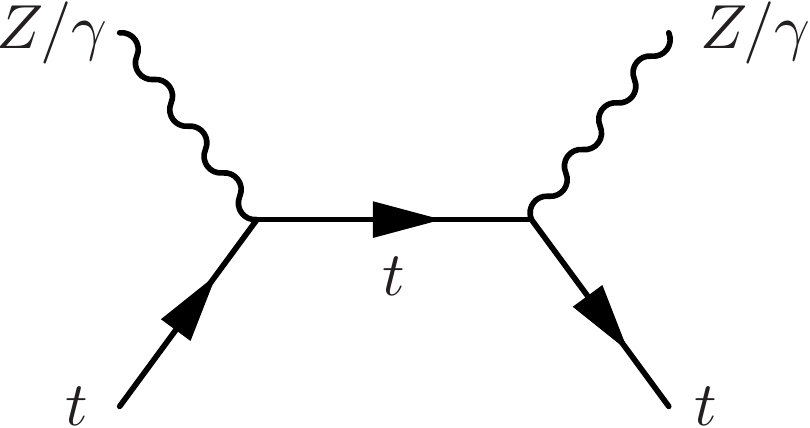}
        }  \subfloat[]{
        \includegraphics[width=0.21\linewidth]{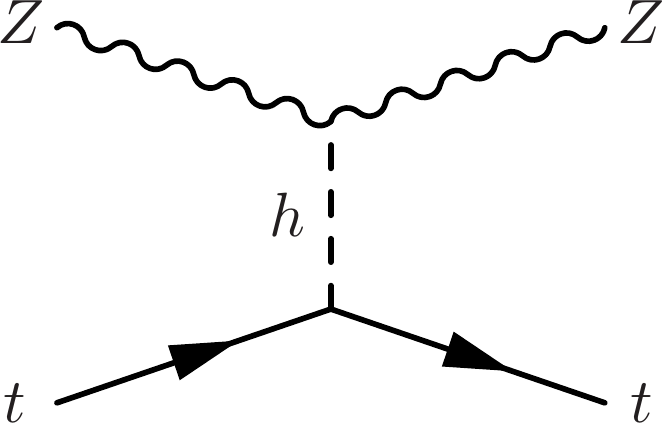}
        }
        \subfloat[]{
        \includegraphics[width=0.25\linewidth]{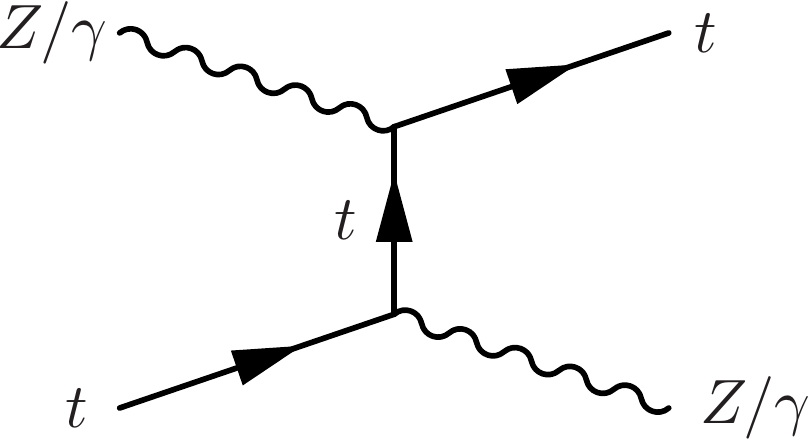}
        }
        \caption{SM diagrams for the $t Z \to t Z$, $t Z \to t \gamma$ and $t \gamma \to t \gamma$ subprocess. \label{tztz_tzta_tata_sm}}
    \end{figure} 
\noindent In these processes the top-Higgs, top-$\gamma$, top-$Z$, and  $Z$-Higgs couplings are probed, as seen in Fig.~\ref{tztz_tzta_tata_sm}. None of them factorise from the process, meaning that they can all participate in unitarity cancellations. In $t\,Z \to t\,Z$, the longitudinal configurations with opposite top helicity, $(\pm,0,\mp,0)$, are proportional to
     \begin{equation}
         \sqrt{-t}\,(\gth \, \gzh - 2 \mt^2 (\gztl-\gztr)^2) \,
     \end{equation}
and probe the Higgs mechanism. This behaviour can also be mapped to the SMEFT by means of the operators $\Op{t \varphi}$, $\Op{\varphi D}$, $\Op{\phi Q}^{\sss (3)}$ and $\Op{\varphi t}$ that  modify the aforementioned couplings and/or the top/$Z$ mass. 

In fact, there are no higher degrees of energy growth in this scattering from ACs. This can be seen from the fact that, since the Higgs interactions is chirality flipping, the Higgs diagram will not affect the like-helicity configurations, $(\pm,0,\pm,0)$ (to zeroth order in $\mt/\sqrt{s}$). Without the Higgs-mediated contribution, the couplings to the gauge bosons factorise from the process and can therefore not participate in unitarity cancellations.
In the SMEFT, this fact is manifested by the lack of $G^{\sss 0}\partial^\mu G^{\sss 0}\bar{t}\gamma_\mu t$ contact terms in Feynman gauge coming from the dimension 6 operators affecting the $t\bar{t}Z$ vertices. This is consistent with the lack of maximal energy growth for the purely longitudinal, like-helicity configurations (see Table~\ref{tab:tztz}). Instead, dimension-5 $G^{\sss 0}G^{\sss 0}\bar{t}_{\sss L}\bar{t}_{\sss R}$ and $G^{\sss 0}Z^\mu\bar{t}_{\sss L,R}\gamma_\mu \bar{t}_{\sss L,R}$ interactions lead to the observed linear growths for the Yukawa and current operators, respectively. Contact terms that mediate the mixed transverse-longitudinal configurations in are also found for the dipoles. The operators like $\Op{\varphi W}$ modify the coupling of the $Z$ to the Higgs but with an effective dimension 5 vertex that affects only the transverse degrees of freedom of the weak boson.

Since the photon does not couple to the Higgs at tree-level, there are no unitarity cancellations in the other two scatterings.  Table~\ref{tab:tzta} and~\ref{tab:tata} confirm that only the new Lorentz structures induce energy growth in $t\,Z\to t\,\gamma$ and $t\gamma\,\to t\,\gamma$, with the maximal growth from the dipole operators also not present in the latter because there is no massive gauge boson participating in the scattering and therefore no longitudinal degrees of freedom.
\subsubsection{$t\bar{t}Z(j)$ \& $t\bar{t}\gamma (j)$\label{subsubsec:ttz_ttzj_tta_ttaj}}
\begin{figure}[h!]
  \centering  $\vcenter{\hbox{\includegraphics[width=0.28\linewidth]{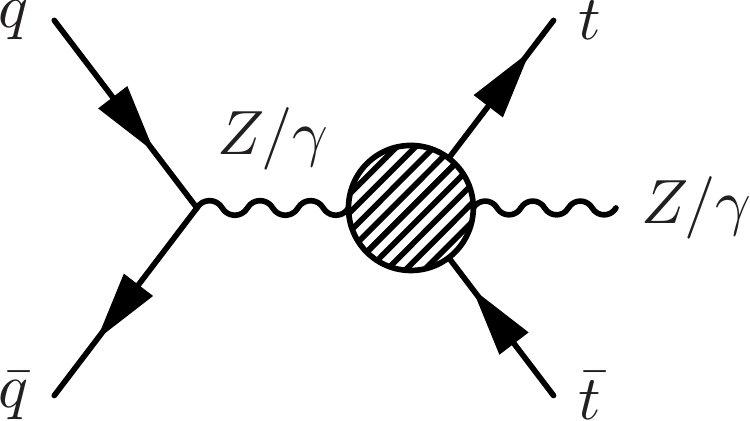}}}$
  \hspace{0.5cm}
$\vcenter{\hbox{\includegraphics[width=0.28\linewidth]{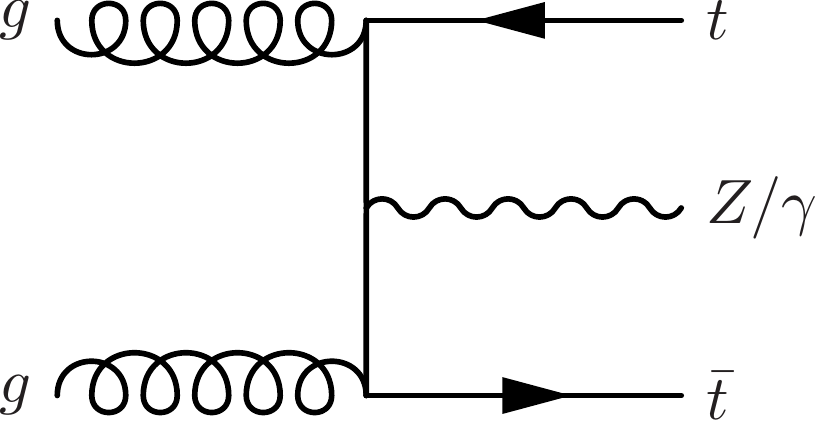}}}$
   \hspace{0.5cm}
$\vcenter{\hbox{\includegraphics[width=0.34\linewidth]{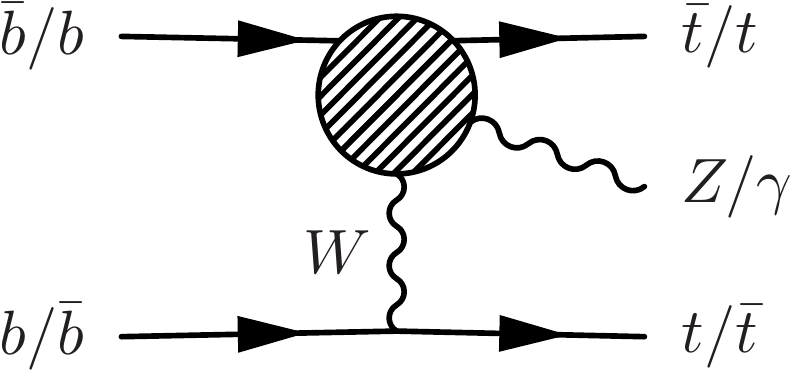}}}$
  \caption{ 
    \emph{Left}: schematic Feynman diagram for the EW-induced $t\bar{t}(Z/\gamma)$ process and its embedding of the $t\,Z\to t\,Z$, $t\,Z\to t\,\gamma$ or $t\,\gamma\to t\,\gamma$ subamplitudes. \emph{Middle}: sample Feynman diagram for QCD-induced $t\bar{t}(Z/\gamma)$, which does not embed the $2\to2$ scatterings. \emph{Right}: sample Feynman diagram for the $b\bar{b}$-induced $t\bar{t}(Z/\gamma)$, which embeds the single top $2\to2$ scatterings.
  \label{fig:diag_ttZ_tta}}
\end{figure} 

$t\bar{t}Z$ and $t\bar{t}\gamma$ each contain two of the three two-top neutral gauge boson subamplitudes. These processes have the significant drawback of being dominantly QCD-induced at hadron colliders through $Z$ or $\gamma$ radiation from the usual $t\bar{t}$ production channels (Figure~\ref{fig:diag_ttZ_tta}). Like $t\bar{t}W$, they embed the $2\to2$ scatterings through a highly off-shell intermediate state that significantly suppresses the EW production rate to around two orders of magnitude below the QCD component. Although the QCD-induced process does not embed the EW scatterings, the operators that 
we study can still affect the total rate of these processes by modifying the SM 
couplings of the top quark to the EW gauge bosons, as  has been studied in Ref.~\cite{Bylund:2016phk}. Therein, a relative impact (of coefficients set to 1 TeV$^{-1}$) on $t\bar{t}Z$ of order 6-10\% is predicted from the current operators coming mainly from the interference term. The dipole interference terms only contribute at the permille level while the squared piece is relatively enhanced and can give an effect of a few percent. One 
therefore requires an $\mathcal{O}(10-50)$ enhancement of the EW-induced processes at high energies in order 
to have a visible impact on the total hadron collider rate that competes with the modification of the QCD induced processes. We define our high energy region by imposing a $p_T$ cut of 500 GeV on both the top and the anti-top. Although the aforementioned current operator contributions to $t\bar{t}Z$ will not grow with energy, one should keep in mind that the small dipole operator contributions at inclusive level will be enhanced by said cut.

In fact, the single top scatterings can also enter in these processes through the $b\bar{b}$ initiated process 
in which a $W$-boson is exchanged in the $t$-channel. Interestingly, channel is found to give the dominant contribution to the EW-induced component 
of $t\bar{t}Z/\gamma$ production, accounting over 80\% of the total rates at the 13 TeV LHC. 
The PDF suppression of the $b\bar{b}$ luminosity is overcome by the favourable 
$t$-channel kinematics of the process compared to the off-shell $s$-channel 
$Z/\gamma$ contributions.
Given than it is a qualitatively different process, probing an entirely new channel, we remove this contribution from our predictions and leave a more detailed study of this topology to future work.

\begin{figure}[h!]
  \centering \includegraphics[height=0.2\paperheight]{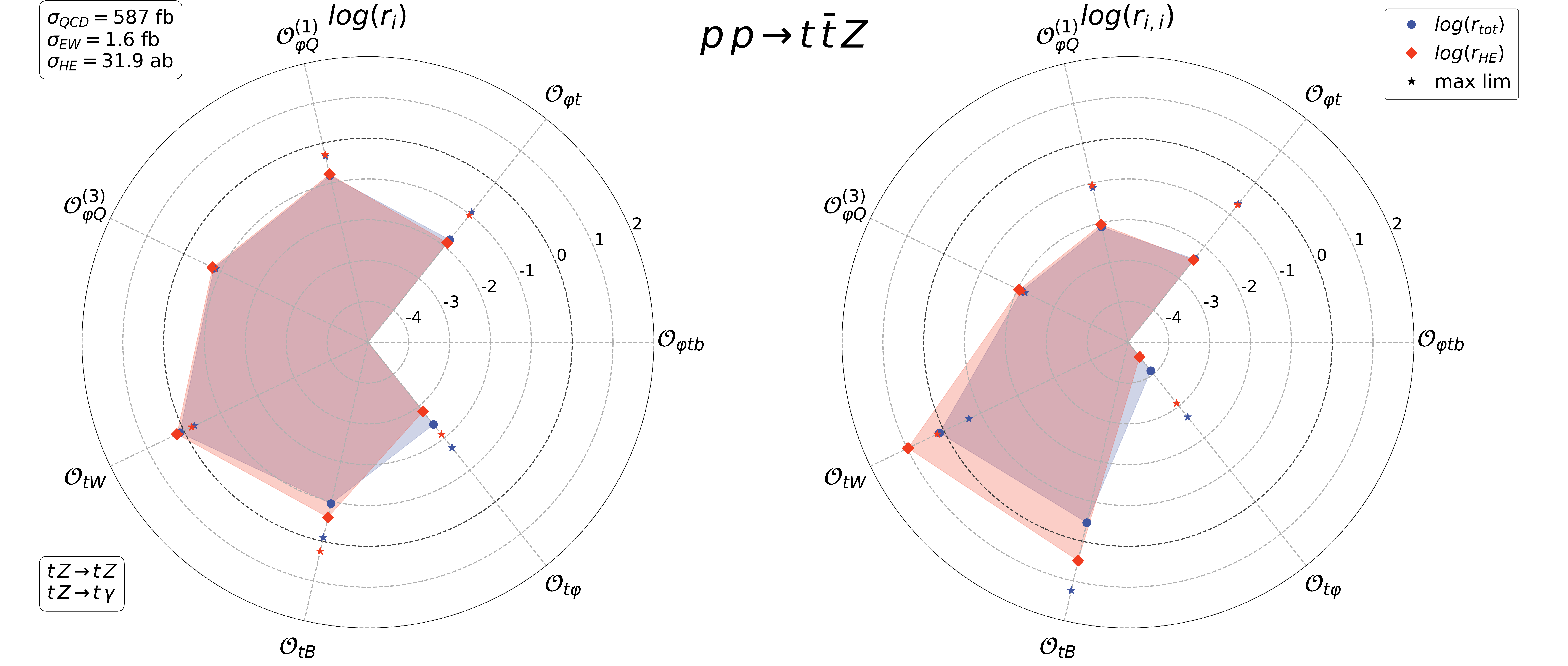}\\[1ex] \includegraphics[height=0.2\paperheight]{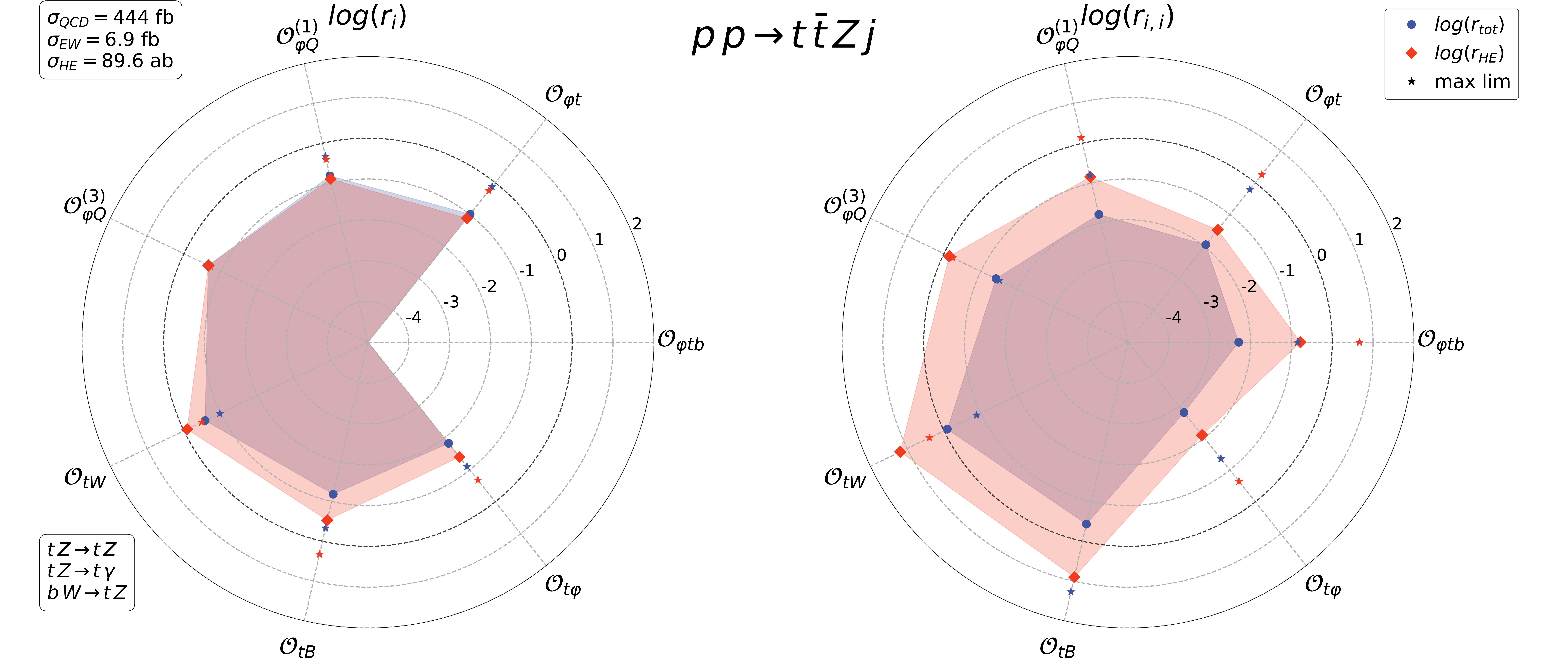}
  \caption{ 
   Radar plot for the $p \, p \to t \, \bar{t} \, Z$ (\emph{upper}) and  $p \, p \to t \, \bar{t} \, Zj$ (\emph{lower}) processes at the 13 TeV LHC, 
 see
  Figure~\ref{fig:radar_twj_LHC13} and Section~\ref{subsubsec:tWj} of the main text for a detailed description. The $b\bar{b}$-initiated component has not been included.
  \label{fig:radar_ttz_LHC13}}
\end{figure}
\begin{figure}[h!]
  \centering \includegraphics[height=0.2\paperheight]{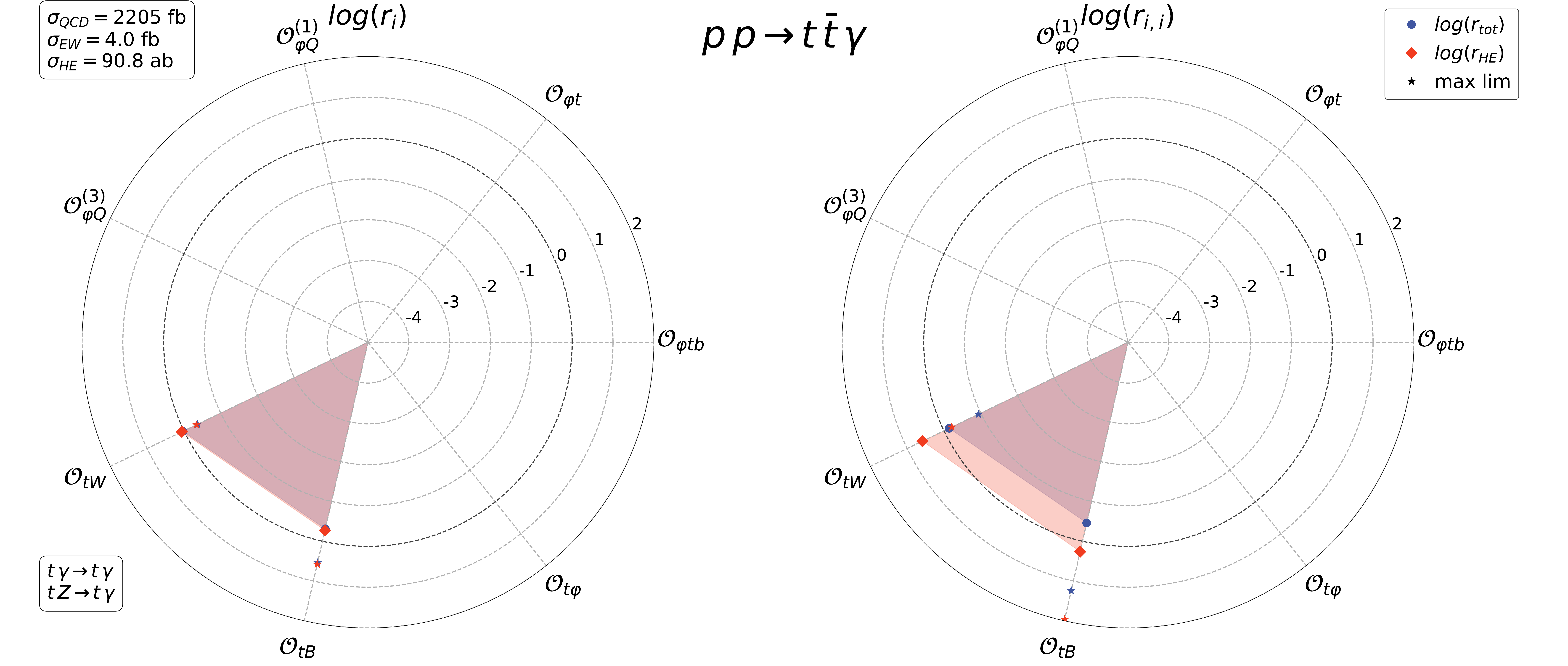}\\[1ex] \includegraphics[height=0.2\paperheight]{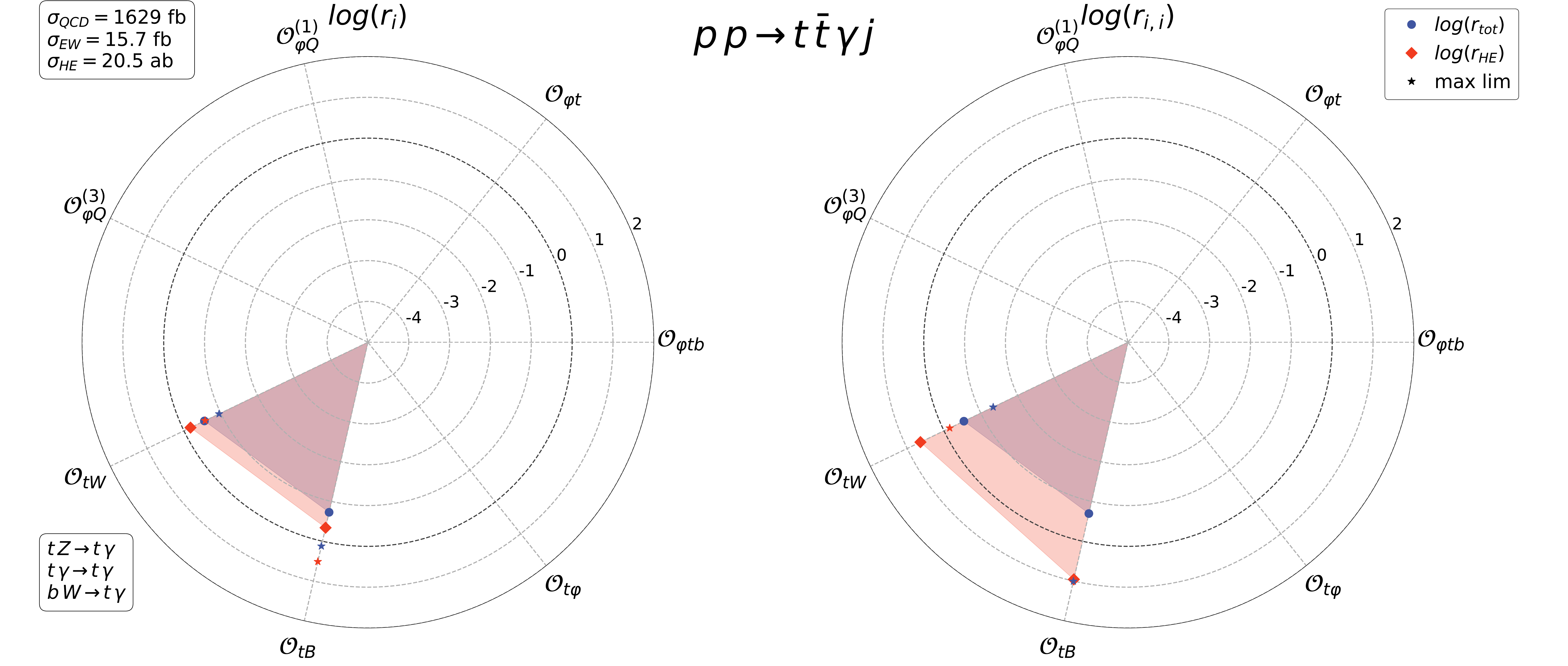}
  \caption{ 
   Radar plot for the $p \, p \to t \, \bar{t} \, \gamma$ (\emph{upper}) and  $p \, p \to t \, \bar{t} \, \gamma j$ (\emph{lower}) processes at the 13 TeV LHC, 
 see
  Figure~\ref{fig:radar_twj_LHC13} and Section~\ref{subsubsec:tWj} of the main text for a detailed description. The $b\bar{b}$-initiated component has not been included.
  \label{fig:radar_tta_LHC13}}
\end{figure}
The upper plot of Figures~\ref{fig:radar_ttz_LHC13} and~\ref{fig:radar_tta_LHC13} summarises the impact of the SMEFT operators on these processes 
at the 13 TeV LHC. The non-$b$ induced, pure EW contribution to the cross section is 1.6 
fb, to be compared with 587 fb in the case of the QCD-induced component.
Although some sensitivity to the operators is present, the required enhancement 
to overcome the large QCD contribution is not observed. In the interference contribution, 
energy growth is essentially absent.
In $t\bar{t}Z$, the interferences of the two quark doublet current operators, $\Opp{\phi Q}{(3)}$ and $\Opp{\phi Q}{(1))}$ have an order 10\% effect while the 
right handed current, $\Op{\phi t}$, and the Yukawa operator, $\Op{t\phi}$ both  have a very mild impact on 
this process. The lack of energy growth in these operator contributions is also present in the quadratic terms, which have a small impact overall. This is even worse than expected by our calculations of the $2\to2$ subamplitudes. Even when saturating the current limits, the net contribution of these pieces are below those of the interference. The dipole operators, which are the only ones that contribute to $t\bar{t}\gamma$, could contribute $\mathcal{O}(10)$ effects and display some energy growth at quadratic-level. We conclude that, as with $t\bar{t}W$, these processes 
are not ideal for probing the high-energy kinematics of the EW-top scatterings. 

\begin{figure}[h!]
  \centering  $\vcenter{\hbox{\includegraphics[width=0.28\linewidth]{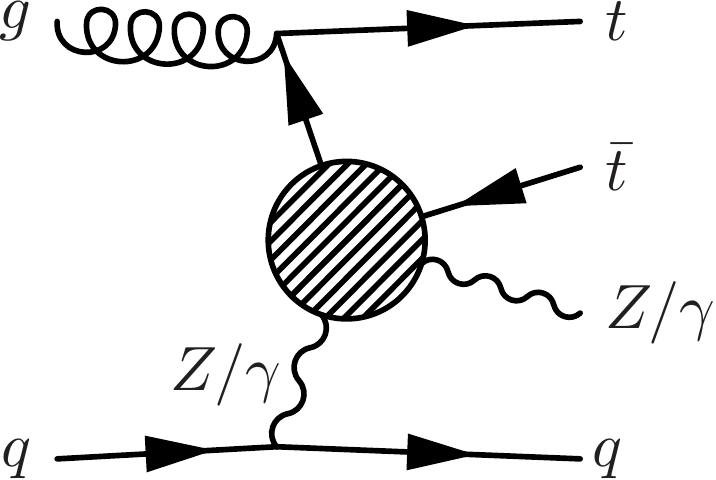}}}$
  \hspace{0.5cm}
$\vcenter{\hbox{\includegraphics[width=0.28\linewidth]{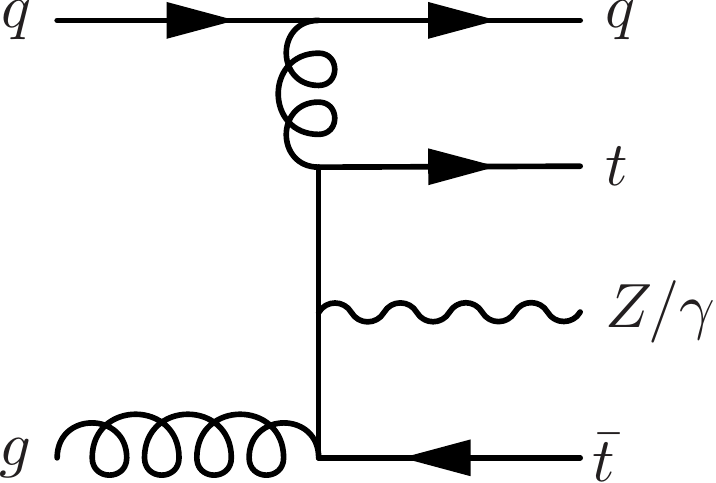}}}$
  \caption{ 
    \emph{Left}: schematic Feynman diagram for the EW-induced $t\bar{t}(Z/\gamma)j$ process and its embedding of the $t\,Z\to t\,Z$, $t\,Z\to t\,\gamma$ or $t\,\gamma\to t\,\gamma$ subamplitudes. \emph{Right}: sample Feynman diagram for QCD-induced $t\bar{t}(Z/\gamma)j$, which does not embed the $2\to2$ scatterings.
  \label{fig:diag_ttZj_ttaj}}
\end{figure} 
We also consider the addition of a jet in analogy with $t\bar{t}W$ (Figure~\ref{fig:diag_ttZj_ttaj}), imposing the same high-energy cuts. We observe an enhancement of a factor of around 4 with respect to the EW $t\bar{t}(Z/\gamma)$.
The relative size of the EW and QCD contributions, as defined by our jet $p_T$ cut, improves by a factor for about 5 in both cases. This difference is not as striking as in the case of $t\bar{t}Wj$ whose EW contribution seems to benefit from a peculiar enhancement with the addition of one jet in the final state. This may be due to the fact that this process is the only one of the $t\bar{t}Xj$ class that can be mediated by a $t$-channel photon between the top and $W$ lines.  As with $t\bar{t}(Z/\gamma)$, the QCD piece also receives similar, energy-constant contributions from the current (Yukawa) operators. The required enhancement of the EW amplitudes in order to compete with these modifications is relaxed to the order of a few.
The charged scattering dependence from the $b\bar{b}$ mediated channel also remains. However, this contribution is now of an expected size compared to the rest of the process contributing less than 10\% of the total $t\bar{t}Z/\gamma j$.

Sensitivity-wise, we observe slightly better, $\mathcal{O}(10\%)$ effects on the inclusive rates at interference-level for the current operators in $t\bar{t}Zj$. In the quadratic contributions, the impact of every operator is enhanced with respect to $t\bar{t}Z$, in some cases already at inclusive level. $t\bar{t}\gamma$ shows the opposite behaviour for the dipoles. One can also notice the remnant dependence on the $b$-initiated contributions in the mild quadratic dependence on $\Op{\phi tb}$. Furthermore, strong energy growth is present at quadratic level in all cases. Considering current limits on the operators, one expects $\mathcal{O}(0.1-1)$ effects to $t\bar{t}Zj$ in the current operators, which begin to compete with the corresponding modifications of the QCD-induced piece. The dipoles, however could contribute extreme enhancements, about ten times bigger than the current operators. 
\subsubsection{$t\bar{t}ZZ$, $t\bar{t}Z\gamma$ \& $t\bar{t}\gamma\gamma$}
\begin{figure}[h!]
  \centering 
\includegraphics[width=0.28\linewidth]{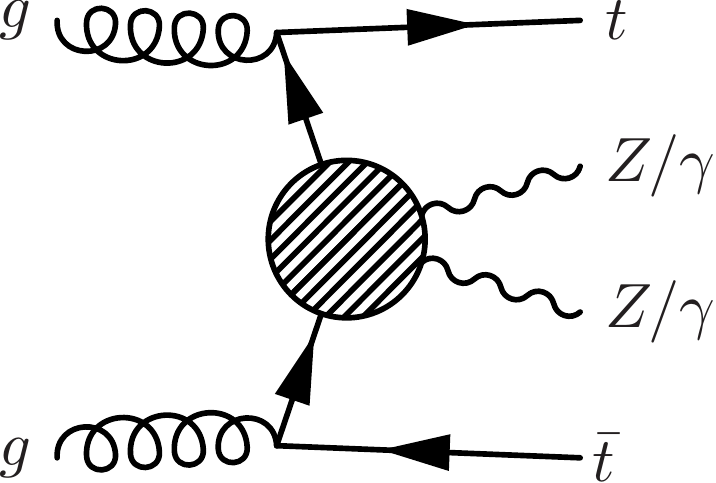}
 \caption{ 
Schematic Feynman diagram for the EW-induced $t\bar{t}(Z/\gamma)(Z/\gamma)$ processes and their embeddings of the $t\,Z\to t\,Z$, $t\,Z\to t\,\gamma$ and $t\,\gamma\to t\,\gamma$ subamplitudes. 
  \label{fig:diag_ttZZ_ttZa_tata}}
 \end{figure}
Remaining in analogy with the $t\,W$ scattering section, we consider the direct production of $t\bar{t}$ in association with a pair of neutral gauge bosons.  At 13 TeV, the $gg$-channel accounts for about one third of the total cross sections, which grows closer to one half at 27 TeV. We do not expect as striking growths in these channels as for the $t\,W\to t\,W$ analogues, particularly for those involving photons. The rates are $1.5(8.9)$, $4.8(26.8)$ and $11(45.6)$ $fb$ for $t\bar{t}ZZ$, $t\bar{t}Z\gamma$ \& $t\bar{t}\gamma\gamma$ at 13(27) TeV respectively. We define the high-energy region as in $t\bar{t}WW$, by making a 500 GeV $p_T$ requirement on both final state gauge bosons.
 
\begin{figure}[h!]
  \centering \includegraphics[height=0.2\paperheight]{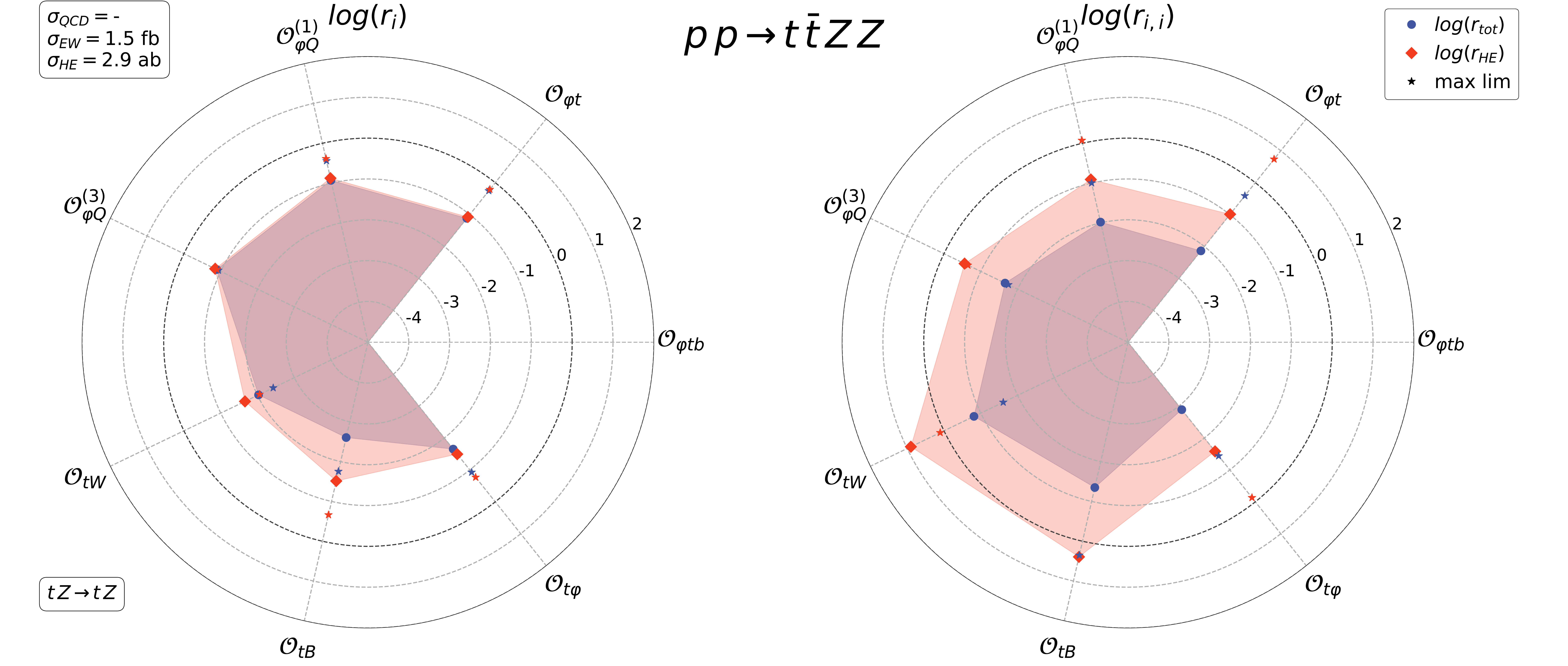}\\[1ex]
\includegraphics[height=0.2\paperheight]{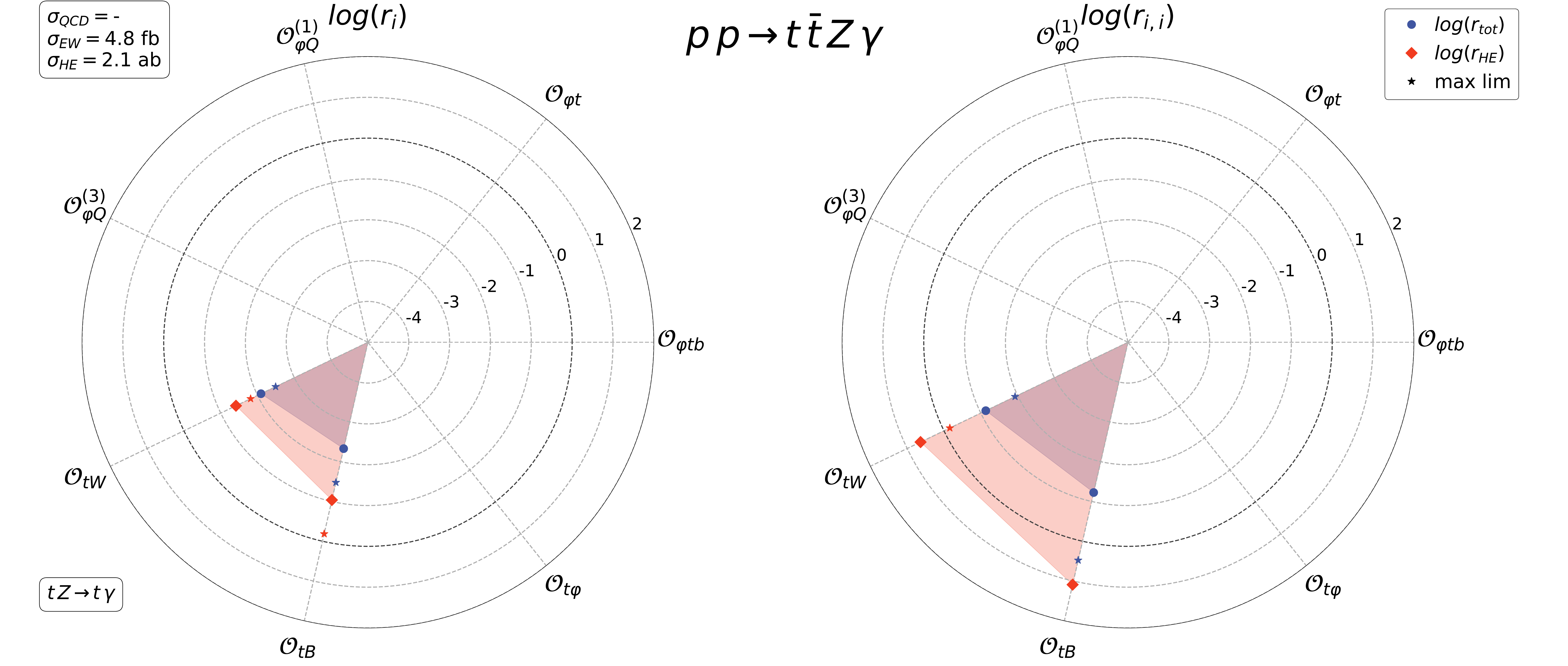}\\[1ex] \includegraphics[height=0.2\paperheight]{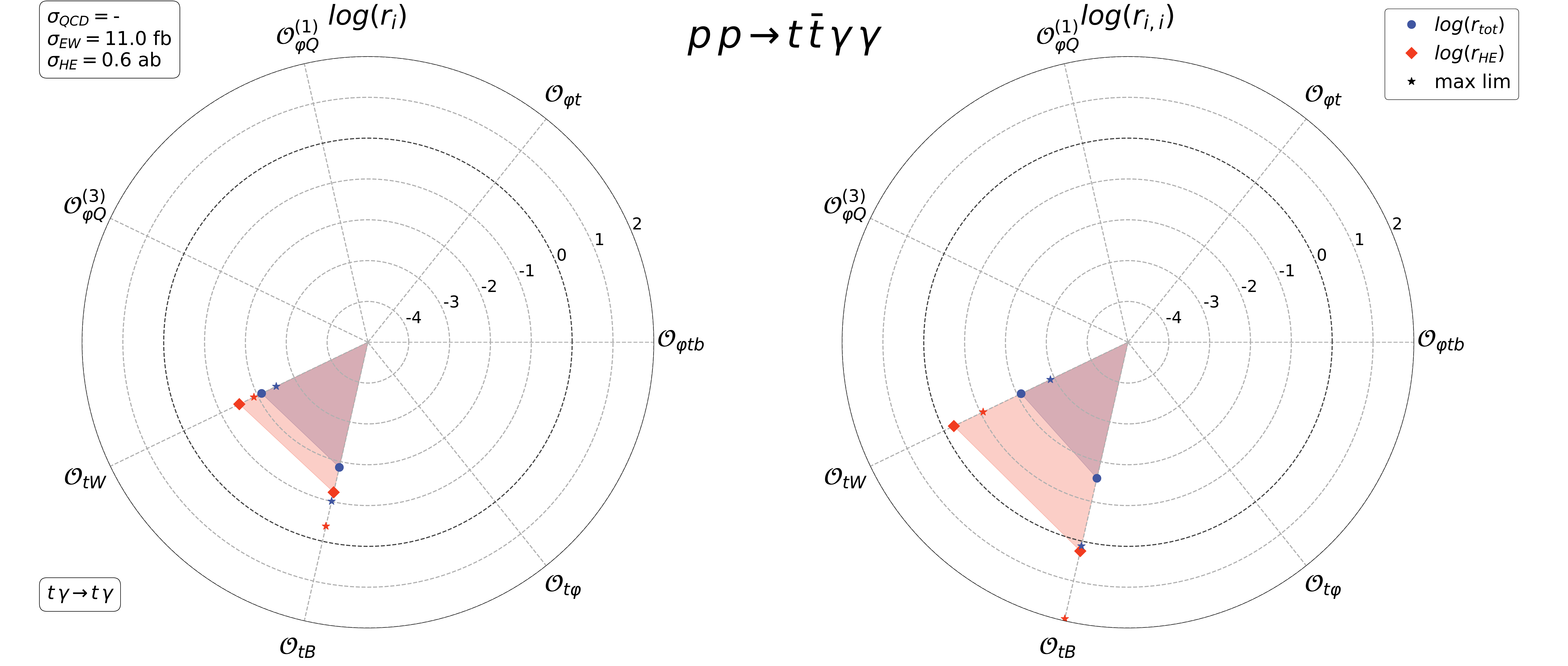}
  \caption{
   Radar plot for the $p \, p \to t \, \bar{t} \,ZZ$, $p \, p \to t \, \bar{t} \,Z\gamma$ and $p \, p \to t \, \bar{t} \,\gamma\gamma$  processes at the 13 TeV LHC,
 see
  Figure~\ref{fig:radar_twj_LHC13} and Section~\ref{subsubsec:tWj} of the main text for a detailed description.
  \label{fig:radar_ttzz_ttza_ttaa_LHC13}}
\end{figure}
Figure~\ref{fig:radar_ttzz_ttza_ttaa_LHC13} confirms the lack of interfering energy growth in $t\bar{t}ZZ$ for the majority of the operators apart from the the dipole operators. The apparent growth in their interference contributions for all three processes is consistent with phase space cancellations undone by the high energy requirement. We see the expected energy growth in the quadratic contributions to $t\bar{t}ZZ$ for all operators while $t\bar{t}Z\gamma$ shows an especially strong growth for the dipole operators. The relative sensitivity at 27 TeV is very similar to the 13 TeV results shown here.
\subsubsection{Vector boson fusion}
The final set of processes that we consider in this section are those that produce $t\bar{t}$ in the VBF topology, that is to say in association with a pair of very forward, light fermions. At the LHC these could be any combination of light quarks in the proton. The major drawback of this channel at the LHC is the overwhelming $t\bar{t}$+jets QCD background which, without any selection requirements, has a cross section 5 orders of magnitude larger than the EW process of interest. It is therefore interesting in this case to consider such a production channel at a future high-energy $e^+e^-$ collider, since the initial and final state light fermions need only to have EW interactions. The lack of the aforementioned QCD background coupled with clean environment of such colliders makes them an ideal testing ground for these processes. Additionally, one can distinguish the fusion of neutral and charged EW gauge bosons by whether we observe neutrinos (missing energy) or electrons in the final state. One however requires significant centre of mass energies to access these types of processes. We therefore choose the CLIC accelerator as a benchmark lepton collider to assess the VBF processes, considering the three proposed centre-of-mass energy configurations of 380, 1500 and 3000 GeV. In order to suppress unwanted contributions from on-shell $t\bar{t}(Z/W)$ where the gauge boson decays to a pair of jets or leptons, we require an invariant mass above 100 GeV for the light fermion pair. Similarly to the $t\bar{t}XY$ processes, 500 GeV $p_T$ cuts are imposed on the top quarks to define the high energy LHC cross section. In the case of CLIC, the lowest energy setup is barely in threshold for this process, meaning that high energy cuts cannot meaningfully be imposed. For the 1500 and 3000 GeV cases, we impose invariant mass cuts on the pair of top quarks of 1000 and 2000 GeV respectively.\\[0.2ex]

\begin{figure}[h!]
  \centering \includegraphics[width=\linewidth]{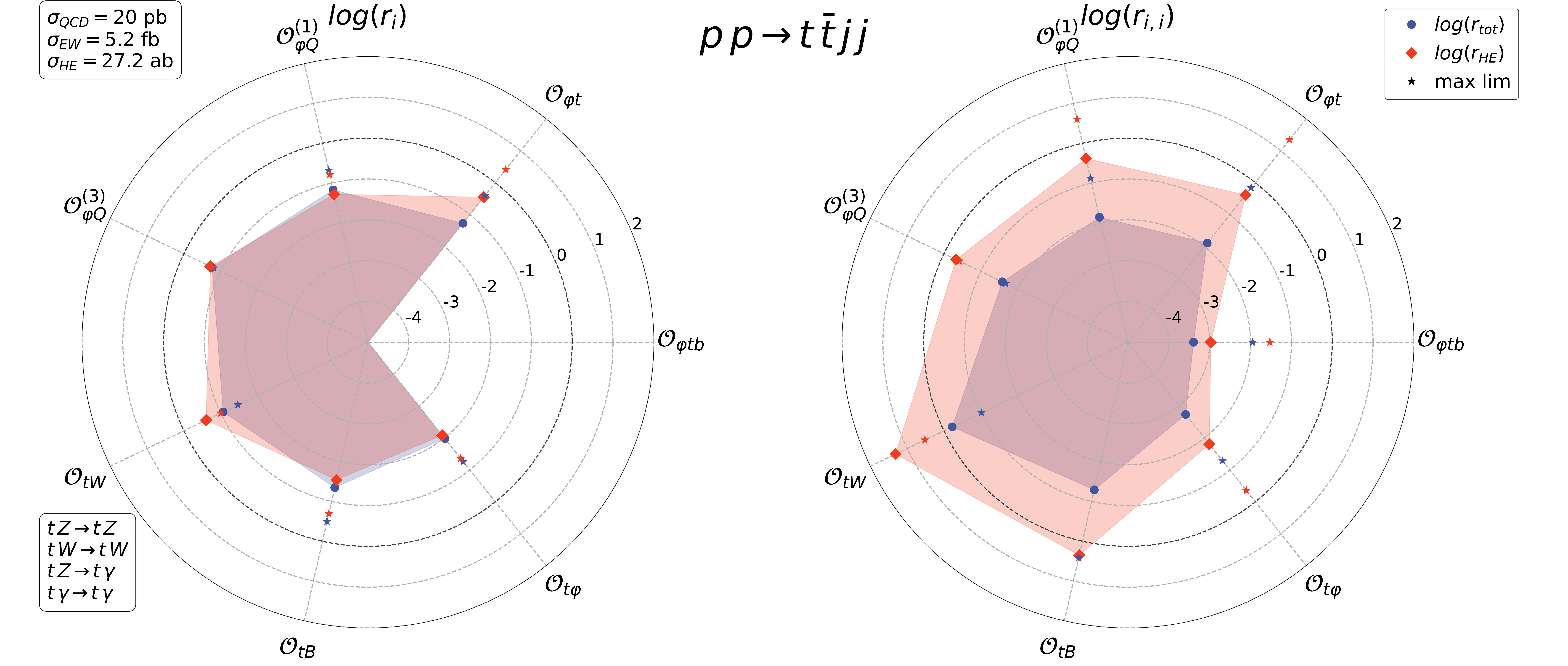}
  \caption{ 
    Radar plot for the VBF $t\bar{t}$ production process at the 13 TeV LHC, see
 Figure~\ref{fig:radar_twj_LHC13} and Section~\ref{subsubsec:tWj} of the main text for a detailed description.
  \label{fig:radar_ttjj_LHC13}}
\end{figure}

\noindent$\mathbf{t\bar{t}jj}$\\
We begin with the proton-proton collider prediction, whose EW component has a cross section of 5.2 fb at 13 TeV. The VBF $t\bar{t}$ component of this process embeds all of the two-top $2\to2$ subprocesses without the Higgs, which cannot couple to the light quark lines. In fact, the single top scattering subprocesses can be embedded into this final state but do not have a VBF topology and enter at the price of two $b$-quark PDFs. Of the relevant neutral subprocesses, $t\,W\to t\,W$ is expected to provide the strongest energy growth. Figure~\ref{fig:radar_ttjj_LHC13} shows that only $\Op{\phi t}$ and $\Op{tW}$ display significant energy growth at interference-level. The quadratic contributions do display the expected growth, although their relative impact is mild, giving $O(1)$ effects in the high energy region. The only exception is $\Op{tW}$, that provides an $O(1)$ effect on the inclusive rate that grows by a factor of 10 at high energies. To estimate the QCD background, we impose simple VBF cuts on the invariant mass of the two jets of $m_{jj} > 400$ GeV and a rapidity separation $\Delta\eta > 3$, obtaining a cross section of around 20 pb, still 1000 times larger than the EW version. Given this overwhelming background, it will be challenging for the effects on the EW cross section to be sufficient to yield additional constraints on the Wilson coefficient space. \\[0.2ex]

\begin{figure}[h!]
  \centering \includegraphics[height=0.2\paperheight]{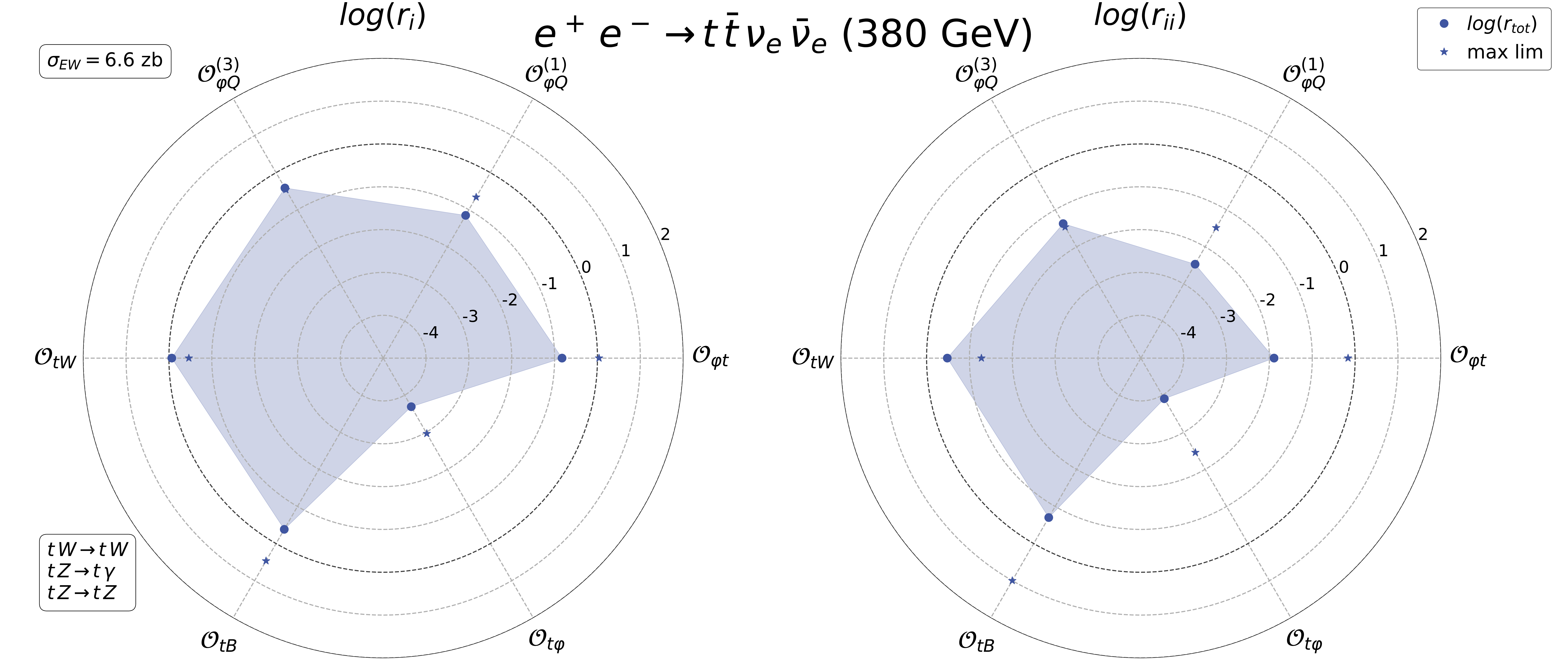}\\[1.5ex]
  \includegraphics[height=0.2\paperheight]{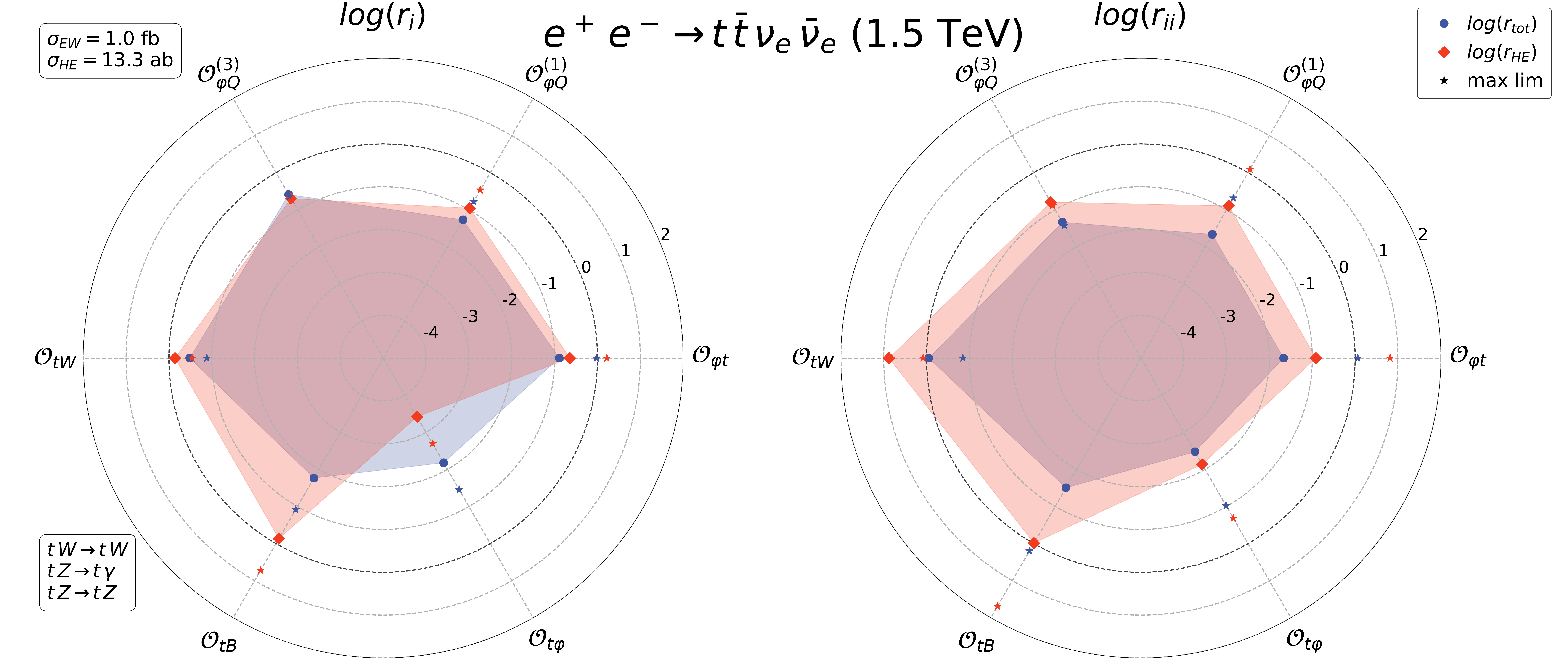}\\[1.5ex]
  \includegraphics[height=0.2\paperheight]{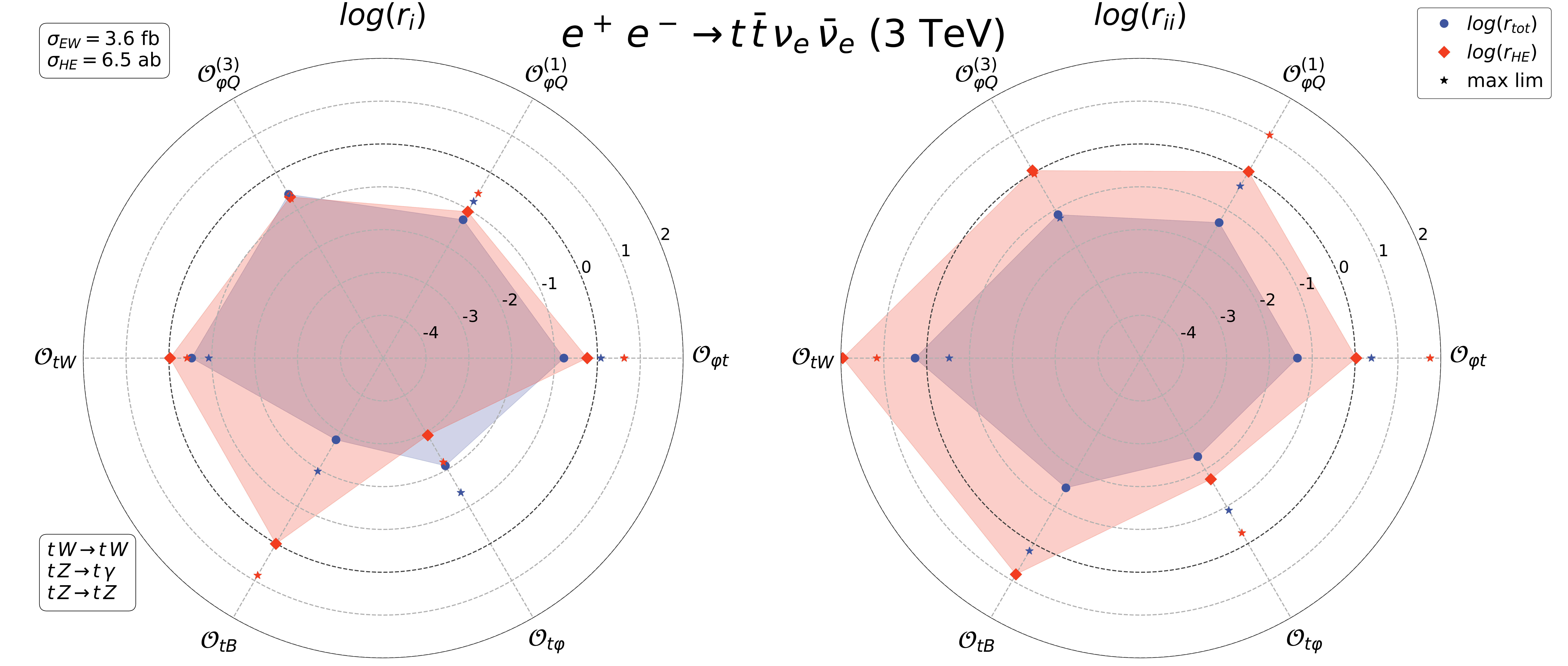}
  \caption{ 
   Radar plot for the $e^+\,e^- \to t\,\bar{t}\,\nu_e\,\bar{\nu}_e$ process at collider centre of mass energies of 380 (\emph{upper}), 1500 (\emph{middle}) and 3000 GeV (\emph{lower}), see
 Figure~\ref{fig:radar_twj_LHC13} and Section~\ref{subsubsec:tWj} of the main text for a detailed description. 
 \label{fig:radar_ttvv_CLIC}}
\end{figure}

\begin{figure}[h!]
  \centering \includegraphics[height=0.2\paperheight]{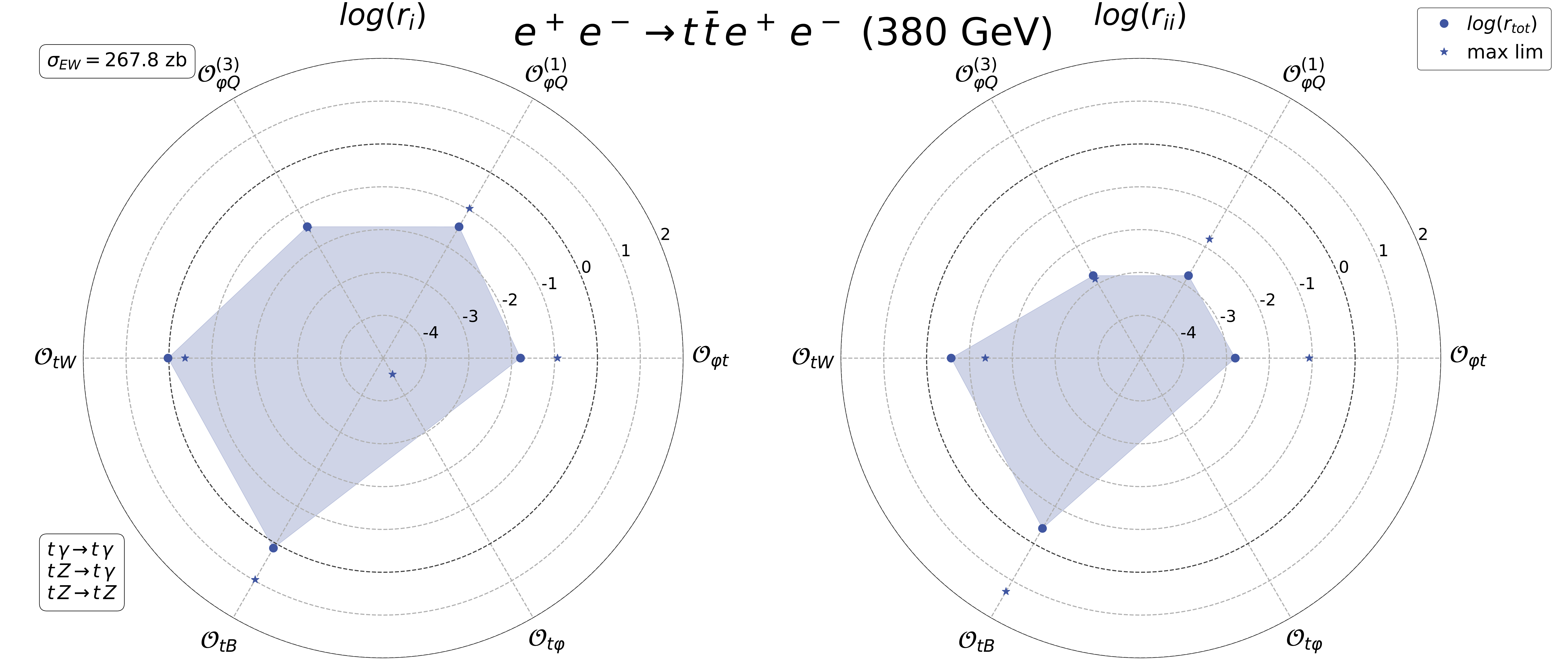}\\[1.5ex]
  \includegraphics[height=0.2\paperheight]{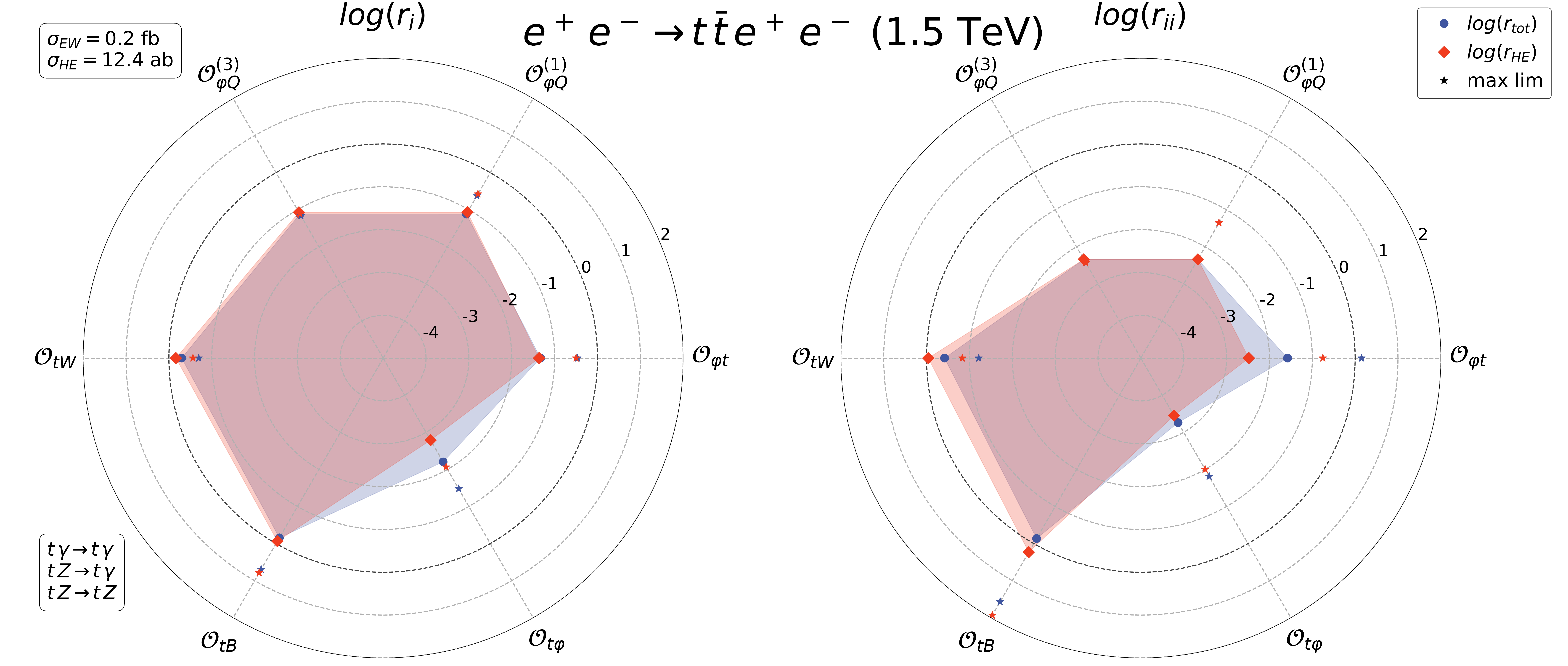}\\[1.5ex]
  \includegraphics[height=0.2\paperheight]{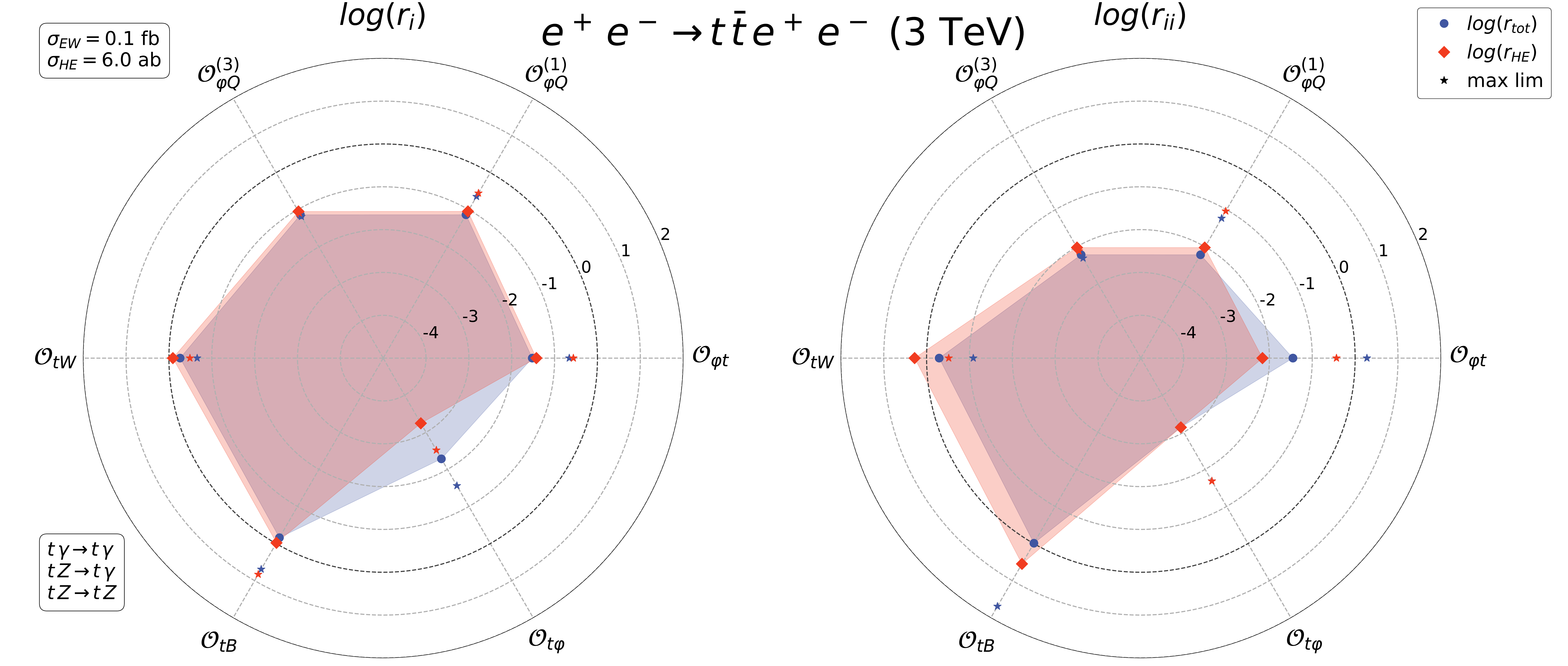}
  \caption{ 
    Radar plot for the $e^+\,e^- \to t\,\bar{t}\,e^+\,e^-$ process at collider centre of mass energies of 380 (\emph{upper}), 1500 (\emph{middle}) and 3000 GeV (\emph{lower}), see
  Figure.~\ref{fig:radar_twj_LHC13} and Section~\ref{subsubsec:tWj} of the main text for a detailed description.
  \label{fig:radar_ttee_CLIC}}
\end{figure}

\noindent$\mathbf{t\bar{t}\nu_e\bar{\nu}_e}$ \textbf{\&} $\mathbf{t\bar{t}e^+e^-}$\\
High energy lepton colliders appear to be the most promising avenue for measuring the VBF $t\bar{t}$ process, simply due to the absence of the QCD $t\bar{t}$+jets background. The final state fermions determine whether the dominant VBF contribution comes from the $2\to2$ subamplitudes involving neutral gauge bosons ($e^+e^-$) or $t\,W\to t\,W$ scattering ($\nu_e\bar{\nu}_e$). The non-VBF contributions from $t\bar{t}(Z\to\bar{\nu}_e\nu_e)$ introduce a weak dependence of the latter process on the neutral gauge bosons scatterings ($t\,Z\to t\,Z$ and $t\,Z\to t\,\gamma$) that is negligible compared to $t\,W\to t\,W$. The scatterings involving only neutral gauge bosons have a weaker expected energy dependence than $t\,W\to t\,W$. Combining this with the larger predicted cross sections for  $t\bar{t}\nu_e\bar{\nu}_e$  reported in Figure~\ref{fig:radar_ttvv_CLIC} compared to those of Figure~\ref{fig:radar_ttee_CLIC} leads to a clear preference for this process in constraining SMEFT operators at high energies. A preliminary sensitivity study of this process to the current and dipole operators was presented in Ref.~\cite{Abramowicz:2018rjq}, promising to go well beyond existing limits.
The figures show qualitatively similar inclusive sensitivities ($O(0.1-1)$) to the 6 operators, with only the hypercharge dipole operator, $\Op{tB}$, having a greater relative impact in $t\bar{t}e^+e^-$. There is evidence for a cancellation in the interference contribution from this operator in $t\bar{t}\bar{\nu}_e\nu_e$. Concerning the high energy behaviour, the expected enhancements from the $t\,W\to t\,W$ scattering are plain to see with the $WW$-fusion process showing larger enhancements both at interference and squared level. That said, the energy growth of the interference terms is somewhat mild, given the expectation of a maximal effect from the current operators in the fully longitudinal $t\,W\to t\,W$ scattering. 

\clearpage
\subsection{With the Higgs\label{subsec:twotop_yeshiggs}}
The set of processes that can be used to investigating scatterings with the Higgs are slightly more limited by the fact that the Higgs does not couple to light fermions. It therefore always appears in the final state for the processes that we consider. Table~\ref{tab:amp_proc_table_twotop_yeshiggs} shows that, as with the scatterings of the previous section, each process is at least accessed by its corresponding $t\bar{t}XY$ mode. 
\begin{table}[h!]
\centering
\begin{tabular}{|p{2cm}|P{1.1cm}|P{1.1cm}|P{1.1cm}|P{1.1cm}|}
     \hline
     
                              &$t\bar{t}h(j)$ & $t\bar{t}Zh$ & $t\bar{t}\gamma h$ &  $t\bar{t}hh$
     \tabularnewline\hline             
                                       
     $t\,Z\to t\,h$           & \cmark        & \cmark       &                     &       
     \tabularnewline\hline                                                                

     $t\,\gamma\to t\,h$      & \cmark        &              &  \cmark             &       
     \tabularnewline\hline                                                                                       
                                                                                 
     $t\,h\to t\,h$           &               &              &                     & \cmark       
     \tabularnewline\hline                                                                
                                                                                                                                    
\end{tabular}

\caption{The set of two-top $2\to2$ scattering amplitudes with Higgs bosons considered in this work mapped to the collider processes in which they are embedded.
\label{tab:amp_proc_table_twotop_yeshiggs}}
\end{table}
A clear favourite also emerges in $t\,Z\to t\,h$ concerning energy growth. Table~\ref{tab:helamp_summary_twotop_yeshiggs} shows very similar growth in this scattering to $t\,W\to t\,W$, with interfering energy growth for all current operators and a sensitivity to the Yukawa operator. The scattering with a photon has the same energy growth profile as its $t\,Z\to t\gamma$ counterpart. Finally, $t\,h\to t\,h$ scattering is only sensitive to the Yukawa operator and operators that modify the Higgs self-interactions with at most linear growths. 
\begin{table}[h!]
{\footnotesize
\setlength{\tabcolsep}{4pt}
\renewcommand{\arraystretch}{1.2}
\begin{center}
  \begin{tabular}{|c|c|c|c|c|c|c|c|c|c|c|c|c|c|}
 \hline
                          & $\Op{\phi D}$ & $\Op{\phi \Box}$ &  $\Op{\phi B}$ & $\Op{\phi W}$  & $\Op{\phi WB}$  & $\Op{W}$ & $\Op{t \phi}$ & $\Op{tB}$ & $\Op{tW}$ & $\Op{\phi Q}^{\sss (1)}$ & $\Op{\phi Q}^{\sss (3)}$ &  $\Op{\phi t}$ & $\Op{\phi tb}$ 
 \tabularnewline\hline

 %
 %
 %
 %
 $t\,Z\to t\,h$           & $E$           & $-$              & $E$            & $E$            & $E$             & $-$      & $E$           & $E^2$     & $E^2$     & \red{$E^2$}              & \red{$E^2$}              & \red{$E^2$}    & $-$           
 \tabularnewline\hline
  
 $t\,\gamma\to t\,h$      & $-$           & $-$              & $E$            & $E$            & $E$             & $-$      & $-$           & $E^2$     & $E^2$     & $-$                      & $-$                      & $-$            & $-$
 \tabularnewline\hline

$t\,h\to t\,h$            & $E$           & $E$              & $-$            & $-$            & $-$             & $-$      & $E$           & $-$       & $-$       & $-$                      & $-$                      & $-$            & $-$           
 \tabularnewline\hline
 
  \end{tabular}
\end{center}
\renewcommand{\arraystretch}{1.}
}

\caption{\label{tab:helamp_summary_twotop_yeshiggs}
Same as Table~\ref{tab:helamp_summary_singletop} for two top scattering amplitudes with the Higgs. See Tables~\ref{tab:tzth}--\ref{tab:thth} for the full helicity amplitude results.
}
\end{table}
\subsubsection{$t\,Z\to t\,h$ \& $t\,\gamma\to t\,h$ scattering \label{subsubsec:tzth_tath}}
\begin{figure}[h!]
    \centering
    \subfloat[]{
    \includegraphics[width=0.25\linewidth]{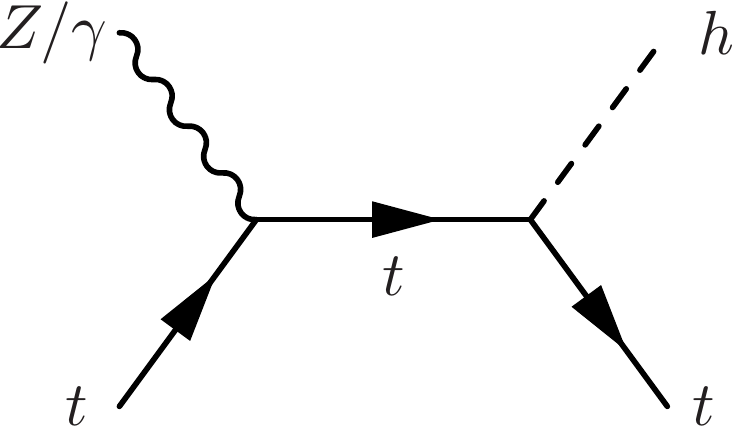}
    }  \subfloat[]{
    \includegraphics[width=0.21\linewidth]{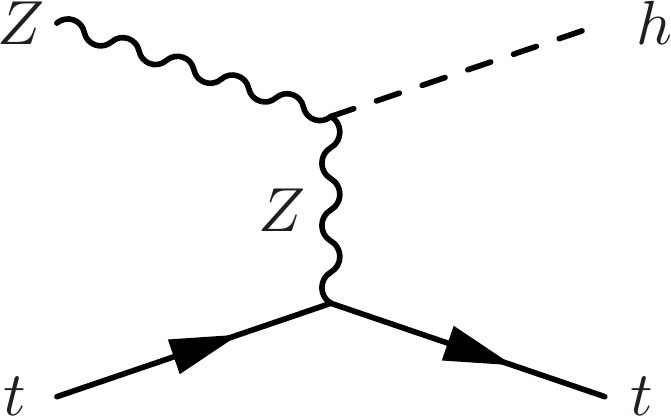}
    }
    \subfloat[]{
    \includegraphics[width=0.25\linewidth]{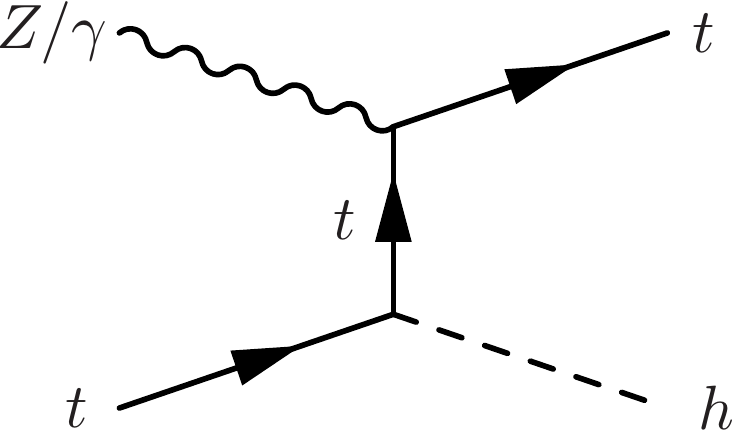}
    }
    \caption{SM diagrams for the $t\,Z \to t\, h$ subprocess. \label{tzth_sm}}
\end{figure} 
    
\noindent 
The relevant couplings for this process are the top-Higgs, the top-$Z$, top-$\gamma$ and the $Z$-Higgs (see Fig.~\ref{tzth_sm}). However, the top-gauge coupling factorises and the photon has not interaction with the Higgs. We therefore only consider the anomalous top-Higgs and $Z$-Higgs couplings as candidates for unitarity cancellations probing the details of the Higgs mechanism in $t\,Z \to t\, h$. These arise in the expected opposite-helicity, longitudinal configurations, $(\pm, 0,\mp,0)$, proportional to
\begin{equation}
    \sqrt{-t}\,(m_t g_{Zh} - 2 m_Z^2 g_{th}) \,.
\end{equation}
As with $b \, W \to t \, h$, discussed in Section~\ref{subsubsec:bwth},
there are no higher degrees of growth from ACs. In the SMEFT the linear growths are attributed to the Yukawa operator $\Op{t \varphi}$, while an analogous effect arises from the operator $\Op{\varphi D}$ that modifies the $Z$ mass and the coupling of the $Z$ to the Higgs boson. All sources of maximal, interfering energy growth in Table~\ref{tab:tzth} come from the dimension-6 contact terms involving a neutral Goldstone boson and dynamical Higgs field, \emph{i.e.} the neutral version of Equation~\eqref{eq:contact_o3phiQ_2}.
\subsubsection{$t\bar{t}h(j)$\label{subsubsec:tth_tthj}}
\begin{figure}[h!]
  \centering  $\vcenter{\hbox{\includegraphics[width=0.28\linewidth]{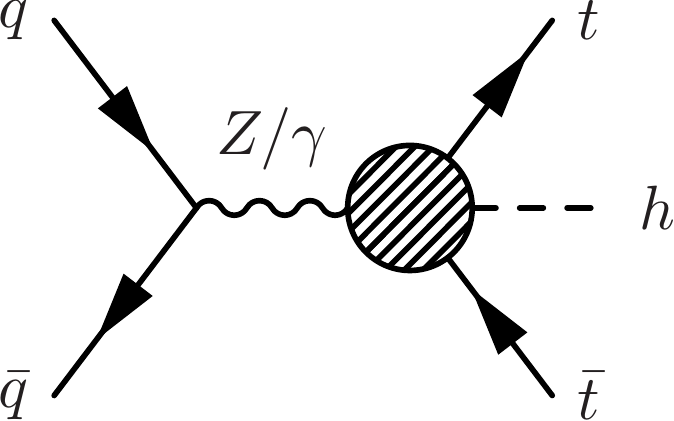}}}$
  \hspace{0.5cm}
$\vcenter{\hbox{\includegraphics[width=0.28\linewidth]{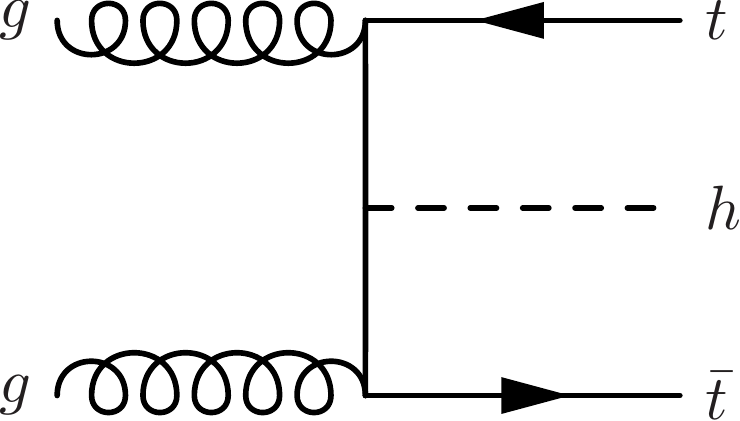}}}$
   \hspace{0.5cm}
$\vcenter{\hbox{\includegraphics[width=0.34\linewidth]{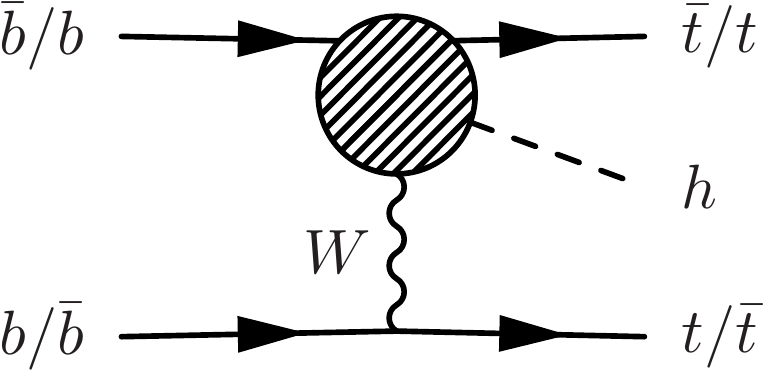}}}$
  \caption{ 
    \emph{Left}: schematic Feynman diagram for the EW-induced $t\bar{t}h$ process and its embedding of the $t\,Z\to t\,h$ and $t\,\gamma\to t\,h$ 
subamplitudes. \emph{Middle}: sample Feynman diagram for QCD-induced $t\bar{t}h$, which does not embed the $2\to2$ scatterings. \emph{Right}: sample Feynman diagram for the $b\bar{b}$-induced $t\bar{t}h$, which embeds the single top $b\,W\to t\,h$ scattering.
  \label{fig:diag_ttH}}
\end{figure} 
The recent discovery of the $t\bar{t}h$ production mode 
completes the set observations of $t\bar{t}$ in association with one EW boson at the LHC. As for all such processes, the EW component 
is overshadowed by the QCD-induced process, shown in Figure~\ref{fig:diag_ttH}. The EW component of this mode is kinematically very similar to the $t\bar{t}Z$ process, with the only difference that the Higgs cannot be radiated from the initial state, meaning that every diagram embeds the relevant scattering. To avoid repetition, we refer the reader to Section~\ref{subsubsec:ttz_ttzj_tta_ttaj} to complement the discussion here. We also find a significant $b\bar{b}$-induced component that accounts for about 70\%  of the total EW rate at 13 TeV, and omit it in the results of this study. The QCD cross section is rescaled by modifications of the top-Higgs interaction, which in the SMEFT corresponds to effects from $\Op{t\phi}$ at the 10\% level~\cite{Maltoni:2016yxb} for $c=1$ TeV$^{-2}$.  The predicted QCD cross section of 399 fb  means that significant enhancements of the 0.7 fb EW rate would be required to show up at hadron colliders. 
\begin{figure}[h!]
  \centering \includegraphics[height=0.2\paperheight]{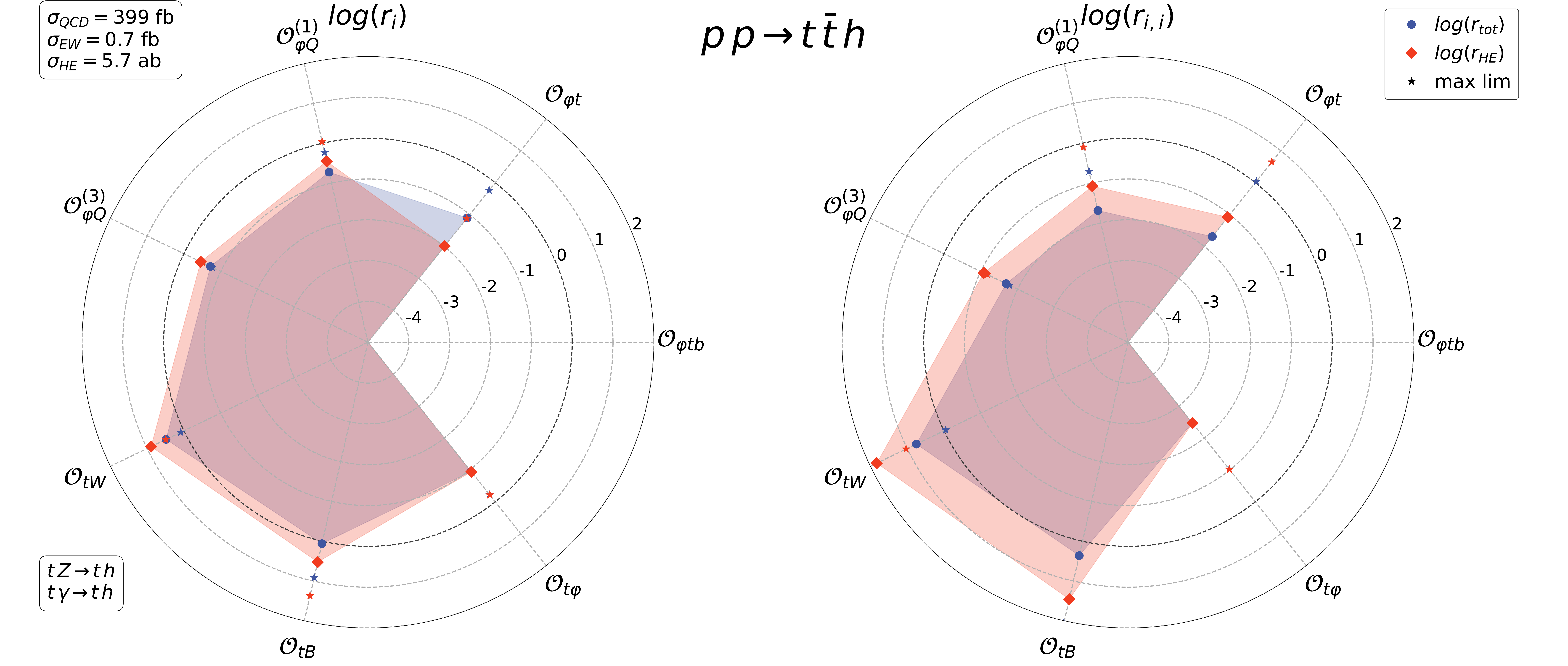}\\[1ex] \includegraphics[height=0.2\paperheight]{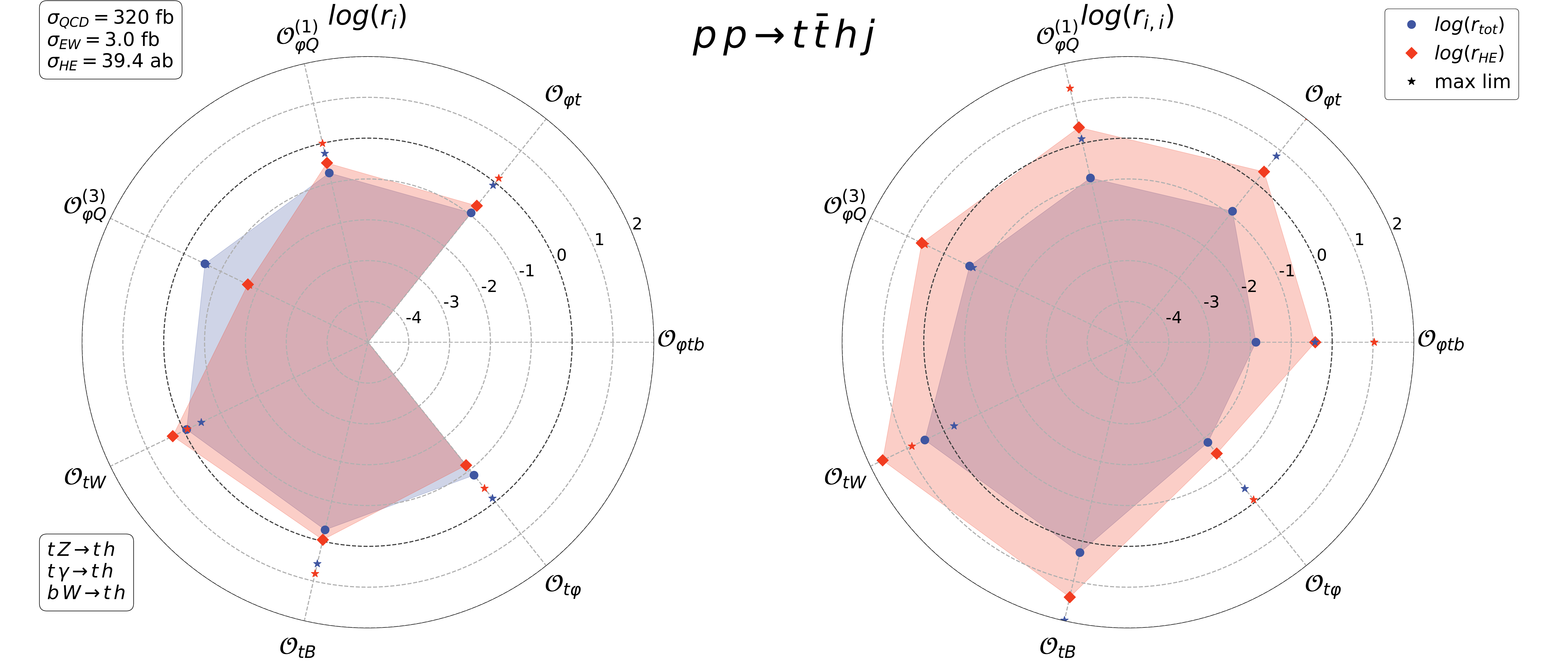}
  \caption{ 
   Radar plot for the $p \, p \to t \, \bar{t} \, h$ (\emph{upper}) and  $p \, p \to t \, \bar{t} \, h j$ (\emph{lower}) processes at the 13 TeV LHC, 
 see
  Figure~\ref{fig:radar_twj_LHC13} and Section~\ref{subsubsec:tWj} of the main text for a detailed description. The $b\bar{b}$-initiated component has not been included.
  \label{fig:radar_tth_LHC13}}
\end{figure}

Figure~\ref{fig:radar_tth_LHC13} shows larger relative enhancements of the EW-induced process compared to $t\bar{t}Z$ (Figure~\ref{fig:radar_ttz_LHC13}), which is likely to be a consequence of the fact that the Higgs cannot be radiated from an initial state quark leg. An interfering energy growth from two of the current operators appears through $t\,Z\to t\,h$ scattering. The third one, $\Op{\phi t}$, shows evidence of a cancellation over the phase space that is worsened by the high-energy cut. The dipole operators in particular yield $\mathcal{O}(1-10)$ effects at inclusive level that grow by a factor of ten at high energies. Sensitivity to the Yukawa operator is also increased with respect to $t\bar{t}Z$ although no growth with energy is observed. Setting the coefficients to their limit values confirms that the required enhancements to overcome the QCD-induced process are not possible, except in the case of the weak dipole operators that could provide as much as a 100-fold enhancement of the EW contribution already at inclusive level. The quadratic contributions from the effective operators display the expect growths in the high energy region. 

Requiring the extra jet in the final state results in a factor 3--4 enhancement of the total rate and a similar sensitivity profile to $t\bar{t}h$. On the other hand, the QCD-induced contributions goes down slightly to 320 $fb$. Interfering growth appears for the currents, with the phase space cancellations moving now to $\Opp{\phi Q}{(3)}$. The energy growth of the quadratic terms, however, is significantly enhanced across the board, with $O(1-10)$ relative enhancements at inclusive level that can grow to $O(100)$ after the high energy cut. Altogether, and inclusive measurement of this process would provide better constraining power than $t\bar{t}Zj$. Both of these processes are good examples of those where the non-EW operators discussed in section~\ref{subsec:SMEFT} affect the QCD-induced component. Although these operators are better constrained from other measurements, the inherent smallness of the EW contribution means that said operators -- even if relatively tightly constrained -- may be relevant in this particular process. It lends further credence to the argument that this process will be a challenging one to analyse in the context of high energy EW top quark scatterings before very significant constraints are established on the non-EW operators.
\subsubsection{$t\bar{t}Zh$ \& $t\bar{t}\gamma h$}
\begin{figure}[h!]
  \centering 
\includegraphics[width=0.28\linewidth]{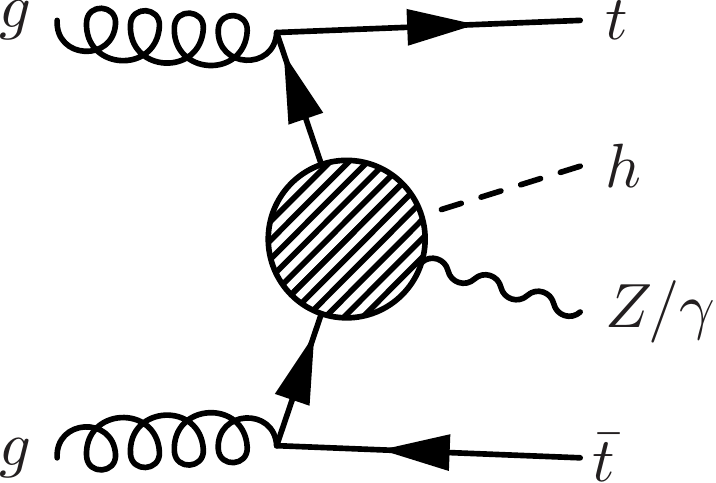}
 \caption{ 
Schematic Feynman diagram for the EW-induced $t\bar{t}(Z/\gamma)h$ processes and their embeddings of the $t\,Z\to t\,h$ and $t\,\gamma\to t\,h$ subamplitudes. 
  \label{fig:diag_ttZh_ttah}}
 \end{figure}
The gluon fusion contribution to these channels is about one half at 13 TeV and two thirds at 27 TeV. This is slightly higher than for, \emph{e.g.}, $t\bar{t} ZZ$, due to the smaller contribution from the $q\bar{q}$-initiated component which does not radiate the Higgs from the initial state. However, the total cross sections are very small, around 1(8) $fb$ for $Zh$ and 3(15) $fb$ for $\gamma h$ at 13(27) TeV. 

 \begin{figure}[h!]
   \centering \includegraphics[height=0.2\paperheight]{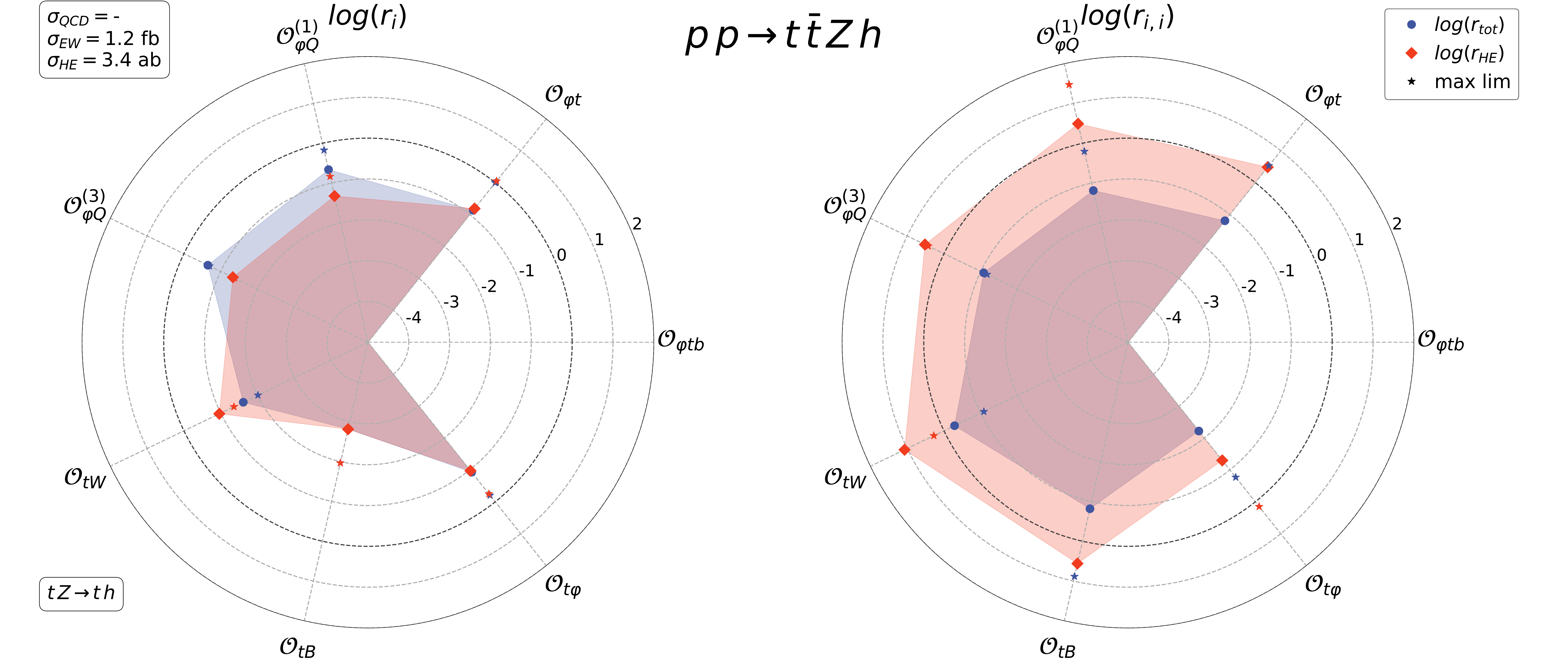}\\[1ex]
 \includegraphics[height=0.2\paperheight]{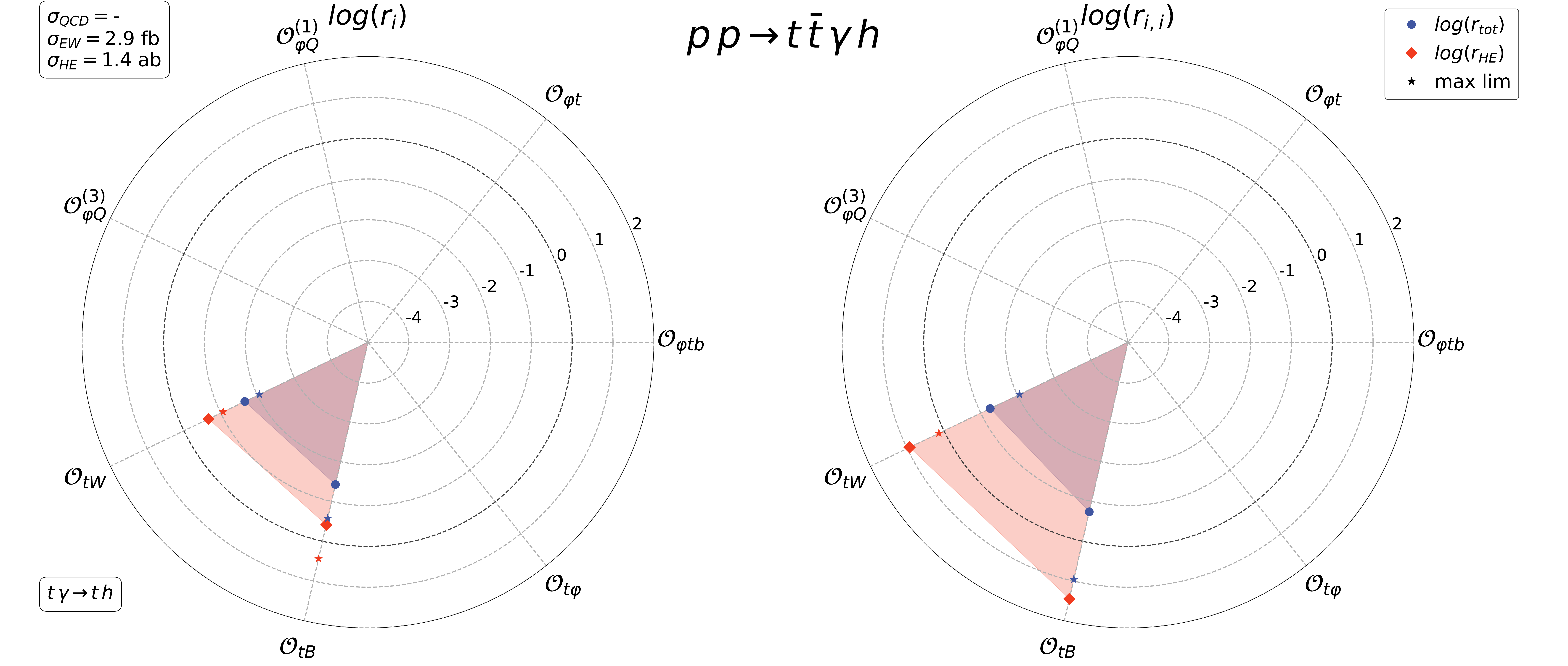}
   \caption{
    Radar plot for the $p \, p \to t \, \bar{t} \,Zh$ and $p \, p \to t \, \bar{t} \,\gamma h$ processes at the 13 TeV LHC,
  see
   Figure~\ref{fig:radar_twj_LHC13} and Section~\ref{subsubsec:tWj} of the main text for a detailed description.
   \label{fig:radar_ttzh_ttah_LHC13}}
 \end{figure}
 
Although one expects energy growing interference from current operators in the $t\bar{t}Zh$ process, Figure~\ref{fig:radar_ttzh_ttah_LHC13} shows that significant phase space cancellations occur that are worsened in the high energy region. This is evidenced by a sign change of the interference contributions with increasing $p_T$ cuts. This is shown in Figures~\ref{fig:summary_plot_a3phidql} and~\ref{fig:summary_plot_aphidql}, where the impact with even higher energy cut, not shown in Figure~\ref{fig:radar_ttzh_ttah_LHC13} is plotted. At quadratic level, a very strong energy growth is observed, consistent with our helicity amplitude computations. The net effect in this channel given current limits can be up to $O(10)$ and $O(100)$ in the  inclusive and high energy phase space, respectively. For the dipole operators, $t\bar{t}\gamma h$ shows a slightly enhanced sensitivity compared to  $t\bar{t}Z\gamma$, with similar cancellations occurring in the interference. Saturating the limits, we see $O(10)$ sensitivity to $\Op{tB}$ already in the total cross section, albeit dominated by the quadratic term. 
\subsubsection{$t\,h\to t\,h$ scattering\label{subsubsec:thth}}
    \begin{figure}[h!]
        \centering
        \subfloat[]{
        \includegraphics[width=0.25\linewidth]{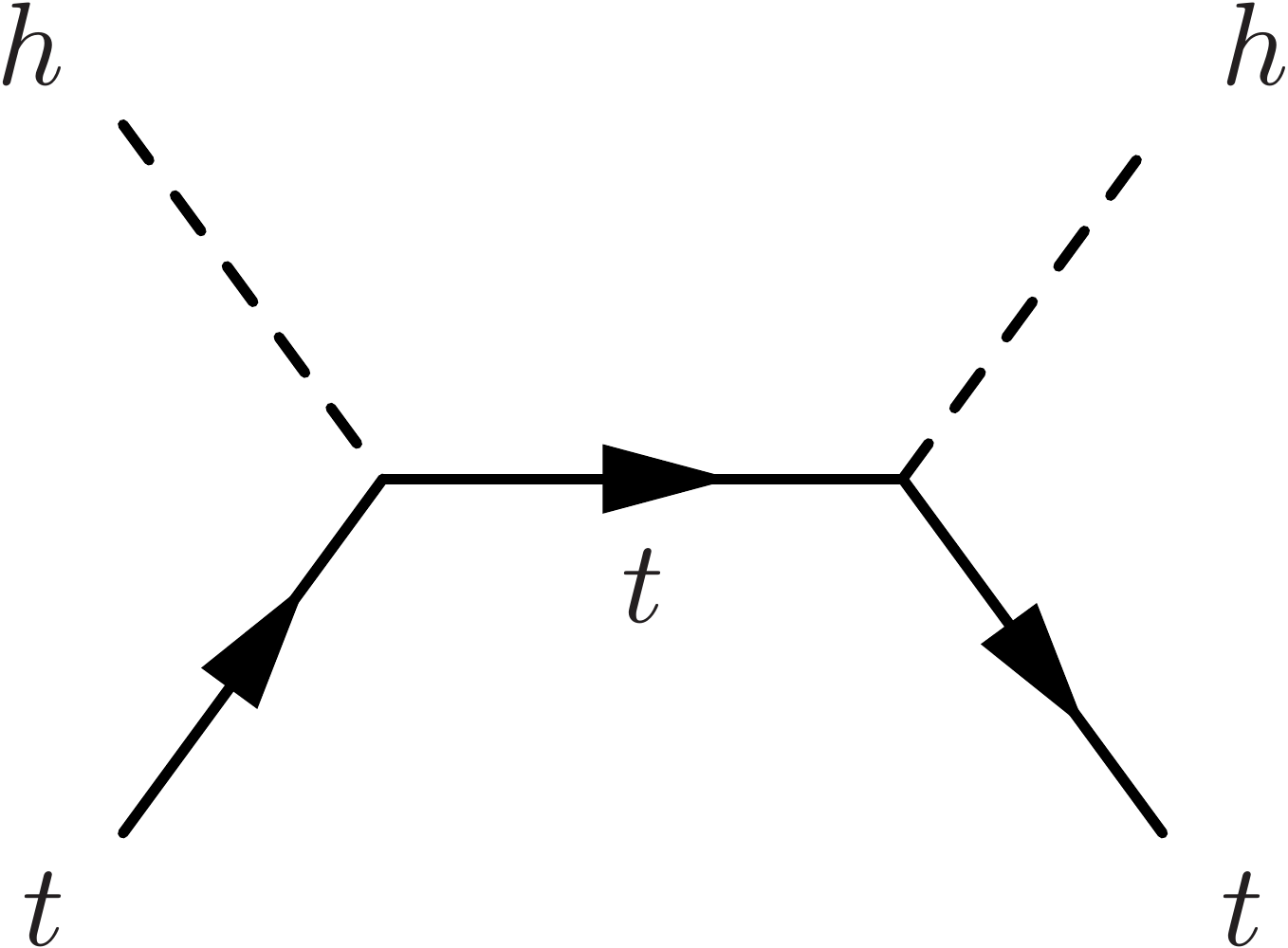}
        }  \subfloat[]{
        \includegraphics[width=0.25\linewidth]{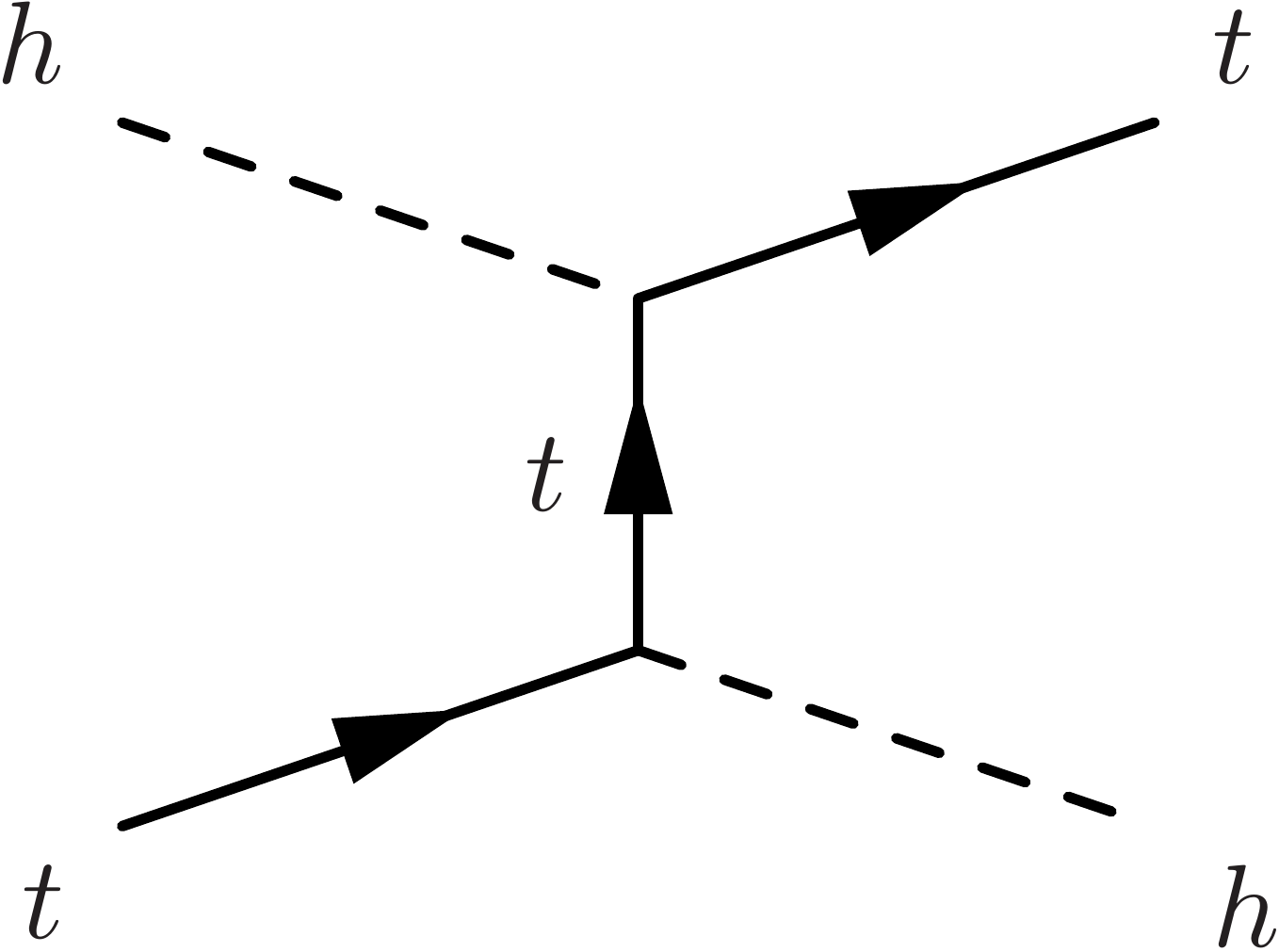}
        }
        \subfloat[]{
        \includegraphics[width=0.25\linewidth]{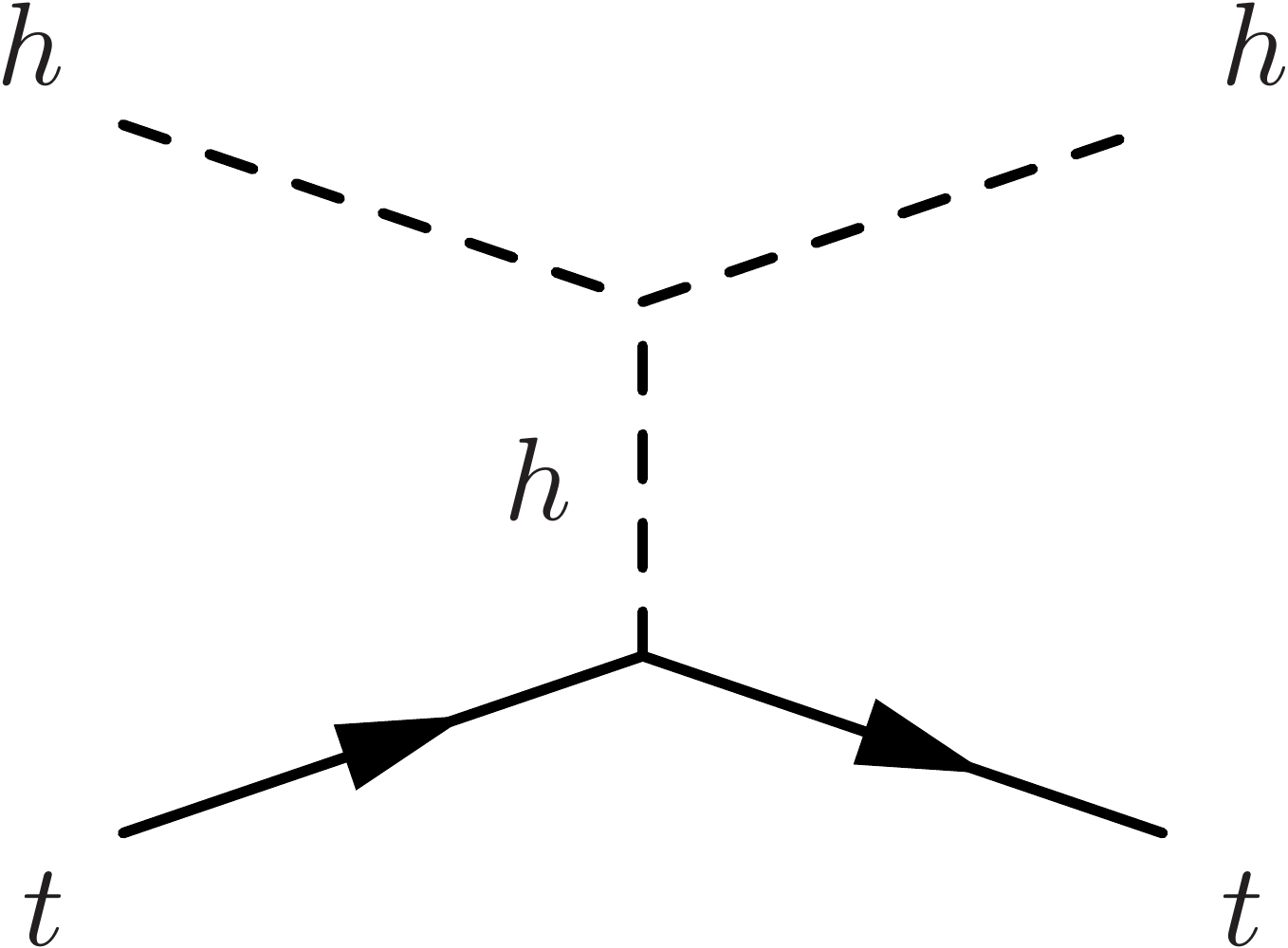}
        }
        \caption{SM diagrams for the $t h \to t h$ subprocess. \label{thth_sm}}
    \end{figure}    
\noindent This process is the simplest in terms of the degrees of freedom involved. The only relevant couplings are the top-Higgs and the triple Higgs interactions (see Fig.~\ref{thth_sm}). Table~\ref{tab:thth}  can be understood in terms of dimension 5 interactions. The operator $\Op{t \phi}$, for example, generates a $tthh$ contact term , while $\Op{\phi D}$ sources a triple Higgs interaction with 2 derivatives. It is not possible to reproduce this behaviour in the AC framework, without the new Lorentz structures for contact terms and higher-derivative couplings.
\subsubsection{$t\bar{t}hh$}
\begin{figure}[h!]
  \centering 
\includegraphics[width=0.28\linewidth]{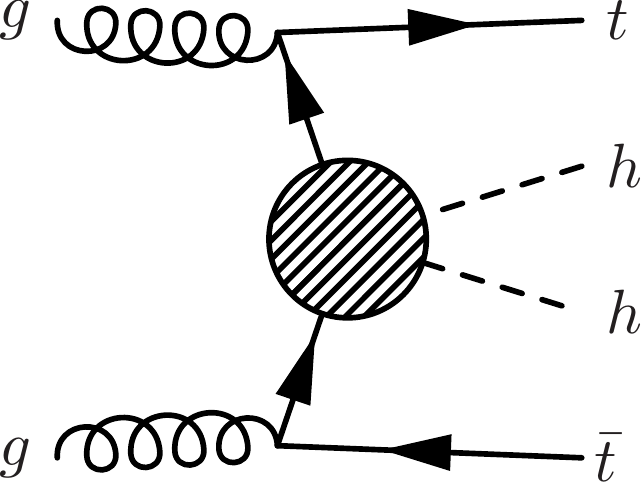}
 \caption{ 
Schematic Feynman diagram for the EW-induced $t\bar{t}hh$ processes and its embedding of the $t\,h\to t\,h$ subamplitude. 
  \label{fig:diag_tthh}}
 \end{figure}
The final process that we consider is $t\bar{t}hh$ production. It has typically been studied as the rarest among several di-Higgs production mechanisms and has also been considered as a way to probe CP-violating top-Higgs interactions~\cite{Liu:2015aka}. It has a similar $gg$-initiated fraction as $t\bar{t}Zh$ and a prediction cross section of 0.7 and 4.5 $fb$ at 13 and 27 TeV, respectively. As far as the top quark operators are concerned, the scattering only depends on the Yukawa operator, meaning that a radar plot is unnecessary. We find that it has a $\sim 30\%$ negative impact on the cross section with a mild energy growth appearing only at quadratic level. Further detail on the sensitivity can be seen in the summary Figure~\ref{fig:summary_plot_atphi}.
\section{Summary\label{sec:summary}}
Our survey of collider processes contains a great deal of information, which we now try to summarise. We have presented the relative impacts of the operators of interest to a large number of EW processes at both $pp$ and $e^+e^-$ colliders whose SM predictions span several orders of magnitude in cross section. These results can serve as a guide for the future directions of top quark measurements targeting SMEFT or modified top-EW interactions in general. However, a number of process-dependent factors must also be taken into account beyond the sensitivity of the EW process, as we have discussed in the individual sections. We have not presented any detailed phenomenological analysis, leaving these for future work. Rather, we have focused on isolating the energy growth from the $2\to2$ subamplitudes in the processes relevant for current and future collider searches. 

The presence of an irreducible, QCD-induced background which does not embed the $2\to2$ amplitudes is often a `diluting' factor to the sensitivity. In processes with an outgoing light jet, the same final state can be obtained by real QCD radiation from an underlying process, i.e. as part of as NLO QCD correction. 
Furthermore, in some cases the modified SM interactions may also contribute to the overall normalisation of the QCD process, without energy growth. The only exception to this is for the dipole operators, which introduce a new Lorentz structure into EW-top interactions. These operators show consistent, maximal energy growth in the EW sub-amplitudes and are also expected to show some milder growth in the QCD components of processes like $t\bar{t}Z$, $t\bar{t}\gamma$ and $tWj$. Given that these contributions tend to lie roughly two orders of magnitude above the EW counterparts, energy growing effects in the EW process will take some time to overcome the basic sensitivity offered by the overall rate of the QCD component. That said, identifying these effects provides a clear pathway for the future, when said measurements become systematics dominated and the large statistics can be exploited to measure differential distributions. 

\begin{figure}[h!]
  \centering 
  \includegraphics[width=\linewidth]{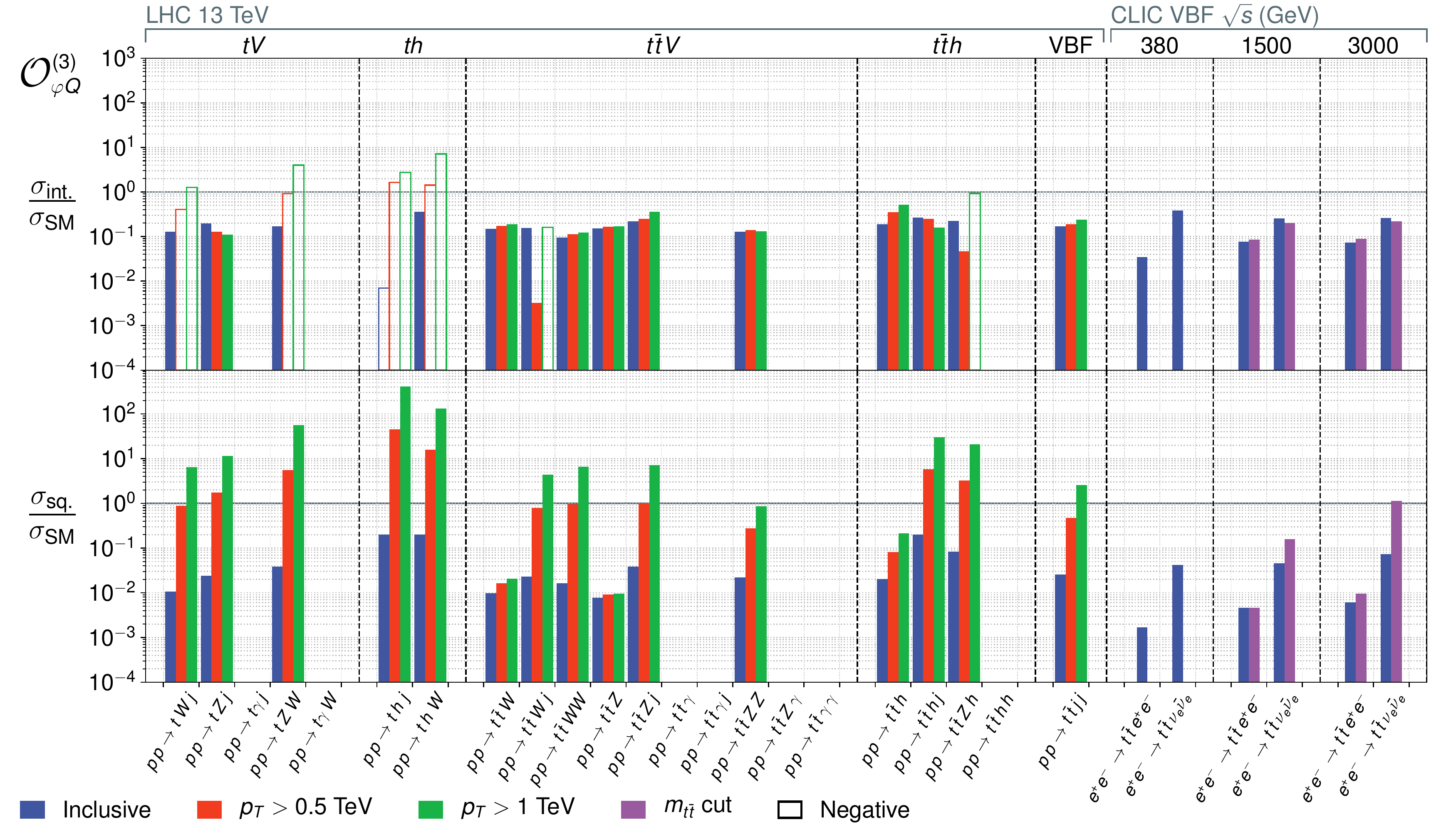}
  \caption{ 
  Summary of relative impact on collider processes for $\Opp{\phi Q}{(3)}$ assuming a Wilson coefficient of 1 TeV$^{-2}$. The upper row shows the interference contribution while the lower row, the quadratic piece. The multiple data points per process denote, from left to right, starting with the inclusive rate, the impact as a function of cuts that access increasingly higher energies. Filled and unfilled bars denote constructive and destructive interference terms respectively, according to the sign conventions in Table~\ref{tab:operators}.
  \label{fig:summary_plot_a3phidql}}
\end{figure}

\begin{figure}[h!]
  \centering 
  \includegraphics[width=\linewidth]{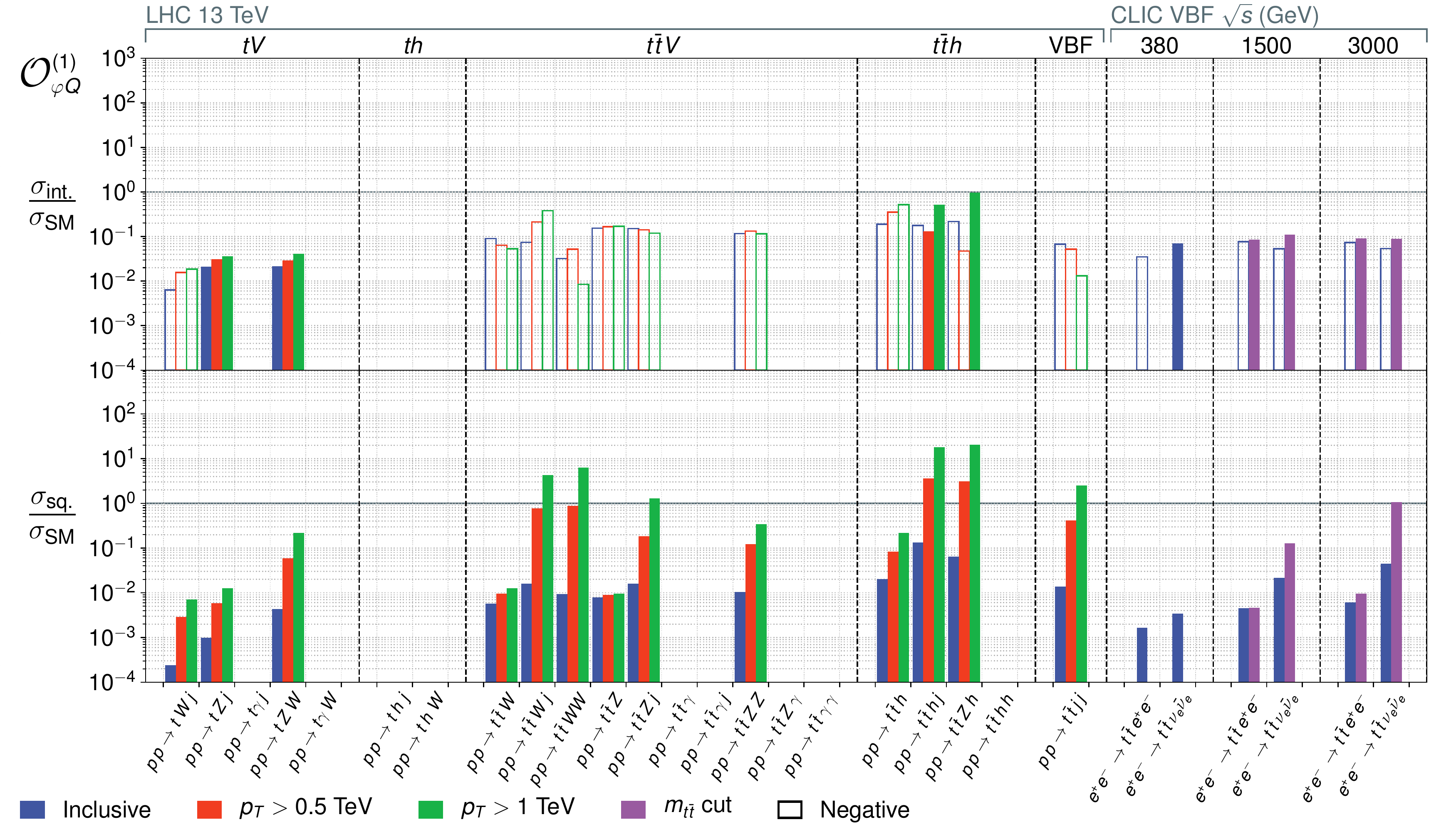}
  \caption{ 
  Same as Figure~\ref{fig:summary_plot_a3phidql} for $\Opp{\phi Q}{(1)}$
  \label{fig:summary_plot_aphidql}}
\end{figure}

\begin{figure}[h!]
  \centering 
  \includegraphics[width=\linewidth]{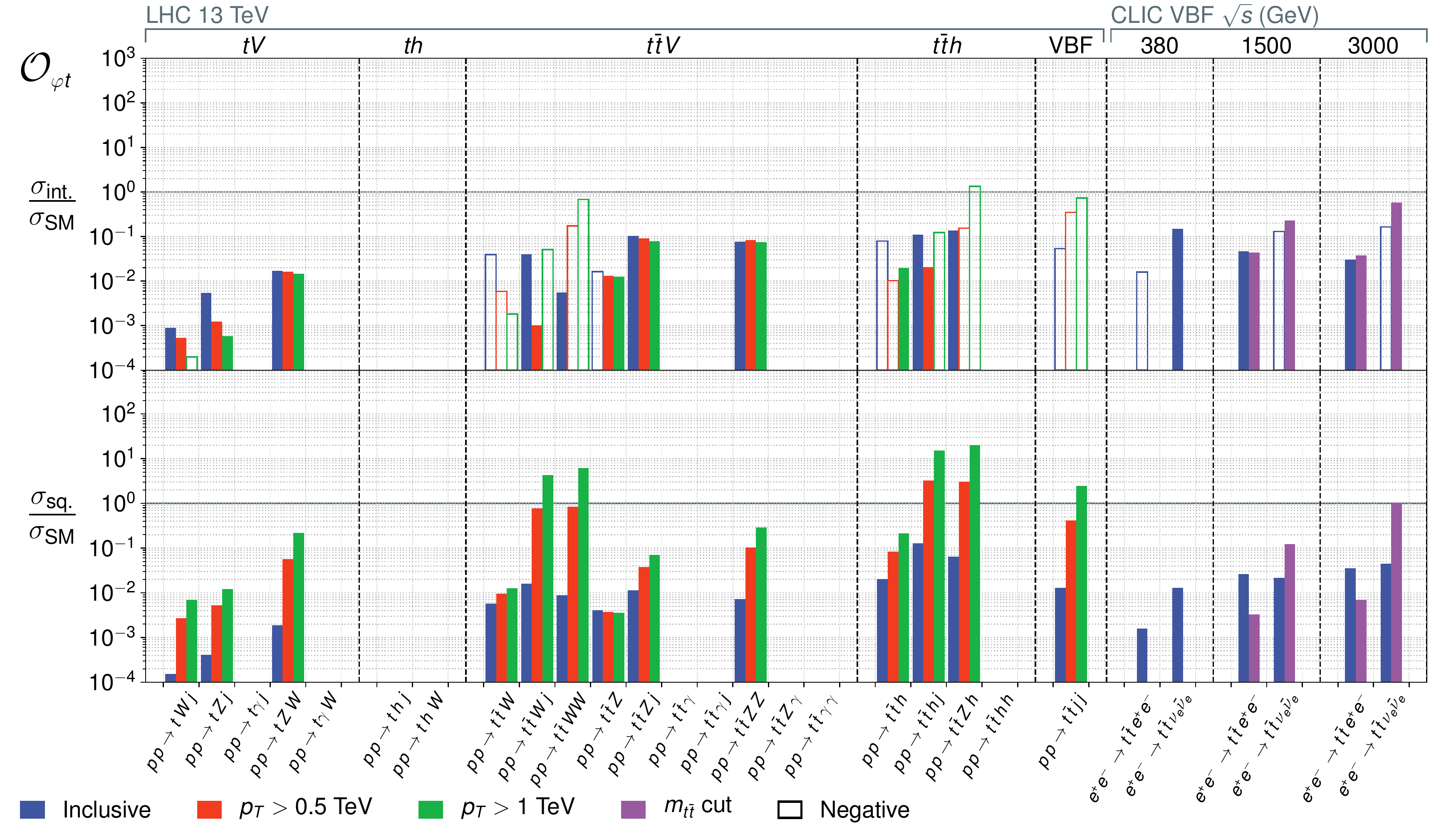}
  \caption{ 
  Same as Figure~\ref{fig:summary_plot_a3phidql} for $\Op{\phi t}$
  \label{fig:summary_plot_aphidtr}}
\end{figure}

\begin{figure}[h!]
  \centering 
  \includegraphics[width=\linewidth]{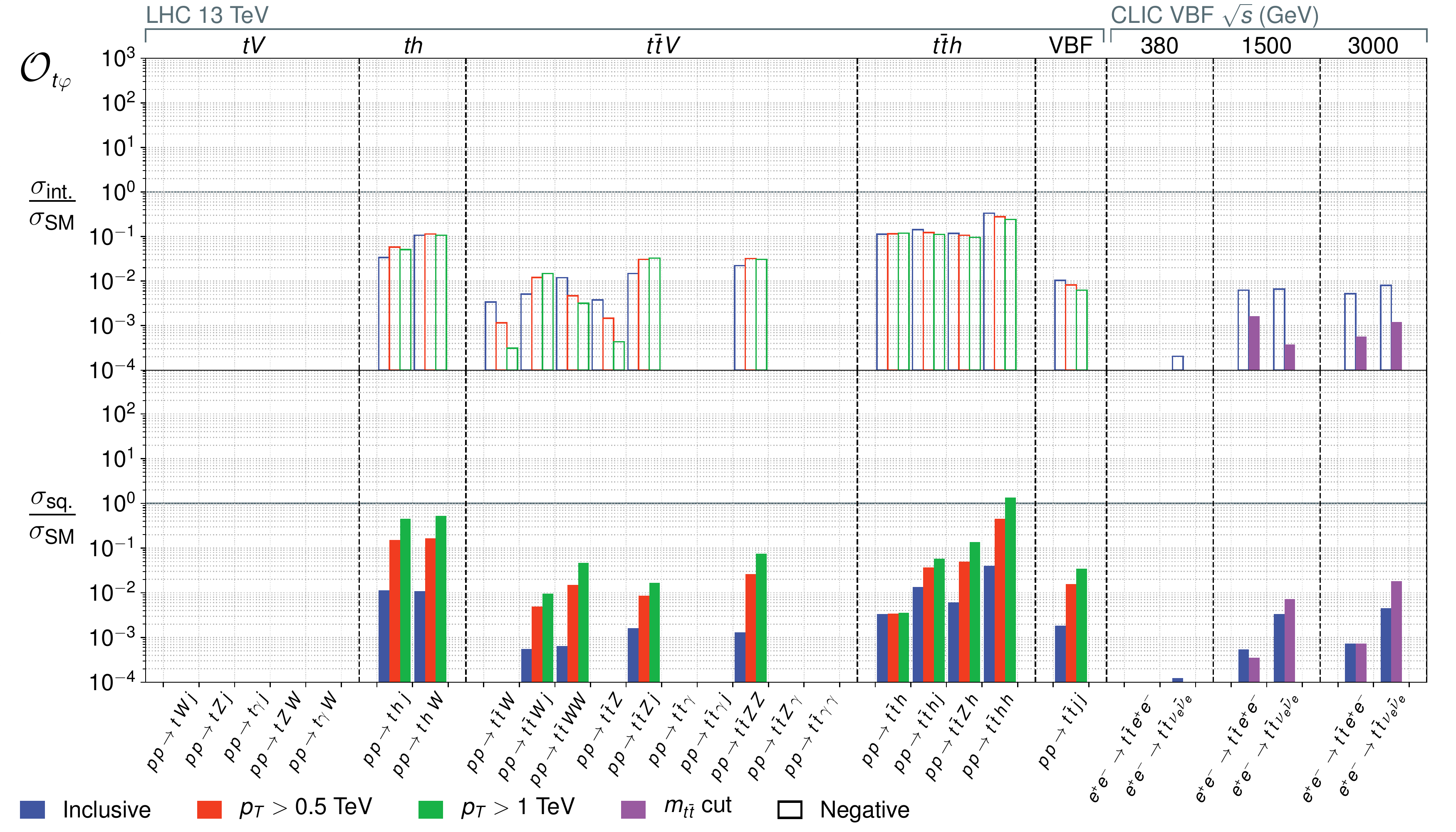}
  \caption{ 
  Same as Figure~\ref{fig:summary_plot_a3phidql} for $\Op{t\phi}$
  \label{fig:summary_plot_atphi}}
\end{figure}

\begin{figure}[h!]
  \centering 
  \includegraphics[width=\linewidth]{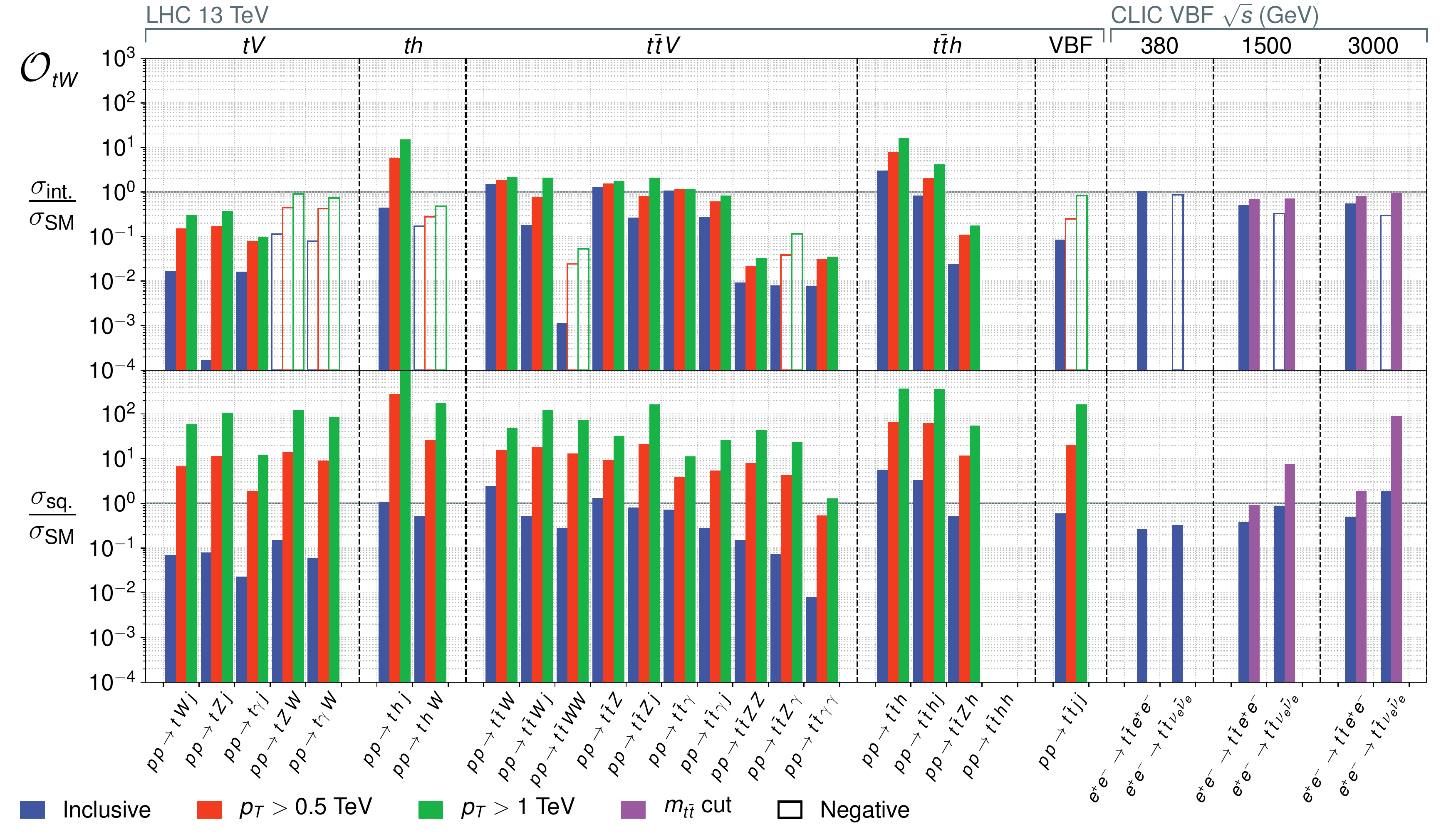}
  \caption{ 
  Same as Figure~\ref{fig:summary_plot_a3phidql} for $\Op{tW}$
  \label{fig:summary_plot_atw}}
\end{figure}

\begin{figure}[h!]
  \centering 
  \includegraphics[width=\linewidth]{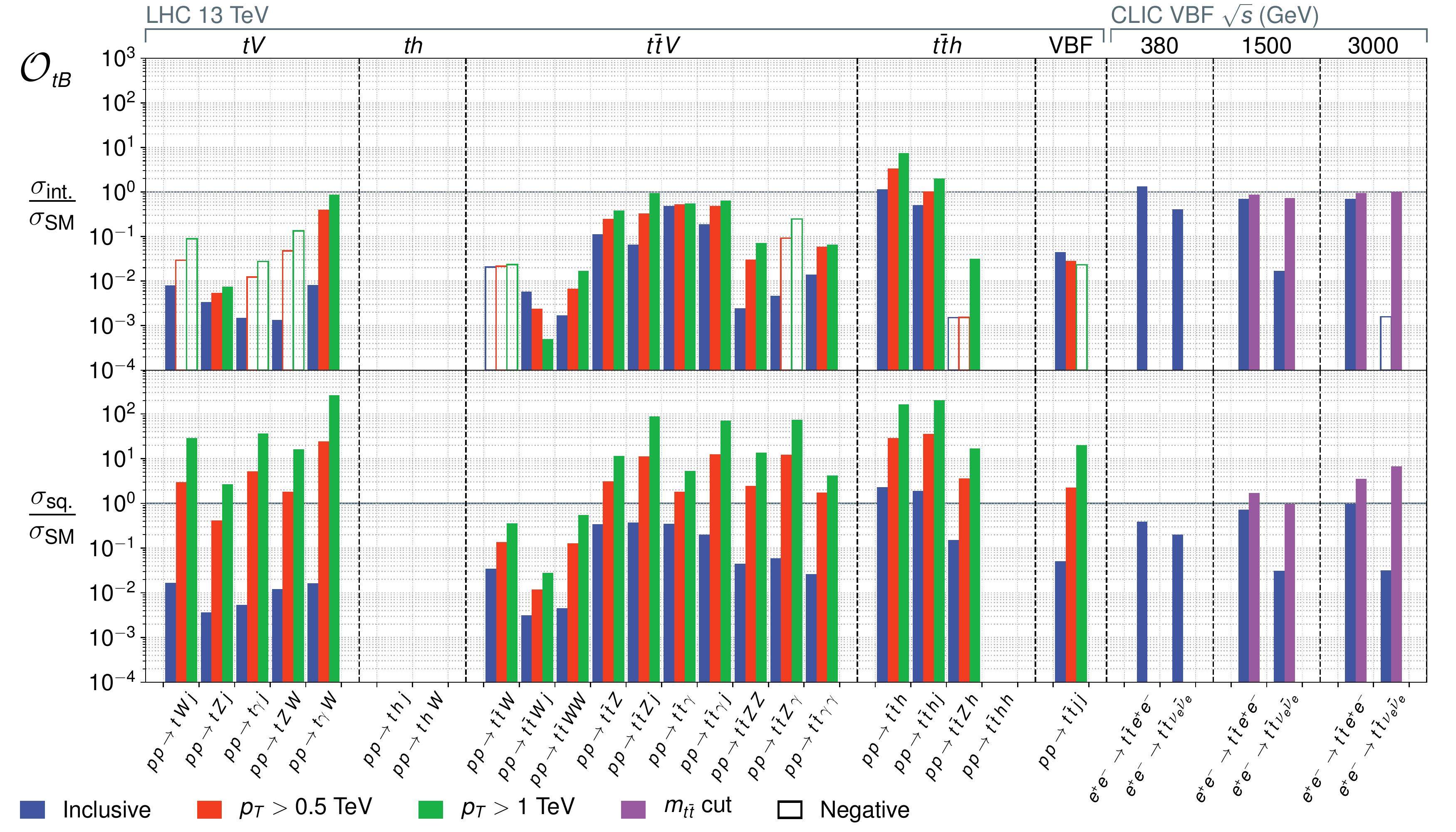}
  \caption{ 
  Same as Figure~\ref{fig:summary_plot_a3phidql} for $\Op{tB}$
  \label{fig:summary_plot_atb}}
\end{figure}

\begin{figure}[h!]
  \centering 
  \includegraphics[width=\linewidth]{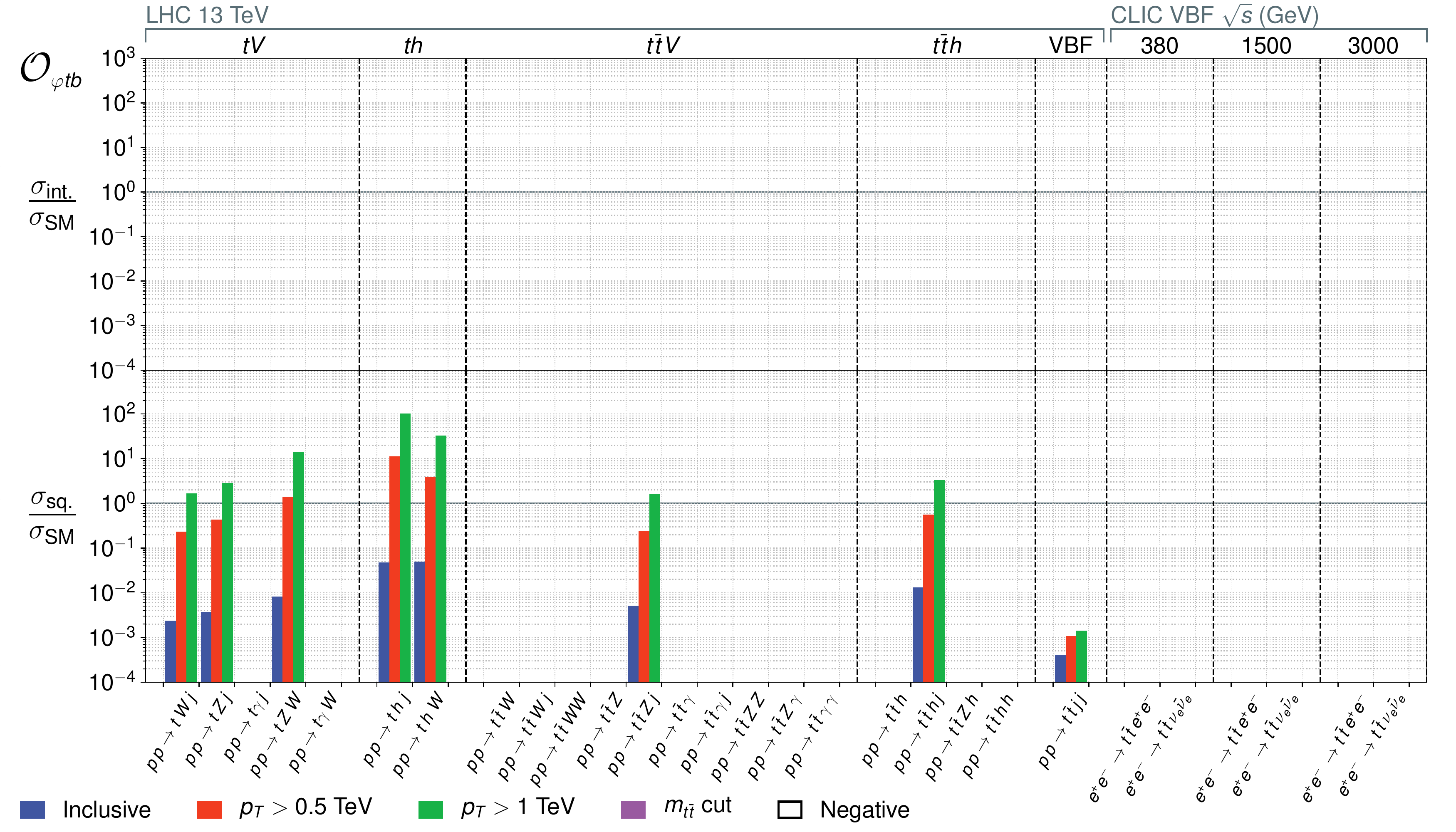}
  \caption{ 
  Same as Figure~\ref{fig:summary_plot_a3phidql} for $\Op{\phi tb}$
  \label{fig:summary_plot_aphitb}}
\end{figure}

The evidence of energy growth in the sub amplitudes is plain to see in the full processes, particularly when considering the quadratic EFT contributions, in which the growth compared to the SM is accentuated. However,  the expected growth in interference contributions is not always present. This is largely due to cancellations over the phase space, as discussed. In this section, we provide an alternative representation of the data in the radar charts, focusing instead on the sensitivities corresponding to each operator. Figures~\ref{fig:summary_plot_a3phidql}--\ref{fig:summary_plot_aphitb} show the interference and squared impacts for each individual operator across all processes that we have studied in this paper. We are able to include two additional sources of information with respect to the previous figures. First, the impacts for a third, higher-energy phase space region are included for the LHC processes, corresponding to the increase of the cut from 500 GeV to 1 TeV. Second, information on the sign of the interference term is also included (filled for constructive and unfilled for destructive). Both pieces of information help us to ascertain whether a given lack of expected energy growth in these terms might be attributed to a phase space cancellation. In many cases, the apparent decrease in the relative contribution due to the first high-energy cut is seen to be due to the contribution changing sign over the phase space. The second high-energy region is then found to have a large relative impact, indicating the actual presence of an energy growth. This behaviour can be seen in, \emph{e.g.}, the $\Opp{\phi Q}{(3)}$ contributions $t\bar{t}Wj$ and $t\bar{t}Zh$ (Figure~\ref{fig:summary_plot_a3phidql}), the $\Opp{\phi Q}{(1)}$ contribution to $t\bar{t}ZH$ (Figure~\ref{fig:summary_plot_aphidql}) or the $\Op{\phi t}$ contributions to $t\bar{t}Wj$, $t\bar{t}h$ and $t\bar{t}hj$ (Figure~\ref{fig:summary_plot_aphidtr}). 

Overall, the picture shows consistency with our $2\to2$ scattering amplitude computations in that, energy growth with respect to the SM is generally encountered when expected in all processes, particularly when considering the quadratic contributions. Interfering growth is more subtle, but more often than not is found when expected. All contributions to configurations including at least one transverse gauge boson do not have an energy growing interference in the high energy limit of the $2\to2$ scatterings. However, there is a complex interplay between the evolution of the various helicity configurations of a process with energy in both the SM and the EFT that determines the eventual importance of an interference term. The case of the dipole operators is an example where the relative impact of the interference term is found to grow even though it is not expected from the $2\to2$ subamplitudes (Figures~\ref{fig:summary_plot_atw} and~\ref{fig:summary_plot_atb}). Our analysis of the EWA suggests that the mediation of these transverse modes may be favoured at high energy when embedding the scatterings into to the collider processes.

\clearpage

\section{Conclusions\label{sec:conclusions}}
We have presented a comprehensive study of energy growing effects in $2\to2$ scattering amplitudes involving top quarks and EW bosons. The source of unitarity violating behaviour due to modified interactions in the EW symmetry breaking sector has been investigated both through the SMEFT and the anomalous couplings frameworks. The impact of SMEFT operators in the Warsaw basis on the helicity amplitudes of 10 relevant scatterings is summarised in the appendix. A subset of the effects have a one-to-one correspondence with the energy growth induced by the anomalous coupling to which a given operator maps. 
We find that, appealing to the Goldstone Equivalence theorem, energy growth can always be connected to contact interactions in Feynman gauge. Unitarity violating behaviour that appears due to rescaling SM interactions can be considered a special case of the general behaviour that we discuss, limited to a subset of purely longitudinal helicity amplitudes. The new, higher-dimensional Lorentz structures introduced by the SMEFT induce yet more energy growth, particularly in non-longitudinal helicity configurations.

We uncovered an interesting difference between the anomalous coupling and SMEFT predictions for the dipole-type interactions in these scattering amplitudes due to the violation of $SU(2)$ gauge invariance of the former. Including just the dipole-type interactions was found to cause energy growth proportional to the third power of energy in the fully-longitudinal configurations, which is higher than the maximum possible growth from dimension-6 operators. In the SMEFT, this component is cancelled by the corresponding $f\bar{f}VV$ contact interactions from the non-Abelian part of the $SU(2)$ gauge field strength. Notice that this difference therefore only appears in processes that can contain this 4-point interaction, meaning that inconsistent predictions do not occur for, \emph{e.g.}, single top production. One should nevertheless interpret constraints on these type of interactions with care in processes with more final states, such as $tZj$ and $tWZ$.

Concerning the energy growing behaviour of the scattering amplitudes, we find that almost all of the operators that we consider lead to maximal ($E^2$)
energy growth in the $2\to2$ scattering amplitudes. By comparing the high energy behaviour of corresponding SM helicity configurations, we also identify the interesting cases in which the interference of the SMEFT contribution with the SM can also grow with energy (implying that the SM contribution does not decrease with energy there). This in found to be a relatively rare scenario, due to the known non-interfering behaviour of dimension-6 $2\to2$ amplitudes, occurring only in the fully-longitudinal gauge boson configurations. These leading contributions in the SMEFT expansion, sourced by so-called current operators, are promising avenues in which the operator space can be investigated at high energies while minimising the impact of formally higher order, dimension-6 squared terms. 
Four out of the ten scattering amplitudes that we studied display such interfering energy growth: $bW\to tZ$, $bW\to th$, $tW\to tW$ and $tZ\to th$. In many cases, SMEFT contributions to a given helicity configuration still have a relative $E^2$ growth with respect to the SM, even if they do not have an energy-growing interference term.

Since the photon couplings are preserved by $U(1)_{\sss EM}$ gauge symmetry, scattering amplitudes involving photons are only affected by the operators containing non-SM Lorentz structures, that source transverse gauge boson polarisations, such as the dipole, triple gauge and gauge-Higgs operators. They therefore do not interfere and do not contain energy growing behaviour not also present in the corresponding scatterings with EW gauge bosons. $t\,h\to t\,h$ scattering is the single instance in which maximal energy growth is not observed. It is only affected by the Yukawa operator and purely bosonic operators containing only the Higgs and derivatives. These can only induce, at most dimension-5 contact interactions, which limits the maximal energy growth to the first power of energy. Higher-point scatterings must be considered to access $E^2$ growth.
 
With our compendium of collider sensitivities, we make a broad survey of the different ways in which these amplitudes might be probed at current and future machines. In general, we find that the expected energy growth from the $2\to2$ study is retained in the full process, especially when looking at the dimension-6 squared pieces. The appraisal of the interference terms is more subtle, due to the possible presence of cancellations over different regions of phase space. Here, the expected interfering energy growth is not always observed.  We hope that this survey will be a starting point for further investigation, having identified several interesting avenues for constraining non-standard interactions in the top-EW sector. 

In particular, a potentially competitive probe of $b\,W\to t\, Z$ scattering emerged in the $tWZ$ process. It is likely that a combination of this and $tZj$ production will be the optimal directions for accessing unitarity violating behaviour. Analogously, $b\,W\to t\, h$ scattering appears to be best probed through a combination of $thj$ and $thW$. The $t\bar{t}$ scattering processes appear somewhat more challenging. We show that the EW $t\bar{t}X$ processes are simply unable to access the high energy phase space region of the $2\to2$ scatterings, due to the off-shell, $s$-channel gauge bosons that mediate them. One is therefore required to go beyond 3-body final states.
The previously studied $t\bar{t}Wj$ process shows promising sensitivity to the squared EFT contributions. The $t\,W\to t\,W$ scattering that it embeds was observed to be the only probe of unitarity violation due to the modification of the $Wtb$ vertex. It also benefits from a peculiar enhancement with respect to $t\bar{t}W$ that we did not observe for the $t\bar{t}Zj$ and $t\bar{t}hj$ analogues. In all cases, the relative impact of the operators was comparable, although slightly better for $t\bar{t}Zj$ and $t\bar{t}hj$. We also investigated $t\bar{t}XY$ and VBF-$t\bar{t}$, typically finding comparable relative sensitivities to $t\bar{t}Xj$ with stronger evidence for the expected interfering energy growth. The latter processes are, however, rate-limited and may require future machines to be sensitive to the SM predictions, especially $VBF$ for which a future $e^+e^-$ collider will most likely be required. Although phenomenologically very interesting, probing $t\,Z\to t\,h$ scattering looks to be a challenging task that may be left to future colliders. The most difficult scattering to access at colliders will be $t\,h\to t\,h$, with only the $t\bar{t}hh$ process able to embed it.

Ultimately, each of the promising processes will merit a dedicated phenomenological study to ascertain the true sensitivity to the SMEFT parameters. This will depend heavily on the details of backgrounds, the presence of QCD-induced versions of the process and process-dependent reconstruction efficiencies in addition to the parametric sensitivities relative to the SM that we have explored in this work.

\section*{Acknowledgements}
This work has received funding from the European Union’s Horizon 2020 research 
and innovation programme as part of the Marie Sk{\l}odowska-Curie Innovative 
Training Network MCnetITN3 (grant agreement no. 722104) and the FNRS 
``Excellence of Science'' EOS be.h Project No. 30820817. KM is supported by a 
Marie Sk{\l}odowska-Curie Individual Fellowship of the European Commission’s 
Horizon 2020 research and innovation programme under contract number 707983. 
\clearpage
\appendix
\section{Helicity amplitudes for top-EW scatterings\label{app:helamp_tables}}
We document the helicity amplitude computations for the ten $f \, B \to f^\prime \, B^\prime$ processes, where $f,f^\prime = b, t$ and $B,B^\prime= h, W, Z, \gamma$ and at least one of the 2 fermions is a top quark.
The {\sc FeynRules}~\cite{Alloul:2013bka} implementation of the SMEFT Lagrangian at dimension 6 based on~\cite{SMEFTatNLO} was used to produce a {\sc FeynArts}~\cite{HAHN2001418} model. The helicity amplitudes were computed  analytically using this model with the help of {\sc FeynCalc} \cite{MERTIG1991345,Shtabovenko:2016sxi}, using the following explicit representations of spinors and polarisation vectors:
    \begin{align}
        u_+(p)&=\sqrt{E+m}\begin{pmatrix}
        \cos{\frac{\theta}{2}} ,& \sin{\frac{\theta}{2}} ,& \frac{|\textbf{p}|\cos{\frac{\theta}{2}}}{E+m} ,& \frac{|\textbf{p}|\sin{\frac{\theta}{2}}}{E+m}
        \end{pmatrix} \, ,
        \\
        u_-(p)&=\sqrt{E+m}\begin{pmatrix}
        -\sin{\frac{\theta}{2}} ,& \cos{\frac{\theta}{2}} ,& \frac{|\textbf{p}|\sin{\frac{\theta}{2}}}{E+m} ,& -\frac{|\textbf{p}|\cos{\frac{\theta}{2}}}{E+m}
        \end{pmatrix} \, ,
        \\
        v_+(p)&=\sqrt{E+m}\begin{pmatrix}
    \frac{|\textbf{p}|\sin{\frac{\theta}{2}}}{E+m} ,& -\frac{|\textbf{p}|\cos{\frac{\theta}{2}}}{E+m} ,& -\sin{\frac{\theta}{2}} ,& \cos{\frac{\theta}{2}}
        \end{pmatrix} \, ,
        \\
        v_-(p)&=\sqrt{E+m}\begin{pmatrix}
        \frac{|\textbf{p}|\cos{\frac{\theta}{2}}}{E+m}  ,& \frac{|\textbf{p}|\sin{\frac{\theta}{2}}}{E+m} ,& \cos{\frac{\theta}{2}} ,& \sin{\frac{\theta}{2}}
        \end{pmatrix} \, ,
        \\
        \varepsilon_\pm(p)&=\frac{1}{\sqrt{2}}\begin{pmatrix}
        0 ,& \cos{\theta} ,& \pm i ,& -\sin{\theta}
        \end{pmatrix} \, ,
        \\
        \varepsilon_0(p)&=\frac{1}{M}\begin{pmatrix}
        |\textbf{p}| ,& E \sin{\theta} ,& 0 ,& E \cos{\theta}
        \end{pmatrix} \, .  
    \end{align}
where m, $E$ and $\textbf{p}$ denote the mass, energy and 3-momentum of a particle and $\theta$ is the polar angle between $\textbf{p}$ and a reference $z$-axis, which we choose to be in the direction of one of the incoming particles in the centre of mass frame of the $2\to2$ scattering. $u$ and $v$ are eigenspinors of the helicity operator, labelled by their eigenvalue, and $\varepsilon_\pm,\,\varepsilon_0$ are the left- and right-circular and longitudinal polarisation vectors. The high-energy limit of the amplitudes in the Mandelstam variables $s~\sim -t\gg v$ is taken for each helicity configuration, keeping sub-leading terms proportional to the EW masses, $\mw,\mz,\mt$. We report our results in table form, containing the SM prediction and the contribution of each operator from Table~\ref{tab:operators} that displays at least one configuration that grows with energy. The energy dependence of the SM prediction is kept schematic (not distinguishing between $s$ and $t$) and only retained down to $1/s$ dependence, with additionally suppressed configurations dropped.
For the SMEFT operators we report energy dependence down to constant energy behaviour ($s^0$), retaining the actual dependence on the Mandelstam variables and dimensionful parameters such as masses, but neglecting overall numerical factors.
\renewcommand{\arraystretch}{1.2}
\begin{table}[h!]
{\footnotesize
\setlength{\tabcolsep}{2pt}
\begin{center}
\begin{tabular}{c|cccccccccc}
$\lambda_{\sss b}$, $\lambda_{\sss W}$, $\lambda_{\sss t}$, $\lambda_{\sss Z}$&SM&$\Op{\phi D}$&$\Op{\phi WB}$&$\Op{W}$&$\Op{tB}$&$\Op{tW}$&$\Op{\phi Q}^{\sss (1)}$&$\Op{\phi Q}^{\sss (3)}$&$\Op{\phi t}$&$\Op{\phi tb}$\tabularnewline
\hline
$-, 0, -, 0$&$s^0$&$s^0$&$s^0$&$s^0$&$-$&$s^0$&$-$&$\sqrt{s (s+t)}$&$-$&$-$\tabularnewline
$-, 0, +, 0$&$\frac{1}{\sqrt{s}}$&$\sqrt{-t} m_t$&$-$&$-$&$\sqrt{-t} m_W$&$\frac{m_W (s+t)}{\sqrt{-t}}$&$\sqrt{-t} m_t$&$ \sqrt{-t} m_t$&$\sqrt{-t} m_t$&$-$\tabularnewline
$+, 0, -, 0$&$-$&$-$&$-$&$-$&$-$&$-$&$-$&$-$&$-$&$-$\tabularnewline
$+, 0, +, 0$&$-$&$-$&$-$&$-$&$-$&$-$&$-$&$-$&$-$&$\sqrt{s (s+t)}$\tabularnewline
\hline
$-, 0, -, -$&$\frac{1}{\sqrt{s}}$&$-$&$\frac{s m_W}{\sqrt{-t}}$&$\frac{m_W^2 (s+t)}{\sqrt{-t} v}$&$\sqrt{-t} m_t$&$\sqrt{-t} m_t$&$-$&$\sqrt{-t} m_W$&$-$&$-$\tabularnewline
$-, -, -, 0$&$\frac{1}{\sqrt{s}}$&$-$&$-$&$\frac{m_W^2 (s+t)}{\sqrt{-t} v}$&$-$&$-$&$-$&$\sqrt{-t}m_W$&$-$&$-$\tabularnewline
$-, 0, +, -$&$s^0$&$s^0$&$s^0$&$s^0$&$-$&$s^0$&$s^0$&$s^0$&$-$&$-$\tabularnewline
$-, -, +, 0$&$\frac{1}{s}$&$s^0$&$s^0$&$s^0$&$s^0$&$\sqrt{s (s+t)}$&$s^0$&$s^0$&$s^0$&$-$\tabularnewline
$-, 0, -, +$&$\frac{1}{\sqrt{s}}$&$-$&$\frac{m_W (s+t)}{\sqrt{-t}}$&$\frac{m_W^2 (s+t)}{\sqrt{-t} v}$&$-$&$-$&$-$&$-$&$-$&$-$\tabularnewline
$-, +, -, 0$&$\frac{1}{\sqrt{s}}$&$-$&$-$&$\frac{m_W^2 (s+t)}{\sqrt{-t} v}$&$-$&$-$&$-$&$-$&$-$&$-$\tabularnewline
$-, 0, +, +$&$\frac{1}{s}$&$s^0$&$-$&$-$&$\sqrt{s (s+t)}$&$\sqrt{s (s+t)}$&$s^0$&$s^0$&$s^0$&$-$\tabularnewline
$-, +, +, 0$&$s^0$&$s^0$&$s^0$&$-$&$-$&$s^0$&$-$&$s^0$&$-$&$-$\tabularnewline
$+, 0, -, \pm$&$-$&$-$&$-$&$-$&$-$&$-$&$-$&$-$&$-$&$s^0$\tabularnewline
$+, -, -, 0$&$-$&$-$&$-$&$-$&$-$&$-$&$-$&$-$&$-$&$s^0$\tabularnewline
$+, 0, +, -$&$-$&$-$&$-$&$-$&$-$&$-$&$-$&$-$&$-$&$-$\tabularnewline
$+, \pm, \mp, 0$&$-$&$-$&$-$&$-$&$-$&$-$&$-$&$-$&$-$&$-$\tabularnewline
$+, 0, +, +$&$-$&$-$&$-$&$-$&$-$&$-$&$-$&$-$&$-$&$\sqrt{-t}m_W$\tabularnewline
$+, +, +, 0$&$-$&$-$&$-$&$-$&$-$&$-$&$-$&$-$&$-$&$\sqrt{-t}m_W$\tabularnewline
\hline
$-, -, -, -$&$s^0$&$s^0$&$s^0$&$s^0$&$s^0$&$s^0$&$s^0$&$s^0$&$-$&$-$\tabularnewline
$-, -, -, +$&$\frac{1}{s}$&$-$&$s^0$&$\frac{m_W \sqrt{s (s+t)}}{v}$&$-$&$-$&$-$&$-$&$-$&$-$\tabularnewline
$-, -, +, -$&$\frac{1}{\sqrt{s}}$&$-$&$-$&$-$&$-$&$\frac{m_W (s+t)}{\sqrt{-t}}$&$-$&$-$&$-$&$-$\tabularnewline
$-, -, +, +$&$-$&$-$&$\sqrt{-t} m_t$&$\frac{\sqrt{-t} m_t m_W}{v}$&$\sqrt{-t} m_W$&$\sqrt{-t} m_W$&$-$&$-$&$-$&$-$\tabularnewline
$-, +, -, -$&$\frac{1}{s}$&$-$&$s^0$&$\frac{m_W \sqrt{s (s+t)}}{v}$&$-$&$-$&$-$&$-$&$-$&$-$\tabularnewline
$-, +, -, +$&$s^0$&$s^0$&$s^0$&$-$&$-$&$-$&$s^0$&$s^0$&$-$&$-$\tabularnewline
$-, +, +, -$&$\frac{1}{\sqrt{s}}$&$-$&$\sqrt{-t} m_t$&$\frac{\sqrt{-t} m_t m_W}{v}$&$-$&$-$&$-$&$-$&$-$&$-$\tabularnewline
$-, +, +, +$&$\frac{1}{\sqrt{s}}$&$-$&$-$&$-$&$-$&$\frac{m_W (s+t)}{\sqrt{-t}}$&$-$&$-$&$-$&$-$\tabularnewline
$+, \,\ast\,, -, \,\ast\,$&$-$&$-$&$-$&$-$&$-$&$-$&$-$&$-$&$-$&$-$\tabularnewline
$+, \pm, +, \pm$&$-$&$-$&$-$&$-$&$-$&$-$&$-$&$-$&$-$&$s^0$\tabularnewline
$+, \pm, +, \mp$&$-$&$-$&$-$&$-$&$-$&$-$&$-$&$-$&$-$&$-$\tabularnewline
\end{tabular}
\end{center}

}
\caption{\label{tab:bwtz}
Helicity amplitudes in the high-energy limit ($s\sim-t\gg v$) for $b\,W^+\to t\,Z$ scattering. Overall multiplicative factors are dropped and only the schematic energy
dependence is retained for the SM contribution. Helicity entries marked by `$\ast$' indicate any combination of $\pm,\mp$. See Section~\ref{subsubsec:bwtz_bwta} for a discussion of the results.
}
\end{table}
\renewcommand{\arraystretch}{1.}
    \renewcommand{\arraystretch}{1.2}
\begin{table}[h!]
\begin{center}
\begin{tabular}{c|cccccc}
$\lambda_{\sss b}$, $\lambda_{\sss W}$, $\lambda_{\sss t}$, $\lambda_\gamma$&SM&$\Op{\phi WB}$&$\Op{W}$&$\Op{tB}$&$\Op{tW}$\tabularnewline
\hline
$-, 0, -, -$&$\frac{1}{\sqrt{s}}$&$\frac{s m_W}{\sqrt{-t}}$&$\frac{m_W^2 (s+ t)}{\sqrt{-t} v}$&$ \sqrt{-t} m_t$&$\sqrt{-t} m_t$\tabularnewline
$-, 0, -, +$&$\frac{1}{\sqrt{s}}$&$\frac{m_W (s+t)}{\sqrt{-t}}$&$\frac{m_W^2 (s+t)}{\sqrt{-t} v}$&$-$&$-$\tabularnewline
$-, 0, +, -$&$s^0$&$s^0$&$s^0$&$-$&$s^0$\tabularnewline
$-, 0, +, +$&$\frac{1}{s}$&$-$&$-$&$ \sqrt{s (s+t)}$&$ \sqrt{s (s+t)}$\tabularnewline
$+, 0, \,\ast\,, \,\ast\,$&$-$&$-$&$-$&$-$&$-$\tabularnewline
\hline
$-, -, -, -$&$s^0$&$s^0$&$s^0$&$s^0$&$s^0$\tabularnewline
$-,\pm, -, \mp$&$\frac{1}{s}$&$s^0$&$\frac{ m_W \sqrt{s (s+t)}}{v}$&$-$&$-$\tabularnewline
$-, \pm, +, \pm$&$\frac{1}{\sqrt{s}}$&$-$&$-$&$-$&$\frac{m_W ( s+ t)}{ \sqrt{-t}}$\tabularnewline
$-, -, +, +$&$-$&$\sqrt{-t} m_t$&$\frac{ \sqrt{-t} m_t m_W}{v}$&$\sqrt{-t} m_W$&$\sqrt{-t} m_W$\tabularnewline
$-, +, -, +$&$s^0$&$s^0$&$-$&$-$&$-$\tabularnewline
$-, +, +, -$&$\frac{1}{\sqrt{s}}$&$\sqrt{-t} m_t$&$\frac{\sqrt{-t} m_t m_W}{v}$&$-$&$-$\tabularnewline
$+, \,\ast\,, \,\ast\,, \,\ast\,$&$-$&$-$&$-$&$-$&$-$\tabularnewline
\end{tabular}
\end{center}
\caption{\label{tab:bwta}
Helicity amplitudes in the high-energy limit ($s\sim-t\gg v$) for $b W^+\to t \gamma$ scattering. Overall multiplicative factors are dropped and only the schematic energy
dependence is retained for the SM contribution. Helicity entries marked by `$\ast$' indicate any combination of $\pm,\mp$. See Section~\ref{subsubsec:bwtz_bwta} for a discussion of the results.
}
\end{table}
\renewcommand{\arraystretch}{1.}

    \renewcommand{\arraystretch}{1.2}
\begin{table}[h!]
\centering
\begin{tabular}{c|cccccc}
$\lambda_{\sss b}$, $\lambda_{\sss W}$, $\lambda_{\sss t}$&SM&$\Op{\phi W}$&$\Op{t \phi}$&$\Op{tW}$&$\Op{\phi Q}^{\sss (3)}$&$\Op{\phi tb}$\tabularnewline
\hline
$-, 0, -$&$s^0$&$s^0$&$s^0$&$s^0$&$\sqrt{s (s+t)}$&$-$\tabularnewline
$-, 0, +$&$\frac{1}{\sqrt{s}}$&$-$&$\sqrt{-t} v$&$\frac{s m_W}{\sqrt{-t}}$&$\sqrt{-t} m_t$&$-$\tabularnewline
$+, 0, -$&$-$&$-$&$-$&$-$&$-$&$\sqrt{-t} m_t$\tabularnewline
$+, 0, +$&$-$&$-$&$-$&$-$&$-$&$ \sqrt{s (s+t)}$\tabularnewline
\hline
$-, -, -$&$\frac{1}{\sqrt{s}}$&$\frac{s m_W}{\sqrt{-t}}$&$-$&$\sqrt{-t} m_t$&$\sqrt{-t} m_W$&$-$\tabularnewline
$-, -, +$&$\frac{1}{s}$&$-$&$s^0$&$ \sqrt{s (s+t)}$&$s^0$&$-$\tabularnewline
$-, +, -$&$\frac{1}{\sqrt{s}}$&$\frac{m_W (s+t)}{\sqrt{-t}}$&$-$&$-$&$-$&$-$\tabularnewline
$-, +, +$&$s^0$&$s^0$&$-$&$s^0$&$s^0$&$-$\tabularnewline
$+, \pm, -$&$-$&$-$&$-$&$-$&$-$&$s^0$\tabularnewline
$+, -, +$&$-$&$-$&$-$&$-$&$-$&$-$\tabularnewline
$+, +, +$&$-$&$-$&$-$&$-$&$-$&$\sqrt{-t} m_W$\tabularnewline
\end{tabular}
\caption{\label{tab:bwth}
Helicity amplitudes in the high-energy limit ($s\sim-t\gg v$) for $b \, W^+ \to t \, h$ scattering.
Overall multiplicative factors are dropped and only the schematic energy
dependence is retained for the SM contribution. Helicity entries marked by `$\ast$' indicate any combination of $\pm,\mp$. See Section~\ref{subsubsec:bwth} for a discussion of the results.
}
\end{table}
\renewcommand{\arraystretch}{1.}


    \renewcommand{\arraystretch}{1.2}
\begin{table}[h!]
{\scriptsize
\setlength{\tabcolsep}{1pt}
\begin{center}
\begin{tabular}{c|cccccccccccc}
$\lambda_{\sss t}$, $\lambda_{\sss W}$, $\lambda_{\sss t}$, $\lambda_{\sss W}$&SM&$\Op{\phi D}$&$\Op{\phi d}$&$\Op{\phi W}$&$\Op{\phi WB}$&$\Op{W}$&$\Op{t \phi}$&$\Op{tB}$&$\Op{tW}$&$\Op{\phi Q}^{\sss (1)}$&$\Op{\phi Q}^{\sss (3)}$&$\Op{\phi t}$\tabularnewline
\hline
$-, 0, -, 0$&$s^0$&$s^0$&$s^0$&$-$&$s^0$&$s^0$&$s^0$&$s^0$&$s^0$&$ \sqrt{s (s+t)}$&$ \sqrt{s (s+t)}$&$-$\tabularnewline
$\pm, 0, \mp, 0$&$\frac{1}{\sqrt{s}}$&$\sqrt{-t} m_t$&$ \sqrt{-t} m_t$&$-$&$-$&$-$&$\sqrt{-t} v$&$\frac{m_W (s+t)}{\sqrt{-t}}$&$\frac{m_W (s+t)}{\sqrt{-t}}$&$\sqrt{-t} m_t$&$\sqrt{-t} m_t$&$\sqrt{-t} m_t$\tabularnewline
$+, 0, +, 0$&$s^0$&$s^0$&$s^0$&$-$&$s^0$&$-$&$s^0$&$s^0$&$-$&$-$&$s^0$&$\sqrt{s (s+t)}$\tabularnewline
\hline
$-, 0, -, -$&$\frac{1}{\sqrt{s}}$&$-$&$-$&$-$&$\frac{s m_W}{ \sqrt{-t}}$&$\frac{m_W^2 (s+ t)}{\sqrt{-t} v}$&$-$&$-$&$-$&$\sqrt{-t} m_W$&$ \sqrt{-t} m_W$&$-$\tabularnewline
$-, -, -, 0$&$\frac{1}{\sqrt{s}}$&$-$&$-$&$-$&$\frac{s m_W}{ \sqrt{-t}}$&$\frac{m_W^2 (s+ t)}{\sqrt{-t} v}$&$-$&$-$&$-$&$ \sqrt{-t} m_W$&$ \sqrt{-t} m_W$&$-$\tabularnewline
$-, 0, +, -$&$s^0$&$s^0$&$s^0$&$s^0$&$s^0$&$s^0$&$s^0$&$s^0$&$s^0$&$s^0$&$s^0$&$s^0$\tabularnewline
$-, -, +, 0$&$-$&$s^0$&$s^0$&$s^0$&$s^0$&$s^0$&$s^0$&$s^0$&$s^0$&$s^0$&$s^0$&$s^0$\tabularnewline
$-, 0, -, +$&$\frac{1}{\sqrt{s}}$&$-$&$-$&$-$&$\frac{ m_W (s+t)}{ \sqrt{-t}}$&$\frac{ m_W^2 (s+t)}{\sqrt{-t} v}$&$-$&$-$&$-$&$-$&$-$&$-$\tabularnewline
$-, +, -, 0$&$\frac{1}{\sqrt{s}}$&$-$&$-$&$-$&$\frac{ m_W (s+t)}{\sqrt{-t}}$&$\frac{ m_W^2 (s+t)}{\sqrt{-t} v}$&$-$&$-$&$-$&$-$&$-$&$-$\tabularnewline
$-, 0, +, +$&$-$&$s^0$&$s^0$&$s^0$&$s^0$&$-$&$s^0$&$s^0$&$\sqrt{s (s+t)}$&$s^0$&$s^0$&$s^0$\tabularnewline
$-, +, +, 0$&$-$&$s^0$&$s^0$&$s^0$&$s^0$&$-$&$s^0$&$s^0$&$s^0$&$s^0$&$s^0$&$s^0$\tabularnewline
$+, 0, -, -$&$-$&$s^0$&$s^0$&$s^0$&$s^0$&$s^0$&$s^0$&$s^0$&$s^0$&$s^0$&$s^0$&$s^0$\tabularnewline
$+, -, -, 0$&$s^0$&$s^0$&$s^0$&$s^0$&$s^0$&$s^0$&$s^0$&$s^0$&$s^0$&$s^0$&$s^0$&$s^0$\tabularnewline
$+, 0, +, -$&$\frac{1}{\sqrt{s}}$&$-$&$-$&$-$&$\frac{ m_W (s+t)}{\sqrt{-t}}$&$-$&$-$&$-$&$\sqrt{-t} m_t$&$-$&$-$&$-$\tabularnewline
$+, -, +, 0$&$\frac{1}{\sqrt{s}}$&$-$&$-$&$-$&$\frac{ m_W (s+t)}{ \sqrt{-t}}$&$-$&$-$&$-$&$ \sqrt{-t} m_t$&$-$&$-$&$-$\tabularnewline
$+, 0, -, +$&$-$&$s^0$&$s^0$&$s^0$&$s^0$&$-$&$s^0$&$s^0$&$s^0$&$s^0$&$s^0$&$s^0$\tabularnewline
$+, +, -, 0$&$-$&$s^0$&$s^0$&$s^0$&$s^0$&$-$&$s^0$&$s^0$&$ \sqrt{s (s+t)}$&$s^0$&$s^0$&$s^0$\tabularnewline
$+, 0, +, +$&$\frac{1}{\sqrt{s}}$&$-$&$-$&$-$&$\frac{s m_W}{ \sqrt{-t}}$&$-$&$-$&$-$&$\sqrt{-t} m_t$&$-$&$-$&$ \sqrt{-t} m_W$\tabularnewline
$+, +, +, 0$&$\frac{1}{\sqrt{s}}$&$-$&$-$&$-$&$\frac{ s m_W}{\sqrt{-t}}$&$-$&$-$&$-$&$\sqrt{-t} m_t$&$-$&$-$&$ \sqrt{-t} m_W$\tabularnewline
\hline
$-, -, -, -$&$s^0$&$-$&$-$&$-$&$s^0$&$s^0$&$-$&$-$&$s^0$&$-$&$s^0$&$-$\tabularnewline
$-, \pm, -, \mp$&$\frac{1}{s}$&$-$&$-$&$s^0$&$s^0$&$\frac{m_W \sqrt{s (s+t)}}{v}$&$-$&$-$&$-$&$-$&$-$&$-$\tabularnewline
$-, +, -, +$&$s^0$&$-$&$-$&$-$&$s^0$&$-$&$-$&$-$&$-$&$s^0$&$-$&$-$\tabularnewline
$-, \pm, \pm, \mp$&$-$&$-$&$-$&$\sqrt{-t} m_t$&$-$&$\frac{\sqrt{-t} m_t m_W}{v}$&$-$&$-$&$-$&$-$&$-$&$-$\tabularnewline
$\pm, -, \mp, -$&$\frac{1}{\sqrt{s}}$&$-$&$-$&$-$&$-$&$-$&$-$&$-$&$\frac{s m_W}{\sqrt{-t}}$&$-$&$-$&$-$\tabularnewline
$\pm, +, \mp, +$&$\frac{1}{\sqrt{s}}$&$-$&$-$&$-$&$-$&$-$&$-$&$-$&$\frac{ m_W (s+t)}{\sqrt{-t}}$&$-$&$-$&$-$\tabularnewline
$\pm, \pm, \mp, \mp$&$-$&$-$&$-$&$\sqrt{-t} m_t$&$-$&$\frac{\sqrt{-t} m_t m_W}{v}$&$-$&$-$&$ \sqrt{-t} m_W$&$-$&$-$&$-$\tabularnewline
$+, -, +, -$&$\frac{1}{s}$&$-$&$-$&$-$&$s^0$&$-$&$-$&$-$&$s^0$&$-$&$-$&$s^0$\tabularnewline
$+, \pm, +, \mp$&$-$&$-$&$-$&$s^0$&$s^0$&$s^0$&$-$&$-$&$-$&$-$&$-$&$-$\tabularnewline
$+, +, +, +$&$\frac{1}{s}$&$-$&$-$&$-$&$s^0$&$-$&$-$&$-$&$s^0$&$-$&$-$&$-$\tabularnewline
\end{tabular}
\end{center}
}
\caption{\label{tab:twtw}
Helicity amplitudes in the high-energy limit ($s\sim-t\gg v$) for $t W^+\to t W^+$ scattering. Overall multiplicative factors are dropped and only the schematic energy
dependence is retained for the SM contribution.  Helicity entries marked by `$\ast$' indicate any combination of $\pm,\mp$. See Section~\ref{subsubsec:twtw} for a discussion of the results.
}
\end{table}
\renewcommand{\arraystretch}{1.}

    \renewcommand{\arraystretch}{1.2}
\begin{table}[h!]
{\footnotesize
\setlength{\tabcolsep}{1pt}
\begin{center}
\begin{tabular}{c|cccccccccccc}
$\lambda_{\sss t}$, $\lambda_{\sss Z}$, $\lambda_{\sss t}$, $\lambda_{\sss Z}$&SM&$\Op{\phi D}$&$\Op{\phi d}$&$\Op{\phi B}$&$\Op{\phi W}$&$\Op{\phi WB}$&$\Op{t \phi}$&$\Op{tB}$&$\Op{tW}$&$\Op{\phi Q}^{\sss (1)}$&$\Op{\phi Q}^{\sss (3)}$&$\Op{\phi t}$\tabularnewline
\hline
$-, 0, -, 0$&$s^0$&$s^0$&$s^0$&$-$&$-$&$-$&$s^0$&$s^0$&$s^0$&$s^0$&$s^0$&$s^0$\tabularnewline
$\pm, 0, \mp, 0$&$\frac{1}{\sqrt{s}}$&$\sqrt{-t} m_t$&$\sqrt{-t} m_t$&$-$&$-$&$-$&$\sqrt{-t} v$&$-$&$-$&$\sqrt{-t} m_t$&$\sqrt{-t} m_t$&$ \sqrt{-t} m_t$\tabularnewline
$+, 0, +, 0$&$s^0$&$s^0$&$s^0$&$-$&$-$&$-$&$s^0$&$s^0$&$s^0$&$s^0$&$s^0$&$s^0$\tabularnewline
\hline
$\pm, 0, \mp, \mp$&$\frac{1}{s}$&$s^0$&$s^0$&$s^0$&$s^0$&$s^0$&$s^0$&$\sqrt{s (s+t)}$&$\sqrt{s (s+t)}$&$s^0$&$s^0$&$s^0$\tabularnewline
$\pm, \pm, \mp, 0$&$\frac{1}{s}$&$s^0$&$s^0$&$s^0$&$s^0$&$s^0$&$s^0$&$\sqrt{s (s+t)}$&$\sqrt{s (s+t)}$&$s^0$&$s^0$&$s^0$\tabularnewline
$\pm, 0, \pm, \mp$&$\frac{1}{\sqrt{s}}$&$-$&$-$&$-$&$-$&$-$&$-$&$\sqrt{-t} m_t$&$\sqrt{-t} m_t$&$-$&$-$&$-$\tabularnewline
$\pm, \mp, \pm, 0$&$\frac{1}{\sqrt{s}}$&$-$&$-$&$-$&$-$&$-$&$-$&$\sqrt{-t} m_t$&$\sqrt{-t} m_t$&$-$&$-$&$-$\tabularnewline
$\pm, 0, \mp, \pm$&$s^0$&$s^0$&$s^0$&$s^0$&$s^0$&$s^0$&$s^0$&$s^0$&$s^0$&$s^0$&$s^0$&$s^0$\tabularnewline
$\pm, \mp, \mp, 0$&$s^0$&$s^0$&$s^0$&$s^0$&$s^0$&$s^0$&$s^0$&$s^0$&$s^0$&$s^0$&$s^0$&$s^0$\tabularnewline
$\pm, 0, \pm, \pm$&$\frac{1}{\sqrt{s}}$&$-$&$-$&$-$&$-$&$-$&$-$&$-$&$-$&$-$&$-$&$-$\tabularnewline
$\pm, \pm, \pm, 0$&$\frac{1}{\sqrt{s}}$&$-$&$-$&$-$&$-$&$-$&$-$&$-$&$-$&$-$&$-$&$-$\tabularnewline
\hline
$-, \pm, -, \pm$&$s^0$&$s^0$&$-$&$-$&$-$&$s^0$&$-$&$s^0$&$s^0$&$s^0$&$s^0$&$-$\tabularnewline
$+, \pm, +, \pm$&$s^0$&$s^0$&$-$&$-$&$-$&$s^0$&$-$&$s^0$&$s^0$&$-$&$-$&$s^0$\tabularnewline
$-, \pm, -, \mp$&$\frac{1}{s}$&$-$&$-$&$s^0$&$s^0$&$s^0$&$-$&$s^0$&$s^0$&$-$&$-$&$-$\tabularnewline
$+, \pm, +, \mp$&$\frac{1}{s}$&$-$&$-$&$s^0$&$s^0$&$s^0$&$-$&$s^0$&$s^0$&$-$&$-$&$-$\tabularnewline
$\pm, \pm, \mp, \mp$&$-$&$-$&$-$&$\sqrt{-t} m_t$&$\sqrt{-t} m_t$&$\sqrt{-t} m_t$&$-$&$\sqrt{-t} m_W$&$\sqrt{-t} m_W$&$-$&$-$&$-$\tabularnewline
$\pm, \mp, \pm, \mp,$&$\frac{1}{\sqrt{s}}$&$-$&$-$&$\sqrt{-t} m_t$&$\sqrt{-t} m_t$&$\sqrt{-t} m_t$&$-$&$-$&$-$&$-$&$-$&$-$\tabularnewline
$\pm, -, \mp, -$&$\frac{1}{\sqrt{s}}$&$-$&$-$&$-$&$-$&$-$&$-$&$\sqrt{-t} m_W$&$ \sqrt{-t} m_W$&$-$&$-$&$-$\tabularnewline
$\pm, +, \mp, +$&$\frac{1}{\sqrt{s}}$&$-$&$-$&$-$&$-$&$-$&$-$&$\sqrt{-t} m_W$&$\sqrt{-t} m_W$&$-$&$-$&$-$\tabularnewline


\end{tabular}
\end{center}
}
\caption{\label{tab:tztz}
Helicity amplitudes in the high-energy limit ($s\sim-t\gg v$) for $t Z\to t Z$ scattering.  Overall multiplicative factors are dropped and only the schematic energy
dependence is retained for the SM contribution.  Helicity entries marked by `$\ast$' indicate any combination of $\pm,\mp$. See Section~\ref{subsubsec:tztz_tzta_tata} for a discussion of the results.
}
\end{table}
\renewcommand{\arraystretch}{1.}

    \renewcommand{\arraystretch}{1.2}
\begin{table}[h!]
\begin{center}
\begin{tabular}{c|cccccc}
$\lambda_{\sss t}$, $\lambda_{\sss Z}$, $\lambda_{\sss t}$, $\lambda_{\sss \gamma}$&SM&$\Op{\phi B}$&$\Op{\phi W}$&$\Op{\phi WB}$&$\Op{tB}$&$\Op{tW}$\tabularnewline
\hline
$\pm, 0, \mp, \mp$&$\frac{1}{s}$&$s^0$&$s^0$&$s^0$&$\sqrt{s(s+t)}$&$\sqrt{s (s+t)}$\tabularnewline
$\pm, 0, \pm, \mp$&$\frac{1}{\sqrt{s}}$&$-$&$-$&$-$&$\sqrt{-t} m_t$&$\sqrt{-t} m_t$\tabularnewline
$\pm, 0, \pm, \pm$&$\frac{1}{\sqrt{s}}$&$-$&$-$&$-$&$-$&$-$\tabularnewline
$\pm, 0, \mp, \pm$&$s^0$&$s^0$&$s^0$&$s^0$&$s^0$&$s^0$\tabularnewline
\hline
$\pm, \pm, \pm, \pm$&$s^0$&$-$&$-$&$s^0$&$s^0$&$s^0$\tabularnewline
$\pm, \mp, \pm, \mp$&$s^0$&$-$&$-$&$s^0$&$s^0$&$s^0$\tabularnewline
$\pm, \pm, \pm, \mp$&$\frac{1}{s}$&$s^0$&$s^0$&$s^0$&$s^0$&$s^0$\tabularnewline
$\pm, \mp, \pm, \pm$&$\frac{1}{s}$&$s^0$&$s^0$&$s^0$&$s^0$&$s^0$\tabularnewline
$\pm, \pm, \mp, \pm$&$\frac{1}{\sqrt{s}}$&$-$&$-$&$-$&$\sqrt{-t} m_W$&$\sqrt{-t} m_W$\tabularnewline
$\mp, \pm, \pm, \pm$&$\frac{1}{\sqrt{s}}$&$-$&$-$&$-$&$\sqrt{-t} m_W$&$\sqrt{-t} m_W$\tabularnewline
$\pm, \pm, \mp, \mp$&$-$&$\sqrt{-t} m_t$&$\sqrt{-t} m_t$&$\sqrt{-t} m_t$&$\sqrt{-t} m_W$&$\sqrt{-t} m_W$\tabularnewline
$\pm, \mp, \mp, \pm$&$\frac{1}{\sqrt{s}}$&$\sqrt{-t} m_t$&$\sqrt{-t} m_t$&$\sqrt{-t} m_t$&$-$&$-$\tabularnewline
\end{tabular}
\end{center}
\caption{\label{tab:tzta}
Helicity amplitudes in the high-energy limit ($s\sim-t\gg v$) for $t Z\to t \gamma$ scattering.  Overall multiplicative factors are dropped and only the schematic energy
dependence is retained for the SM contribution.  Helicity entries marked by `$\ast$' indicate any combination of $\pm,\mp$. See Section~\ref{subsubsec:tztz_tzta_tata} for a discussion of the results.
}
\end{table}
\renewcommand{\arraystretch}{1.}

    \renewcommand{\arraystretch}{1.2}
\begin{table}[h!]
\begin{center}
\begin{tabular}{c|cccccc}
$\lambda_{\sss t}$, $\lambda_{\sss \gamma}$, $\lambda_{\sss t}$, $\lambda_{\sss \gamma}$&SM&$\Op{\phi B}$&$\Op{\phi W}$&$\Op{\phi WB}$&$\Op{tB}$&$\Op{tW}$\tabularnewline
\hline
$\pm, \pm, \pm, \pm$&$s^0$&$-$&$-$&$s^0$&$s^0$&$s^0$\tabularnewline
$\pm, \pm, \pm, \mp$&$\frac{1}{s}$&$s^0$&$s^0$&$s^0$&$s^0$&$s^0$\tabularnewline
$\pm, \mp, \pm, \pm$&$\frac{1}{s}$&$s^0$&$s^0$&$s^0$&$s^0$&$s^0$\tabularnewline
$\pm, \pm, \mp, \pm$&$\frac{1}{\sqrt{s}}$&$-$&$-$&$-$&$\sqrt{-t} m_W$&$\sqrt{-t} m_W$\tabularnewline
$\mp, \pm, \pm, \pm$&$\frac{1}{\sqrt{s}}$&$-$&$-$&$-$&$\sqrt{-t} m_W$&$\sqrt{-t} m_W$\tabularnewline
$\pm, \pm, \mp, \mp$&$-$&$\sqrt{-t} m_t$&$\sqrt{-t} m_t$&$\sqrt{-t} m_t$&$\sqrt{-t} m_W$&$\sqrt{-t} m_W$\tabularnewline
$\pm, \mp, \pm, \mp$&$s^0$&$-$&$-$&$s^0$&$-$&$-$\tabularnewline
$\pm, \mp, \mp, \pm$&$\frac{1}{\sqrt{s}}$&$\sqrt{-t} m_t$&$\sqrt{-t} m_t$&$\sqrt{-t} m_t$&$-$&$-$\tabularnewline
\end{tabular}
\end{center}
\caption{\label{tab:tata}
Helicity amplitudes in the high-energy limit ($s\sim-t\gg v$) for $t \gamma\to t \gamma$ scattering. Overall multiplicative factors are dropped and only the schematic energy
dependence is retained for the SM contribution.  Helicity entries marked by `$\ast$' indicate any combination of $\pm,\mp$. See Section~\ref{subsubsec:tztz_tzta_tata} for a discussion of the results.
}
\end{table}
\renewcommand{\arraystretch}{1.}

    \renewcommand{\arraystretch}{1.2}
\begin{table}[h!]
{\footnotesize
\setlength{\tabcolsep}{1pt}
\begin{center}
\begin{tabular}{c|ccccccccccc}
$\lambda_{\sss t}$, $\lambda_{\sss Z}$, $\lambda_{\sss t}$&SM&$\Op{\phi D}$&$\Op{\phi B}$&$\Op{\phi W}$&$\Op{\phi WB}$&$\Op{t \phi}$&$\Op{tB}$&$\Op{tW}$&$\Op{\phi Q}^{\sss (1)}$&$\Op{\phi Q}^{\sss (3)}$&$\Op{\phi t}$\tabularnewline
\hline
$-, 0, -$&$s^0$&$s^0$&$s^0$&$s^0$&$s^0$&$s^0$&$s^0$&$s^0$&$\sqrt{s (s+t)}$&$\sqrt{s (s+t)}$&$s^0$\tabularnewline
$-, 0, +$&$\frac{1}{\sqrt{s}}$&$\sqrt{-t} m_t$&$-$&$-$&$-$&$\sqrt{-t} v$&$\frac{s m_W}{\sqrt{-t}}$&$\frac{s m_W}{\sqrt{-t}}$&$\sqrt{-t} m_t$&$\sqrt{-t} m_t$&$-$\tabularnewline
$+, 0, -$&$\frac{1}{\sqrt{s}}$&$\sqrt{-t} m_t$&$-$&$-$&$-$&$\sqrt{-t} v$&$\frac{s m_W}{\sqrt{-t}}$&$\frac{ s m_W}{\sqrt{-t}}$&$-$&$-$&$\sqrt{-t} m_t$\tabularnewline
$+, 0, +$&$s^0$&$s^0$&$s^0$&$-$&$s^0$&$s^0$&$s^0$&$s^0$&$s^0$&$s^0$&$\sqrt{s (s+t)}$\tabularnewline
\hline
$-, -, -$&$\frac{1}{\sqrt{s}}$&$-$&$\frac{s m_W}{\sqrt{-t}}$&$\frac{s m_W}{\sqrt{-t}}$&$\frac{ s m_W}{\sqrt{-t}}$&$-$&$ \sqrt{-t} m_t$&$\sqrt{-t} m_t$&$\sqrt{-t} m_W$&$\sqrt{-t} m_W$&$-$\tabularnewline
$-, -, +$&$-$&$s^0$&$s^0$&$-$&$s^0$&$s^0$&$\sqrt{s (s+t)}$&$ \sqrt{s (s+t)}$&$s^0$&$s^0$&$s^0$\tabularnewline
$-, +, -$&$\frac{1}{\sqrt{s}}$&$-$&$\frac{m_W (s+t)}{\sqrt{-t}}$&$\frac{m_W (s+t)}{\sqrt{-t}}$&$\frac{m_W (s+t)}{\sqrt{-t}}$&$-$&$\sqrt{-t} m_t$&$\sqrt{-t} m_t$&$-$&$-$&$-$\tabularnewline
$-, +, +$&$s^0$&$s^0$&$s^0$&$s^0$&$s^0$&$s^0$&$s^0$&$s^0$&$s^0$&$s^0$&$s^0$\tabularnewline
$+, -, -$&$s^0$&$s^0$&$s^0$&$-$&$s^0$&$s^0$&$s^0$&$s^0$&$s^0$&$s^0$&$s^0$\tabularnewline
$+, -, +$&$\frac{1}{\sqrt{s}}$&$-$&$\frac{m_W (s+t)}{\sqrt{-t}}$&$-$&$\frac{m_W (s+t)}{\sqrt{-t}}$&$-$&$\sqrt{-t} m_t$&$\sqrt{-t} m_t$&$-$&$-$&$-$\tabularnewline
$+, +, -$&$\frac{1}{s}$&$s^0$&$s^0$&$s^0$&$s^0$&$s^0$&$\sqrt{s (s+t)}$&$\sqrt{s (s+t)}$&$s^0$&$s^0$&$s^0$\tabularnewline
$+, +, +$&$\frac{1}{\sqrt{s}}$&$-$&$\frac{s m_W}{\sqrt{-t}}$&$-$&$\frac{s m_W}{\sqrt{-t}}$&$-$&$ \sqrt{-t} m_t$&$ \sqrt{-t} m_t$&$-$&$-$&$\sqrt{-t} m_W$\tabularnewline
\end{tabular}
\end{center}
}
\caption{\label{tab:tzth}
Helicity amplitudes in the high-energy limit ($s\sim-t\gg v$) for $t Z\to t h$ scattering.  Overall multiplicative factors are dropped and only the schematic energy
dependence is retained for the SM contribution.  Helicity entries marked by `$\ast$' indicate any combination of $\pm,\mp$. See Section~\ref{subsubsec:tzth_tath} for a discussion of the results.
}
\end{table}
\renewcommand{\arraystretch}{1.}

    \renewcommand{\arraystretch}{1.2}
\begin{table}[h!]
\begin{center}
\begin{tabular}{c|cccccc}
$\lambda_{\sss t}$, $\lambda_{\sss \gamma}$, $\lambda_{\sss t}$&SM&$\Op{\phi B}$&$\Op{\phi W}$&$\Op{\phi WB}$&$\Op{tB}$&$\Op{tW}$\tabularnewline
\hline
$-, -, -$&$\frac{1}{\sqrt{s}}$&$\frac{s m_W}{\sqrt{-t}}$&$\frac{s m_W}{\sqrt{-t}}$&$\frac{s m_W}{\sqrt{-t}}$&$\sqrt{-t} m_t$&$\sqrt{-t} m_t$\tabularnewline
$-, -, +$&$\frac{1}{s}$&$s^0$&$-$&$s^0$&$\sqrt{s(s+t)}$&$\sqrt{s (s+t)}$\tabularnewline
$-, +, -$&$\frac{1}{\sqrt{s}}$&$\frac{m_W (s+t)}{\sqrt{-t}}$&$\frac{m_W (s+t)}{\sqrt{-t}}$&$\frac{m_W (s+t)}{\sqrt{-t}}$&$\sqrt{-t} m_t$&$\sqrt{-t} m_t$\tabularnewline
$-, +, +$&$s^0$&$s^0$&$s^0$&$s^0$&$s^0$&$s^0$\tabularnewline
$+, -, -$&$s^0$&$s^0$&$-$&$s^0$&$s^0$&$s^0$\tabularnewline
$+, -, +$&$\frac{1}{\sqrt{s}}$&$\frac{m_W (s+t)}{\sqrt{-t}}$&$-$&$\frac{m_W (s+t)}{\sqrt{-t}}$&$\sqrt{-t} m_t$&$\sqrt{-t} m_t$\tabularnewline
$+, +, -$&$\frac{1}{s}$&$s^0$&$s^0$&$s^0$&$\sqrt{s(s+t)}$&$\sqrt{s (s+t)}$\tabularnewline
$+, +, +$&$\frac{1}{\sqrt{s}}$&$\frac{s m_W}{\sqrt{-t}}$&$-$&$\frac{s m_W}{\sqrt{-t}}$&$\sqrt{-t} m_t$&$\sqrt{-t} m_t$\tabularnewline
\end{tabular}
\end{center}
\caption{\label{tab:tath}
Helicity amplitudes in the high-energy limit ($s\sim-t\gg v$) for $t \gamma\to t h$ scattering. Overall multiplicative factors are dropped and only the schematic energy
dependence is retained for the SM contribution. Helicity entries marked by `$\ast$' indicate any combination of $\pm,\mp$. See Section~\ref{subsubsec:tzth_tath} for a discussion of the results.
}
\end{table}
\renewcommand{\arraystretch}{1.}

\clearpage
    \renewcommand{\arraystretch}{1.2}
\begin{table}[h!]
\begin{center}
\begin{tabular}{c|cccc}
$\lambda_{\sss t}$, $\lambda_{\sss t}$,&SM&$\Op{\phi D}$&$\Op{\phi d}$&$\Op{t \phi}$\tabularnewline
\hline
$\pm, \pm$&$s^0$&$s^0$&$s^0$&$s^0$\tabularnewline
$\pm, \mp$&$\frac{1}{\sqrt{s}}$&$\sqrt{-t} m_t$&$\sqrt{-t} m_t$&$\sqrt{-t} v$\tabularnewline
\end{tabular}
\end{center}
\caption{\label{tab:thth}
Helicity amplitudes in the high-energy limit ($s\sim-t\gg v$) for $t h\to t h$ scattering. Overall multiplicative factors are dropped and only the schematic energy
dependence is retained for the SM contribution. Helicity entries marked by `$\ast$' indicate any combination of $\pm,\mp$. See Section~\ref{subsubsec:thth} for a discussion of the results.
}
\end{table}
\renewcommand{\arraystretch}{1.}

\bibliography{refs.bib}

\end{document}